\documentclass[final]{siamltex}
\usepackage{amsmath}
\usepackage{amsfonts}
\usepackage{amssymb}

\usepackage{graphicx}

\usepackage{url}
\usepackage{xcolor}
\usepackage{fullpage}
\usepackage{bm}
\usepackage{mathrsfs}

\renewcommand{\hat}{\widehat}

\newcommand{\om}{\omega}

\newcommand{\la}{\lambda}

\newcommand{\cA}{\mathcal{A}}
\newcommand{\cS}{\mathcal{S}}

\newcommand{\cR}{\mathcal{R}}
\newcommand{\EE}{\mathbb{E}}

\newcommand{\brho}{\bm{\rho}}

\newcommand{\bd}{\bm{d}}
\newcommand{\bu}{\bm{u}}

\newcommand{\vy}{\vec{\bm{y}}}
\newcommand{\vx}{\vec{\bm{x}}}
\newcommand{\vz}{\vec{\bm{z}}}
\newcommand{\by}{\bm{y}}
\newcommand{\bz}{\bm{z}}
\newcommand{\bx}{\bm{x}}
\newcommand{\ve}{\vec{\bm{e}}}

\newcommand{\vH}{\vec{\bm{H}}}
\newcommand{\w}{{\it w}}
\newcommand{\cbZ}{\overline{\bm{\zeta}}}
\newcommand{\ctZ}{\widetilde{\bm{\zeta}}}
\newcommand{\be}{\beta}

\newcommand{\tom}{\widetilde \omega}
\newcommand{\tbx}{\widetilde {\bx}}
\newcommand{\tbz}{\widetilde{\bz}}
\newcommand{\cbz}{\overline{\bz}}
\newcommand{\tz}{\widetilde{z}}
\newcommand{\cz}{\overline{z}}

\makeatletter
\newsavebox\myboxA
\newsavebox\myboxB
\newlength\mylenA

\newcommand*\xoverline[2][1.3]{%
    \sbox{\myboxA}{$\m@th#2$}%
    \setbox\myboxB\null
    \ht\myboxB=\ht\myboxA%
    \dp\myboxB=\dp\myboxA%
    \wd\myboxB=#1\wd\myboxA
    \sbox\myboxB{$\m@th\overline{\copy\myboxB}$}
    \setlength\mylenA{\the\wd\myboxA}
    \addtolength\mylenA{-\the\wd\myboxB}%
    \ifdim\wd\myboxB<\wd\myboxA%
       \rlap{\hskip 0.3\mylenA\usebox\myboxB}{\usebox\myboxA}%
    \else
        \hskip -0.3\mylenA\rlap{\usebox\myboxA}{\hskip 0.5\mylenA\usebox\myboxB}%
    \fi}
\makeatother

\begin{document}

\title{Imaging in random media with convex optimization}
\author{Liliana Borcea \and Ilker Kocyigit \footnotemark[1]}
\renewcommand{\thefootnote}{\fnsymbol{footnote}}
\footnotetext[1]{Department of Mathematics, University of Michigan,
  Ann Arbor, MI 48109-1043. \\\hspace{0.3in}Email: borcea@umich.edu \&
  ilkerk@umich.edu} \maketitle \date{today}
\begin{abstract}
We study an inverse problem for the wave equation where localized wave
sources in random scattering media are to be determined from time
resolved measurements of the waves at an array of receivers. The
sources are far from the array, so the measurements are affected by
cumulative scattering in the medium, but they are not further than a
transport mean free path, which is the length scale characteristic of
the onset of wave diffusion that prohibits coherent imaging. The
inversion is based on the Coherent Interferometric (CINT) imaging
method which mitigates the scattering effects by introducing an
appropriate smoothing operation in the image formation. This smoothing
stabilizes statistically the images, at the expense of their
resolution.  We complement the CINT method with a convex ($l_1$)
optimization in order to improve the source localization and obtain
quantitative estimates of the source intensities.  We analyze the
method in a regime where scattering can be modeled by large random
wavefront distortions, and quantify the accuracy of the inversion in
terms of the spatial separation of individual sources or clusters of
sources. The theoretical predictions are demonstrated with numerical
simulations.
  
\end{abstract}

\section{{Introduction}}
\label{sect:intro} 
Waves measured by a collection of nearby sensors, called an array of
receivers, carry information about their source and the medium through
which they travel. We consider a typical remote sensing regime with
sources of small (point-like) support, and study the inverse problem
of determining them from the array measurements.

When the waves travel in a known and non-scattering (e.g. homogeneous) medium, the
sources can be localized with reverse time migration
\cite{Biondi,bleistein2001mathematics} also known as backprojection
\cite{munson1983tomographic}.  This estimates the source locations as
the peaks of the image formed by superposing the array recordings
delayed by travel times from the receivers to the imaging points. The
accuracy of the estimates depends on the array aperture, the distance of the sources from the array, 
and the temporal support of the signals emitted by the sources. It may be improved under certain conditions by using
$l_1$ optimization, which seeks to invert the linear mapping from
supposedly sparse vectors of the discretized source amplitude on some
mesh, to the array measurements. The fast growing literature of
imaging with $l_1$ optimization in homogeneous media includes
compressed sensing studies such as
\cite{fannjiang2010compressed,fannjiang2010compressive}, synthetic
radar imaging studies like
\cite{anitori2013design,borcea2016synthetic}, array imaging studies
like \cite{chai2013robust}, and the resolution study
\cite{borcea2015resolution}.

In this paper we assume that the waves travel in heterogeneous media
with fluctuations of the wave speed caused by numerous
inhomogeneities. The amplitude of the fluctuations is small, meaning
that a single inhomogeneity is a weak scatterer. However, there are
many inhomogeneities that interact with the waves on their way from
the sources to the receivers, and their scattering effect
accumulates. Because in applications it is impossible to know the
inhomogeneities in detail, and these cannot be estimated from the
array measurements as part of the inversion, the fluctuations of the
wave speed are uncertain. We model this uncertainty with a random
process, and thus study inversion in random media. In this stochastic
framework, the actual heterogeneous medium in which the waves
propagate  is one realization of the random model. 
The data measured at the array are uncertain and the question is how to mitigate 
the uncertainty to get images that are  
robust with respect to arbitrary  medium realizations, i.e., they are statistically stable.

The mitigation of uncertainty in the wave propagation model becomes
important when the sources are further than a few scattering mean free
paths from the array. The scattering mean free path is the length
scale on which the waves randomize \cite{van1999multiple}, meaning
that their fluctuations from one medium realization to another are
large in comparison with their coherent (statistical expectation)
part.  The random wave distortions registered at the array are very
different from additive and uncorrelated noise assumed usually in
inversion. They are more difficult to mitigate and lead to poor and
unreliable source reconstructions by coherent methods like reverse
time migration or standard $l_1$ optimization. The Coherent
Interferometric (CINT) method
\cite{borcea2006adaptive,borcea2011enhanced} is designed to deal
efficiently with such random distortions, as long as there is some
residual coherence in the array measurements. This holds when the
sources are separated from the array by distances (ranges) that are
large with respect to the scattering mean free path, but do not exceed
a transport mean free path, which is the distance at which the waves
forget their initial direction \cite{van1999multiple}. The transport
mean free path defines the range limit of applicability of coherent
inversion methods. Beyond it only incoherent methods based on
transport or diffusion equations \cite{arridge2009optical} can be
used.

In this paper we assume a scattering regime where the CINT method is
useful. It forms images by superposing cross-correlations of the array
measurements, delayed by travel times between the receivers and the
imaging points. As shown in
\cite{borcea2006adaptive,borcea2007asymptotics,borcea2011enhanced},
the cross-correlations must be computed locally, in appropriate time
windows, and over limited receiver offsets. This introduces a
smoothing in the CINT image formation, which is essential for
stabilizing statistically the images, at the expense of
resolution. The larger the random distortions of the array
measurements, the more smoothing is needed and the worse the
resolution \cite{borcea2006adaptive,borcea2007asymptotics}. Thus, it
is natural to ask if it is possible to improve the source localization
by using the prior information that the sources have small support.

We show that under generic conditions, the CINT imaging function is
approximately a discrete convolution of the vector of source
intensities discretized on the imaging mesh, with a blurring kernel.
To reconstruct the sources we seek to undo the convolution using
convex ($l_1$) optimization.  We present an analysis of the method in
a scattering regime where the random medium effects on the array
measurements can be modeled by large wavefront distortions, as assumed
in adaptive optics \cite{beckers1993adaptive}.  We derive from first
principles the CINT blurring kernel in this regime, and state the
inversion problem as an $l_1$ optimization. We also quantify the
quality of the reconstruction with error estimates that depend on the
separation of the sources, or of clusters of sources,
but are  independent of the source placement on or off the imaging mesh, as long as the sources 
are sufficiently far apart. 
The analysis
shows that we can expect almost exact reconstructions when the sources
are further apart than the CINT resolution limits. This is similar to
the super-resolution results in \cite{candes2014towards}, that show
that one dimensional discrete convolutions can be undone by convex
optimization, assuming that the minimum distance between the points in
the support of the unknown vectors is $2/f_c$, where $f_c$ is the
largest ``frequency'' in the Fourier transform of the convolution
kernel. When the sources are clustered together, the $l_1$
reconstruction is not guaranteed to be close to the true vector of
source intensities in the point-wise sense. However, we show that its
support is in the vicinity of the clusters, and its entries are
related to the average source intensities there.

The paper is organized as follows: We begin in section \ref{sect:form}
with the formulation of the inverse problem as an $l_1$
optimization. The analysis of the method is in sections
\ref{sect:anal} - \ref{sect:res}, and we demonstrate its performance
with numerical simulations in section \ref{sect:num}. We end with a
summary in section \ref{sect:sum}.

\section{Formulation of the inverse problem}
\label{sect:form}

\begin{figure}[t]
\vspace{-0.15in}
\centerline{\includegraphics[width = 0.7
    \textwidth]{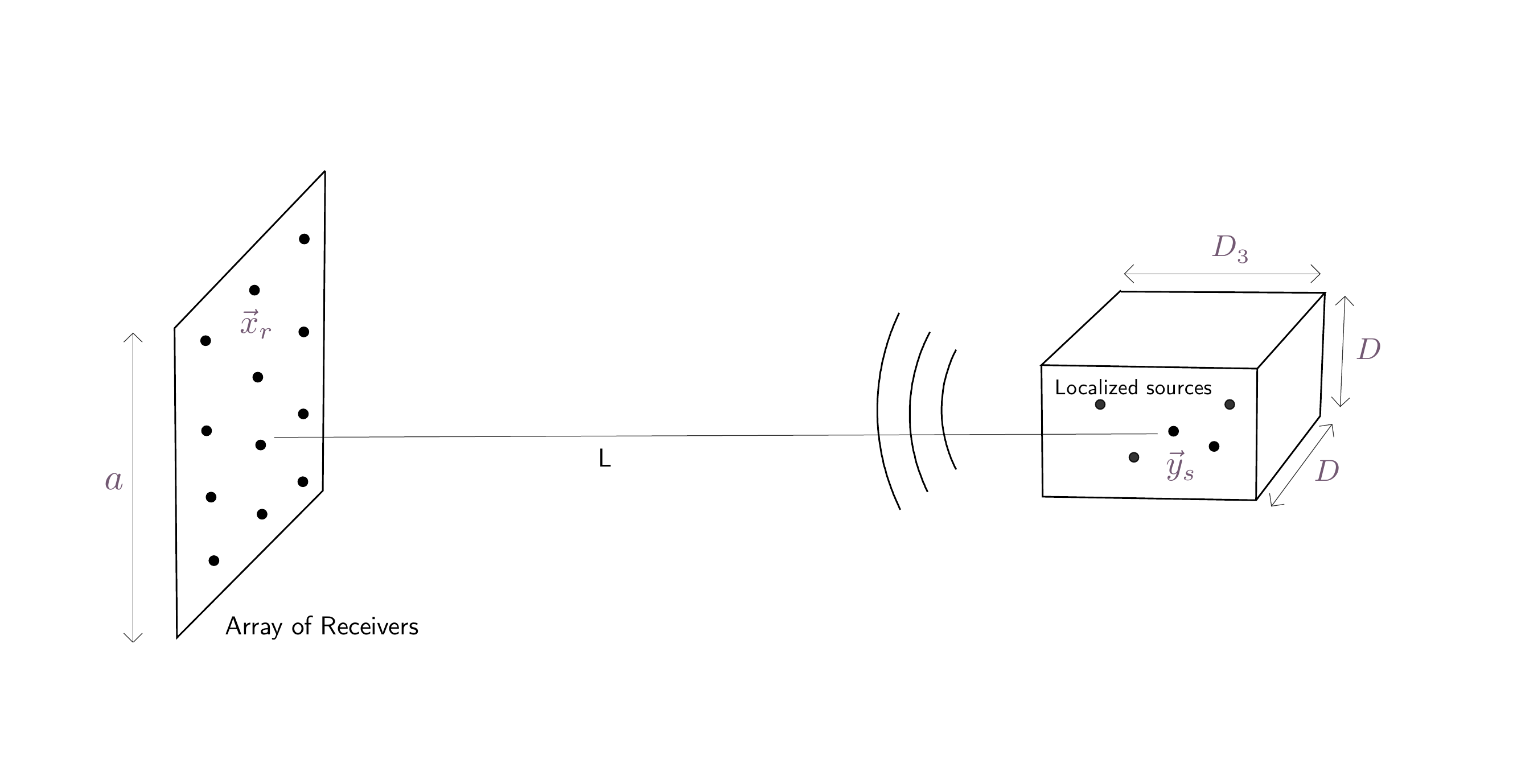}}
\vspace{-0.45in}
\caption{Schematic for the inversion setup with an array of receivers
  that is planar square of side $a$. The range direction is orthogonal
  to the array aperture. The sources are at ranges of order $L$ in an
  imaging region $\mathfrak{D}$ modeled as a rectangular prism with
  size $D_3$ in range and $D$ in cross-range. }
\label{fig:setup}
\end{figure}

Consider the inversion setup illustrated in Figure \ref{fig:setup},
where $N_s$ sources located at $\vy_s$, for $s = 1, \ldots, N_s$, emit
signals $f_s(t)$ that generate sound waves recorded at a remote array
of receivers placed at $\vx_r$, for $r = 1, \ldots, N_r$. For
simplicity we assume that the array aperture is planar and square, of
side $a$. This allows us to introduce a system of coordinates centered
at the array, with range direction orthogonal to the array, and
cross-range plane parallel to the array. In this system of coordinates
we have $ \vx_r = (\bx_r,0)$, with cross-range vectors $\bx_r =
(x_{r,1},x_{r,2})$ satisfying $|x_{r,1}|,|x_{r,2}| \le a/2$.  The
sources are at $\vy_s = (\by_s,y_{s,3})$, with range coordinates
$y_{s,3}$ of order $L$, satisfying $L \gg a$, and two dimensional
cross-range vectors $\by_s$.

In general, the signals $f_s(t)$ emitted by the sources may be pulses, chirps or
even noise-like, with Fourier transforms
\begin{equation}
\hat f_s(\om) = \int_{-\infty}^\infty dt \, f_s(t) e^{i \om t}
\label{eq:f2}
\end{equation} 
supported in the frequency interval $[\om_o-\pi B,\om_o + \pi B]$,
where $\om_o$ is the central frequency and $B$ is the
bandwidth.  We denote the recorded waves by $p(\vx_r,t)$, and use the
linearity of the wave equation to write
\begin{equation}
p(\vx_r,t) = \int_{\om_o-\pi B}^{\om_o + \pi B} \frac{d \om}{2 \pi} \,
\hat p(\vx_r, \om) e^{-i \om t}, \qquad \hat p(\vx_r,\om) =
\sum_{s=1}^{N_s} \hat f_s(\om) \hat G(\vx_r,\vy_s,\om) + \hat
n(\vx_r,\om), 
\label{eq:f1}
\end{equation}
for $r = 1, \ldots, N_r$. Here the Green's function $\hat G$ models
the propagation of time harmonic waves in the medium, and $\hat n$
denotes additive and uncorrelated noise. The inverse problem is to
determine the sources from these array measurements.

\subsection{Imaging in homogeneous media}
The Green's function in media with constant speed $c_o$ is 
\begin{equation}
  \hat G_o(\vx_r,\vy_s,\om) = \frac{\exp[i \om \tau(\vx_r,\vy_s)]}{4
    \pi |\vx_r-\vy_s|},
  \label{eq:Go}
\end{equation}
where 
\begin{equation}
\tau(\vx_r,\vy_s) = {|\vx_r-\vy_s|}/{c_o}
\label{eq:tau}
\end{equation}
is the travel time from the source at $\vy_s$ to the receiver at
$\vx_r$. The measurements are of the form
\begin{equation}
p_o(\vx_r,t) = \sum_{s=1}^{N_s} \frac{f_s\left(t -
  \tau(\vx_r,\vy_s)\right)}{4 \pi |\vx_r-\vy_s|} + n(\vx_r,t),
\label{eq:f4}
\end{equation}
and in reverse time migration they are synchronized using travel time
delays with respect to a presumed source location at a search point
$\vy$, and then superposed to form an image
\begin{align}
\mathcal{J}_o(\vy) = \sum_{r=1}^{N_r}
p_o\left(\vx_r,\tau(\vx_r,\vy)\right).
\label{eq:f5}
\end{align}
The resolution limits of the imaging function \eqref{eq:f5} are well
known. The cross-range resolution is of order $\la_o L/a$, where
$\la_o = 2 \pi c_o/\om_o$ is the central wavelength, and the range
resolution is inverse proportional to the temporal support of
$f_s(t)$, which determines the precision of the travel time
estimation.  If the signals $f_s(t)$ are pulses, their temporal
support is of order $1/B$, and the range resolution is of order
$c_o/B$.  If they are chirps or other long signals that are known,
then they must be compressed in time by cross-correlation (matched
filtering) with the time reversed $f_s(t)$ to achieve the $c_o/B$
range resolution \cite{Curlander}.  If the signals are unknown and
noise-like, then imaging must be based on cross-correlations of the
array measurements, like in CINT.

We refer to
\cite{borcea2015resolution} for the formulation of the inverse source
problem as an $l_1$ optimization, and recall from there that the
resolution limits $\la_o L/a$ and $c_o/B$ also play a role in the
successful recovery of the presumed sparse source support.

\subsection{Coherent interferometric imaging}
The imaging function $\mathcal{J}_o(\vy)$ does not work well in random
media at ranges $L$ that exceed a few scattering mean free paths. This
is because the measurements have large random distortions that are
very different than the additive noise $n(\vx_r,t)$, and cannot be
reduced by simply summing over the receivers as in \eqref{eq:f5}. To
mitigate these distortions we image with the CINT function
\begin{align} 
\mathcal{J}(\vy) = \int_{-\infty}^\infty d \om \int_{-\infty}^\infty d
\tom \, \hat \phi\left(\frac{\tom}{\Omega}\right) \sum_{r=1}^{N_r}
\sum_{r' = 1}^{N_r} \psi\left(\frac{|\bx_r-\bx_{r'}|}{X(\om)}\right)
\hat p (\vx_r,\om + \tom/2) \overline{\hat p(\vx_{r'},\om-\tom/2)}
\nonumber \\ \times \exp[-i (\om + \tom/2) \tau(\vx_r,\vy) + i (\om -
  \tom/2) \tau(\vx_{r'},\vy)],
\label{eq:f6}
\end{align}
where $\hat \phi$ and $\psi$ are smooth window functions of dimensionless
argument and support of order one, and the domain of integration is
restricted by the finite bandwidth that supports the measurements,
\[
\om \pm \tom/2 \in [\om_o-\pi B,\om_o + \pi B].
\]
As in reverse time migration, the travel times are used in
\eqref{eq:f6} to synchronize the waves due to a presumed source at the
search location $\vy$. However, the image is formed by superposing
cross-correlations of the array measurements $p(\vx_r,t)$, instead of the
measurements themselves. The cross-correlations are convolutions of
$p(\vx_r,t)$ with the time reversed $p(\vx_{r'},t)$, for $r, r' = 1,
\ldots, N_r$. The time reversal appears as complex conjugation in the
frequency domain, denoted with the bar in \eqref{eq:f6}. The time
window
\begin{equation}
  \label{eq:defphi}
\phi\left(\Omega t\right) = \frac{1}{2 \pi
  \Omega}\int_{-\infty}^\infty d \om\,\hat
\phi\left(\frac{\om}{\Omega}\right)e^{-i \om t}
\end{equation}
and spatial window $\psi(|\bx|/X)$ ensure that the cross-correlations
are computed locally, over receiver offsets that do not exceed the
distance $X$, and over time offsets of order $1/\Omega$.  These
threshold parameters account for the fact that scattering in random
media decorrelates statistically the waves at frequencies separated by
more than $\Omega_d$, the decoherence frequency, and points separated
by more than $X_d$, the decoherence length. We refer to
\cite{borcea2007asymptotics,borcea2011enhanced} and the next section
for more details. Here it suffices to recall that \eqref{eq:f6} is
robust\footnote{Robustness refers to statistical stability of the
  image with respect to the realizations of the random medium.  It
  means that the standard deviation of $\eqref{eq:f6}$ is small with
  respect to its expectation near the peaks.}  when $X \lesssim X_d$
and $\Omega \lesssim \Omega_d$, and that the image $\mathcal{J}(\vy)$
has a cross-range resolution of order $\la_o L /X$ and a range
resolution of order $c_o/\Omega$. The best focus occurs at $X \approx
X_d$ and $\Omega \approx \Omega_d$, so the decoherence parameters
$X_d$ and $\Omega_d$ can be estimated with optimization, as explained
in \cite{borcea2006adaptive}. 

To state the inverse problem as a convex optimization, we make the
simplifying assumption that the sources emit the same known pulse
$f(t)$, so that
\begin{equation}
\hat f_s(\om) = \hat f(\om) \rho(\vy_s), \qquad s = 1, \ldots, N_s,
\label{eq:f7}
\end{equation}
for an unknown, complex valued amplitude $\rho(\vy)$.  Using \eqref{eq:f7} in
\eqref{eq:f1} and substituting in \eqref{eq:f6} we obtain 
\begin{equation}
\mathcal{J}(\vy) \approx \sum_{s=1}^{N_s} \sum_{s'=1}^{N_s}
\rho(\vy_s) \overline{\rho(\vy_{s'})}\, \kappa(\vy, \vy_s,\vy_{s'}),
\label{eq:f8}
\end{equation}
with kernel 
\begin{align}
\kappa(\vy,\vy_s,\vy_{s'}) =& \sum_{r=1}^{N_r} \sum_{r' = 1}^{N_r}
\psi\left(\frac{|\bx_r-\bx_{r'}|}{X}\right) \int_{-\infty}^\infty d
\om \int_{-\infty}^\infty d \tom \,
\hat \phi\left(\frac{\tom}{\Omega}\right) \hat f\left(\om +
         {\tom}/{2}\right) \overline{\hat
           f\left(\om-{\tom}/{2}\right)} \nonumber
         \\ & \hspace{-0.5in}\times \hat
         G\left(\vx_r,\vy_s,\om+{\tom}/{2}\right)\overline{\hat
           G(\vx_{r'},\vy_{s'},\om-\tom/2)}\exp[-i (\om + \tom/2)
           \tau(\vx_r,\vy)] + i (\om - \tom/2)\tau(\vx_{r'},\vy)],
\label{eq:f9}
\end{align}
where the approximation is because we neglect the additive
noise\footnote{Additive noise is considered in all the numerical
  simulations in section \ref{sect:num}, but for simplicity we neglect
  it in the analysis.}. 
  
  In the analysis of the next section we take the Gaussian pulse
\begin{equation}
\hat f(\om) = \left(\frac{\sqrt{2 \pi}}{B}\right)^{1/2} \exp \left[
  -\frac{(\om-\om_o)^2}{4 B^2}\right],
\label{eq:f10}
\end{equation}
normalized by
\[
\|f \|_{_2} = \left[\int_{-\infty}^\infty dt \, |f(t)|^2
  \right]^{1/2}= \left[\int_{-\infty}^\infty \frac{d \om}{2 \pi}\,
  |\hat f(\om)|^2\right]^{1/2} = 1.
\]
This choice allows us to obtain an
explicit expression of the kernel \eqref{eq:f9}.  Naturally, in
practice the pulses may not be Gaussian, and the sources may emit
different signals. The method described here still applies to such
cases, with $\rho(\vy_s)$ replaced in \eqref{eq:f8} by $\|f_s\|_{_2}$,
and the substitution
\[
\hat f(\om + \tom/2) \overline{\hat f(\om-\tom/2)} \leadsto 
\frac{\hat f_s(\om + \tom/2)}{\|f_s\|} \frac{\overline{\hat
  f_{s'}(\om-\tom/2)}}{\|f_{s'}\|_{_2}}
\]
in \eqref{eq:f9}. Since $f_s$ is unknown in general, we may only
estimate the kernel $\kappa$ up to unknown, constant multiplicative
factors. This still allows the determination of the location of the
sources, but does not give good estimates of their intensities.

\subsection{The optimization problem}
\label{sect:OPT}
Let us consider a reconstruction mesh with $N_z$ points denoted
generically by $\vz$, and name $\brho$ the column vector with entries
given by the unknown $\rho(\vz)$. We sample the CINT image
$\mathcal{J}(\vy)$ at $N_y < N_z$ points, and gather these samples in
the ``data'' vector $\bd$.  It is natural to take $N_y < N_z$, because
we seek to super-resolve the CINT image, which is blurred by the
kernel \eqref{eq:CINTK}.

At first glance it appears that we may use equation \eqref{eq:f8} to
formulate the inversion as an optimization problem for recovering the
rank one matrix $\brho \brho^\star$, as in
\cite{candes2013phaselift,chai2010array}.  However, as shown in the
next section, in strong random media where CINT is needed, the kernel
$\kappa(\vy,\vz,\vz')$ is large only when $\vz$ and $\vz'$ are nearby.
In fact, for reasonable mesh sizes on which we can expect to obtain
unique reconstructions, the kernel satisfies
\begin{equation}
  \kappa(\vy,\vz,\vz') \approx \left\{ \begin{array}{ll}
    \kappa(\vy,\vz,\vz) \quad &\mbox{if} ~ \vz'=\vz, \\ 0 & \mbox{if}~
    \vz \ne \vz'.   \end{array}\right.
  \label{eq:diagKappa}
\end{equation}
Thus, only the diagonal entries $|\rho(\vz)|^2$ of $\brho \brho^\star$
play a role. These are the source intensities and we denote by $\bu
\in \mathbb{R}^{N_z}$ the vector of unknowns formed by them.  Equation
\eqref{eq:f8} gives
\begin{equation}
\bm{\mathcal{M}}\, \bu \approx \bd,
\label{eq:f11}
\end{equation}
where $\bm{\mathcal{M}}$ is the $N_y \times N_z$ ``measurement''
matrix with entries $\kappa(\vy,\vz,\vz).$ We formulate the inversion
as the $l_1$ optimization problem
\begin{equation}
\min_{\bu \in \mathbb{R}^{N_z}} \|\bu\|_{_1} ~ ~ \mbox{such
  that} ~~ \|\bm{\mathcal{M}}\, \bu - \bd\|_{_2} \le
~\mbox{tolerance},
\label{eq:f12}
\end{equation}
where the tolerance accounts for additive noise
effects and the random fluctuations of the CINT imaging function,
which are small for large enough array aperture $a$  and bandwidth $B$, as shown in 
\cite{borcea2007asymptotics,borcea2011enhanced}.

We prove in section \ref{sect:CINT} that the left hand side in
\eqref{eq:f11} is approximately a discrete convolution. The $l_1$
optimization is useful in this context, and recovers well sources that
are well separated, as expected from the results in
\cite{candes2014towards}. We rediscover such results in this paper
using a different analysis. We also consider cases where the sources
are clustered together, and show that although we cannot expect good
reconstructions in the point-wise sense, the $l_1$ minimizer is
supported in the vicinity of the clusters, and estimates the average
source intensity there.
\section{Setup of the analysis}
\label{sect:anal}
We introduce in this section the random wave speed model and the
scaling assumptions which define the relations between the wavelength
$\la_o$, the typical size $\ell$ of the inhomogeneities in the medium,
the standard deviation $\sigma$ of the fluctuations of the wave speed,
the array aperture $a$, the range $L$, and the extent of the imaging
region. The scaling allows us to describe the scattering effects of
the random medium as large wavefront distortions. This is a simple
wave propagation model that is convenient for analysis, and captures
qualitatively all the important features of imaging with CINT. That is
to say, equation \eqref{eq:f11} holds in general scattering regimes, and
the CINT kernel $\kappa$ has a similar form, but the mathematical
expression of the decoherence length $X_d$ and frequency $\Omega_d$,
which quantify the blurring by the kernel, are expected to change.
The expressions of $X_d$ and $\Omega_d$ in terms of $\la_o, \sigma,
\ell$ and $L$ are needed for analysis, but they are unlikely to be
known in practice. This is why one should determine the decoherence
parameters directly form the data or adaptively, during the CINT image
formation, as in \cite{borcea2006adaptive}.

The model  of the wave speed $c(\vx)$ is
\begin{equation}
c(\vx) = c_o \left[ 1 + \sigma \mu \left(\frac{\vx}{\ell}\right)
  \right]^{-1/2},
\label{eq:a1}
\end{equation}
where $\mu$ is a mean zero, stationary random process of dimensionless
argument. We suppose that $\mu$ is bounded almost surely and denote by
$\cR$ its auto-correlation, assumed isotropic and Gaussian for
convenience 
\begin{equation}
\cR(\vx) = \EE \left[ \mu(\vx + \vx') \mu(\vx)\right] = e^{-
  |\vx|^2/2}.
\label{eq:a2}
\end{equation}
Then, $\sigma \ll 1$ quantifies the small amplitude of the
fluctuations of $c(\vx)$, and $\ell$ is the correlation length, which
characterizes the typical size of the inhomogeneities in the medium.

We explain in section \ref{sect:sc1} how far the waves should
propagate in media modeled by \eqref{eq:a1}, so that the cumulative
scattering effects can be described by large wavefront distortions. We
are interested in imaging with finite size arrays at long distances,
so the waves propagate in the range direction, within a cone (beam) of
small opening angle. This is the paraxial regime defined in section
\ref{sect:sc2}. We describe in section \ref{sect:sc3} the wave
randomization quantified by the scattering mean free path, and the
statistical decorrelation quantified by the decoherence length and
frequency. We end the section with a summary of the scaling
assumptions.
 
\subsection{Wave scattering regime}
\label{sect:sc1}
We use a geometrical optics (Rytov) wave propagation model that holds
in high frequency regimes with separation of scales
\begin{equation}
\label{eq:a3} \la_o \ll \ell \ll L,
\end{equation}
and standard deviation $\sigma$ of the fluctuations satisfying 
\begin{equation}
\sigma \ll \min \left\{ \left(\frac{\ell}{L}\right)^{3/2},
\frac{\sqrt{\ell \la_o}}{L}\right\}.
\label{eq:a4}
\end{equation} 
It is shown in \cite{rytov1989principle} that the first bound in
\eqref{eq:a4} ensures that the rays remain straight and the variance
of the amplitude of the Green's function is negligible, so we can use
the same geometrical spreading factor as in the homogeneous medium.
The second bound in \eqref{eq:a4} ensures that only first order (in
$\sigma$) corrections of the travel time matter, so the propagation
model is
\begin{equation}
\hat G(\vx,\vy,\om) \approx \hat G_o(\vx,\vy,\om) \exp \left[ i \om  
\tau(\vx,\vy)\frac{\sigma}{2} \int_0^1 d \vartheta \, \mu
  \left(\frac{(1-\vartheta) \vy + \vartheta \vx}{\ell}\right) \right].
\label{eq:a6}
\end{equation}

Let us write the random phase correction in \eqref{eq:a6} as
\begin{equation}
  \om \delta \tau(\vx,\vy) = \frac{(2 \pi)^{1/4}}{2} \sigma k
  \sqrt{\ell |\vx-\vy|} \, \nu(\vx,\vy),
  \label{eq:a7p}
\end{equation}
where $k = \om/c_o$ is the wavenumber and
\begin{equation}
\nu(\vx,\vy) = \frac{1}{(2 \pi)^{1/4}} \sqrt{\frac{|\vx-\vy|}{\ell}}
\int_0^1 d \vartheta \, \mu \left(\frac{(1-\vartheta) \vy + \vartheta
  \vx}{\ell}\right)
\label{eq:a7}
\end{equation}
is defined by the integral of the fluctuations along the straight ray
between $\vy$ and $\vx$. It is shown in \cite[Lemma
  3.1]{borcea2011enhanced} that $\nu(\vx,\vy)$ converges in
distribution as $\ell/|\vx-\vy| \sim \ell/L \to 0$ to a Gaussian
process. Its mean
\begin{equation}
\EE \left[ \nu(\vx,\vy)\right] = 0,
\end{equation}
and variance 
\begin{equation}
\EE \left[ \nu^2(\vx,\vy) \right] = \frac{|\vx-\vy|}{\ell \sqrt{2
    \pi}}\int_0^1 d \vartheta \int_0^1 d \vartheta' \, \cR \left(
\frac{(\vartheta'-\vartheta) (\vx-\vy)}{\ell}\right) \approx 1,
\label{eq:a8}
\end{equation}
are calculated from definition \eqref{eq:a7} and the expression
\eqref{eq:a2} of $\cR$. The approximation is for $\ell/L$ small.

We conclude from \eqref{eq:a7p}, \eqref{eq:a8} and $k = O( k_o)$,
with $k_o = 2 \pi /\la_o$, that the random phase fluctuations in
\eqref{eq:a6} have standard deviation of order $\sigma \sqrt{\ell
  L}/\la_o$. When this is small, the random medium effects are
negligible and any coherent imaging method works well. We are
interested in the case of large fluctuations, so we ask that
\begin{equation}
\label{eq:a9} 
\sigma \gg \frac{\la_o}{\sqrt{\ell L}}.
\end{equation}
This is consistent with \eqref{eq:a4} when 
\[
\frac{\la_o/\sqrt{\ell L}}{(\ell/L)^{3/2}} = \frac{\la_o L}{\ell^2}
\ll 1, \qquad \frac{\la_o/\sqrt{\ell L}}{\sqrt{\ell \la_o}/{L} } =
\frac{\sqrt{\la_o L}}{\ell} \ll 1,
\]
so we tighten our assumption \eqref{eq:a3} on the correlation length
as
\begin{equation}
  \sqrt{\la_o L} \ll \ell \ll L.
  \label{eq:a10}
\end{equation}

\subsection{The paraxial regime}
\label{sect:sc2}
Suppose that the sources are contained in the search (imaging) region
\begin{equation}
  \label{eq:imreg}
\mathfrak{D} =
\left[-D/2,D/2\right]\times\left[-D/2,D/2\right]\left[-D_3/2,D_3/2\right],
\end{equation}
which is a rectangular prism of sides $D$ in cross-range and
$D_3$ in range, as illustrated in Figure \ref{fig:setup}.  When $D$
and the array aperture $a$ are small with respect to the range scale
$L$, the rays connecting the sources and the receivers are contained
within a cone (beam) of small opening angle, and we can use the
paraxial approximation to simplify the calculations.

The paraxial regime is defined by the scaling relations
\begin{equation}
  \la_o \ll D \lesssim a \ll L, \qquad D_3 \ll L, \qquad
  \frac{a^4}{\la_o L^3} \ll 1, \qquad \frac{a^2 D_3}{\la_o L^2} \ll 1,
  \label{eq:a11}
\end{equation}
so that for $\vy = (\by,y_3)$ and $\vx = (\bx,0)$ we get
\begin{align}
k |\vx-\vy| &= k \left( y_3 + \frac{|\bx|^2}{2 L}- \frac{\bx \cdot
  \by}{L} + \frac{|\by|^2}{2 L} \right) + O \left( \frac{a^4}{\la_o
  L^3}\right) + O\left(\frac{a^2 D_3}{\la_o L^2}\right) \nonumber
\\ &\approx k \left( y_3 + \frac{|\bx|^2}{2 L}- \frac{\bx \cdot
  \by}{L} + \frac{|\by|^2}{2 L} \right),
\label{eq:a12}
\end{align}
and 
\begin{equation}
  \frac{1}{4 \pi |\vx-\vy|} = \frac{1}{4 \pi L} \left[ 1 +
    O\left(\frac{D_3}{L}\right) + O\left(\frac{a^2}{L^2} \right)
    \right]\approx \frac{1}{4 \pi L}.
  \label{eq:a13}
\end{equation}
These approximations are proved in appendix \ref{ap:DerPar}, and the
deterministic factor in \eqref{eq:a6} becomes
\begin{equation}
  \hat G_o(\vx,\vy,\om) \approx \frac{1}{4 \pi L} \exp \left[ i k
    \left(y_3 + \frac{|\bx|^2}{2 L}- \frac{\bx \cdot \by}{L} +
    \frac{|\by|^2}{2 L} \right) \right].
  \label{eq:a14}
\end{equation}

\subsection{Randomization and statistical decorrelation of the waves}
\label{sect:sc3}
Here we quantify the scattering mean free path $\cS$, the length scale
on which the waves randomize (lose coherence), and the decoherence
length $X_d$ and frequency $\Omega_d$, which describe the statistical
decorrelation of the waves due to scattering. These important scales
appear in the definition of the CINT blurring kernel defined in
section \ref{sect:CINT}.

\vspace{0.05in}
\begin{proposition}
  \label{prop.1}
The expectation of the Green's function \eqref{eq:a6} is given by
\begin{equation}
  \EE\big[ \hat G(\vx,\vy,\om)\big] \approx \hat G_o (\vx,\vy,\om)
  e^{- \frac{|\vx-\vy|}{\cS(\om)}} \approx 0,
  \label{eq:prop1.1}
\end{equation}
where $\cS(\om)$ is the scattering mean free path defined by
\begin{equation}
  \cS(\om) = \frac{8}{\sqrt{2 \pi} \sigma^2 k^2 \ell} \ll L.
  \label{eq:prop1.2}
\end{equation}
\end{proposition}

\noindent This result, proved in appendix \ref{ap:moments}, shows that
the wavefront distortions due to scattering in the random medium do
not average out.  The expectation of $\hat{G}$ is not the
same as the Green's function in the homogeneous medium, but decays
exponentially with the distance of propagation on the scale
$\cS(\om)$, the scattering mean free path. The scaling assumption
\eqref{eq:a9} and definition \eqref{eq:prop1.2} give
\[
\frac{|\vx-\vy|}{\cS(\om)} = O \left(\frac{L}{\cS(\om)}\right) = O(\sigma^2
k^2 \ell L) \gg 1,
\]
which is why the expectation in \eqref{eq:prop1.1} is almost zero.
The standard deviation of the fluctuations is approximately 
\[
\mbox{std}[\hat G(\vx,\vy,\om)] \approx \sqrt{\Big|\hat G_o(\vx,\vy,\om)\Big|^2 - 
\Big|\EE\big[ \hat G(\vx,\vy,\om)\big] \Big|^2} \approx \Big|
\hat G_o(\vx,\vy,\om)\Big|,
\]
where we used that $|\hat G(\vx,\vy,\om)| \approx |\hat G_o(\vx,\vy,\om)|$. Thus,  
the random fluctuations of the waves dominate their coherent part (the
expectation) at the ranges considered in our analysis,
\[
\frac{ \Big| \EE \big[ \hat G(\vx,\vy,\om) \big] \Big|}{\mbox{std}[\hat G(\vx,\vy,\om)]}
\approx e^{-\frac{|\vx-\vy|}{S(\om)}} \approx 0,
\]
and the wave is randomized. Reverse time
migration or standard $l_1$ optimization methods cannot mitigate these
large random distortions, as we illustrate with numerical
simulations. This is why we base our inversion on the CINT method.

\vspace{0.05in}
\begin{proposition}
\label{prop.2}
Consider two points $\vx = (\bx,0)$ and $\vx' = (\bx',0)$ in the array
aperture and two points $\vy = (\by,y_3)$ and $\vy' = (\by',y_3')$ in
the imaging region. Assume that the bandwidth $B$ is small with
respect to the central frequency $\om_o$, so that $|\om-\om_o|,
|\om'-\om_o| \le \pi B \ll \om_o$. Then, the second moments of
\eqref{eq:a6} are
\begin{align}
   \EE\left[ \hat G(\vx,\vy,\om) \overline{\hat
       G(\vx',\vy',\om')}\right] \approx \hat G_o(\vx,\vy,\om)
   \overline{\hat G_o(\vx',\vy',\om')} e^{-
     \frac{|y_3-y_3'|}{\cS}- \frac{(\om-\om')^2}{2 \Omega_d^2}
     - \frac{|\by-\by'|^2 + (\by-\by')\cdot(\bx-\bx') +
       |\bx-\bx'|^2}{2 X_d^2}},
   \label{eq:prop2.1}
\end{align}
with short notation $\cS = \cS(\om_o)$, and decoherence frequency
$\Omega_d$ and decoherence length $X_d$ defined by
\begin{equation}
  \Omega_d = \frac{2 \om_o}{(2 \pi)^{5/4}} \left(\frac{\la_o}{\sigma
    \sqrt{\ell L}}\right) \ll \om_o, \qquad X_d =\sqrt{3} \ell\,
  \frac{ \Omega_d}{\om_o} \ll \ell.
 \label{eq:prop2.2}
\end{equation}
\end{proposition}

\noindent Moment formula \eqref{eq:prop2.1} is proved in appendix
\ref{ap:moments}, and the inequalities in \eqref{eq:prop2.2} are due
to assumption \eqref{eq:a9}. The first exponential factor in
\eqref{eq:prop2.1} accounts for the randomization due to the travel
time fluctuations between the two ranges. In our scaling $|y_3-y_3'|
\lesssim D_3$, and by the last inequality in \eqref{eq:a11}, and the
definition of the scattering mean free path, we have
\begin{equation}
  \frac{|y_3-y_3'|}{\cS(\om_o)} = O \left(D_3
  \frac{\sigma^2\ell}{\la_o^2} \right) \ll O\left(\frac{\la_o
    L^2}{a^2} \frac{\sigma^2 \ell}{\la_o^2} \right) \ll
  O\left(\frac{\ell^2}{a^2} \right),
  \label{eq:a15}
\end{equation}
where we used the bound \eqref{eq:a4} on $\sigma$. It is shown in
\cite[Section 4]{borcea2011enhanced} that the standard deviation of
the CINT image is small with respect to the expectation of its peak
value (i.e., the imaging function is statistically stable) when
\begin{equation}
  a > \ell.
  \label{eq:a16}
\end{equation}
Stability is essential for imaging to succeed, so we ask that the
array aperture satisfy \eqref{eq:a16}, and conclude from
\eqref{eq:a15} that the second moments \eqref{eq:prop1.1} simplify
as\footnote{Although this moment formula is derived here using the
  model \eqref{eq:a6}, the result is generic and can be obtained in
  other scattering regimes. The only difference is the definition of
  $X_d$ and $\Omega_d$. See for example \cite{papanicolaou2007self}.}
\begin{align}
   \EE\left[ \hat G(\vx,\vy,\om) \overline{\hat
       G(\vx',\vy',\om')}\right] &\approx \hat G_o(\vx,\vy,\om)
   \overline{\hat G_o(\vx',\vy',\om')}e^{ - \frac{(\om-\om')^2}{2
       \Omega_d^2} - \frac{|\by-\by'|^2 + (\by-\by')\cdot(\bx-\bx') +
       |\bx-\bx'|^2}{2 X_d^2}},
   \label{eq:a17}
\end{align}
with $\hat G_o$ given in \eqref{eq:a14}.

The exponential decay in \eqref{eq:a17} models the statistical
decorrelation of the waves due to scattering. In our context, the
spatial decorrelation, modeled by the decay in $\bx-\bx'$ and
$\by-\by'$, can be explained by the fact that rays connecting sources
to far apart receivers traverse through different parts of the random
medium. Because $\mu$ does not have long range correlations, the
fluctuations of the travel time along such different rays are
statistically uncorrelated. The waves at far apart frequencies are
uncorrelated because they interact differently with the random
medium. This gives the decay in $\om-\om'$ in equation \eqref{eq:a17}.

Definition \eqref{eq:f9} of the CINT kernel  involves the
superposition of $\hat G(\vx,\vy,\om) \overline{\hat
  G(\vx',\vy',\om')}$ over the array elements and frequencies. If the
array aperture $a$ is large with respect to $X_d$, the superposition stabilizes
statistically because we sum many uncorrelated entries, as in the law
of large numbers. This is why CINT is robust with respect to the
uncertainty of the fluctuations of the wave speed, as shown in
\cite{borcea2007asymptotics,borcea2011enhanced}.

\subsection{Summary of the scaling assumptions}
\label{sect:sc4}
We gather here the scaling assumptions stated throughout the section,
and complement them with extra assumptions on the bandwidth and the
size of the imaging region.  We refer to appendix \ref{ap:verify} for
the verification of their consistency.

The wavelength $\la_o$ is the smallest length scale, and the range $L$
is the largest.  The assumptions \eqref{eq:a16} and \eqref{eq:a11} on
the aperture are
\begin{equation}
  \ell < a \ll (\la_o L^3)^{1/4}.
  \label{eq:as1}
\end{equation}
The upper bound ensures that we can use the paraxial approximation and
the lower bound gives $a > \ell \gg X_d$, so that the CINT image is
statistically stable.

Assumption \eqref{eq:a10} combined with \eqref{eq:as1} gives that the
correlation length of the wave speed fluctuations should satisfy
\begin{equation}
\sqrt{\la_o L} \ll \ell \ll (\la_o L^3)^{1/4} \ll L.
\label{eq:as2}
\end{equation}
The standard deviation $\sigma$ of the fluctuations is bounded above
as in \eqref{eq:a4}, and below as in \eqref{eq:a9},
\begin{equation}
\frac{\la_o}{\sqrt{\ell L}} \ll
\sigma \ll \frac{\sqrt{\ell \la_o}}{L}.
\label{eq:as3}
\end{equation}

The cross-range and range sizes $D$ and $D_3$ of the imaging region
should be large with respect to the CINT resolution limits of $\la_o
L/X$ in cross-range and $c_o/\Omega$ in range (see next section), so
we can observe the image focus. We take the threshold parameters
\begin{equation}
X/X_d = O(1), \qquad \Omega/\min\{\Omega_d,B\} = O(1),
\label{eq:as4pp}
\end{equation}
and recalling the scaling assumptions \eqref{eq:a11} that allow us to
use the paraxial approximation, we obtain
\begin{equation}
\frac{c_o}{\Omega} \ll D_3 \ll \frac{\la_o L^2}{a^2} \quad 
\mbox{and} \quad \frac{\la_o L}{X} \ll D \lesssim a.
\label{eq:as5}
\end{equation}

In general, the CINT image is statistically stable if in addition to
having $a \gg X_d$, which follows from \eqref{eq:prop2.2} and
\eqref{eq:as1}, the bandwidth $B$ is larger than the decoherence
frequency $\Omega_d$. However, for the propagation model \eqref{eq:a6}
considered in this section, where the effect of the random medium
consists only of wavefront distortions and no delay spread (reverberations), the
bandwidth $B$ does not play a role in the stabilization of CINT, as
shown in \cite[Section 4.4.4]{borcea2011enhanced}.  Thus, we study imaging 
in both narrowband and broadband regimes:

\vspace{0.05in} \noindent 
\textbf{The narrowband regime} is defined by  $B$ satisfying
\begin{equation}
\om_o\left(\frac{a}{L}\right)^2 \ll B \ll \om_o \min\left\{1, \frac{\la_o L}{a X_d} 
\right\}= \om_o \frac{\la_o L}{a X_d}.
\label{eq:as4}
\end{equation}
As verified in Appendix \ref{ap:verify}, 
\begin{equation}
\frac{\la_o L}{X_d} \ll \ell < a, \qquad \frac{\la_o L}{a X_d} \ll
\frac{\Omega_d}{\om_o},
\label{eq:as4p}
\end{equation}
so $B \ll \Omega_d$.  This choice leads to a simpler expression of the
CINT blurring kernel, but since $\Omega$ is of the order of $B$, the
range resolution is the same as in the homogeneous medium, and cannot
be improved with optimization unless the sources are very far apart in
range. However, the optimization can improve the cross-range focusing.

\vspace{0.05in}\noindent
\textbf{The broadband regime} is defined by 
\begin{equation}
\Omega_d \ll B \ll \om_o,
\label{eq:as4bb}
\end{equation}
so we may seek to improve the CINT resolution in both range and cross-range. The 
expression of the CINT kernel is more complicated in this case, but it simplifies slightly 
when 
\begin{equation}
\frac{\la_o}{\sqrt{\ell L}} \ll \frac{\la_o^{2/3} \ell^{1/6}}{L^{5/6}} \ll \sigma \ll \frac{\sqrt{\ell \la_o}}{L}, \qquad \frac{\ell}{a} = O(1).
\label{eq:as3bb}
\end{equation}
We present the analysis that uses these conditions, which say that the fluctuations in the random medium are even stronger 
than in \eqref{eq:as3}, but the correlation length is not much smaller than $a$. Extensions to larger apertures are possible,  
although the analysis is more complicated.
\section{The CINT blurring kernel}
\label{sect:CINT}
Here we derive the CINT convolution model.  To obtain an explicit
expression of the kernel \eqref{eq:f9}, we use the Gaussian pulse
\eqref{eq:f10} and the Gaussian threshold windows
\begin{equation}
\hat \phi\left(\frac{\tom}{\Omega}\right) = e^{-\frac{\tom^2}{2 \Omega^2}},
\qquad \psi\left(\frac{|\tbx|}{X}\right) = e^{-\frac{|\tbx|^2}{2
    \Omega^2}},
\label{eq:R1}
\end{equation}
with $X$ and $\Omega$ satisfying
\begin{equation}
{X}/{X_d} = O(1), \qquad \Omega = \left\{ \begin{array}{ll} B \quad & \mbox{in narrowband regime} \\ \\
O(\Omega_d) & \mbox{in broadband regime}.
\end{array} \right.
\label{eq:R2}
\end{equation}
As stated previously, and shown in \cite{borcea2006adaptive}, $X_d$
and $\Omega_d$ can be estimated adaptively, by optimizing the focusing
of the CINT image. This is why we can assume that $X_d$ is known
approximately.  The same holds for $\Omega_d$, if the bandwidth is big
enough. The expression of the CINT kernel is simpler in the narrowband
scaling \eqref{eq:as4}, where $B \ll \Omega_d$, as shown in section \ref{sect:CINT1}, and we take $\Omega =
B$.  The broadband regime is discussed in section
\ref{sect:CINT2}.

Typically, the receivers are separated by distances of order $\la_o$,
so that they behave collectively as an array. Since $\la_o \ll a$,
we have $ N_r = O\left(a^2/\la_o^2\right) \gg 1,  $ and we can
approximate the sums in \eqref{eq:f9} by integrals
\[
\sum_{r=1}^N \leadsto \frac{N_r}{a^2} \int_{\cA} d \bx, 
\]
where $\cA$ denotes the array aperture, the square of side $a$. To
avoid specifying the finite aperture in the integrals, and to simplify the calculations,  we use a Gaussian apodization
factor 
\begin{equation}
\psi_\cA(\bx) = \exp\left[-\frac{|\bx|^2}{2(a/6)^2}\right],
\label{eq:R3}
\end{equation}
which is negligible outside the disk of radius $a/2$.

\subsection{The CINT kernel in the narrowband regime}
\label{sect:CINT1}
The calculation of the kernel \eqref{eq:f9} is in appendix
\ref{sect:apCINT}, and we state the results in the next proposition.
\vspace{0.05in}
\begin{proposition}
  \label{prop.3}
  Let $\vz = (\bz,z_3)$, $\vz'=(\bz',z_3')$ and $\vy = (\by,y_3)$ be
  three points in the imaging region and define the center and
  difference vectors
\[
\frac{\vz + \vz'}{2} = (\cbz,\cz_3), \qquad \vz-\vz'
= (\tbz,\tz_3).
\]
Under the assumptions \eqref{eq:as1}-\eqref{eq:as4}, the 
CINT kernel \eqref{eq:f9} is approximated by
\begin{align}
\kappa(\vy,\vz,\vz') &\approx C\exp\left[-\frac{|\cbz-\by|^2}{2 R^2} - \frac{
    (\cz_3-y_3)^2}{2 R_3^2}
  \right]\, \mathfrak{M}(\vy,\vz,\vz'),
\label{eq:CINTK}
\end{align}
where $C$ is a constant, and 
\begin{equation}
R = \frac{L}{k_o X_e}, \qquad R_3 = \frac{c_o}{\Omega_e}, \label{eq:prop3.1}
\end{equation}
with   $X_e$ and $\Omega_e$  defined by 
\begin{equation}
\frac{1}{X_e^2} = \frac{1}{X_d^2} + \frac{1}{X^2} +
\frac{1}{4(a/6)^2}, \qquad \frac{1}{\Omega_e^2} = \frac{1}{\Omega_d^2}
+ \frac{1}{\Omega^2} + \frac{1}{4 B^2}.
\label{eq:R4}
\end{equation}
The factor  $\mathfrak{M}(\vy,\vz,\vz')$   is complex, with absolute value 
\begin{align}
|\mathfrak{M}(\vy,\vz,\vz')| =  \exp \left[-\frac{\tz_3^2}{2
 \widetilde{R}^2} - \frac{|\tbz|^2}{2} \left(
  \frac{1}{\gamma X_d^2} +  \frac{1}{\widetilde{R}^2}  \right) \right],
\label{eq:prop3.2}
\end{align}
where 
\begin{equation}
\frac{1}{\gamma} = 1 - \frac{X_e^2}{4 X_d^2} > \frac{3}{4}, \quad \widetilde{R}_3 = \frac{c_o}{B}, 
\quad \widetilde{R}  = 6 \sqrt{2} \frac{L}{k_o a}.
\end{equation}
\end{proposition}

The parameter $R$ defined in \eqref{eq:prop3.1} is the CINT
cross-range resolution limit, the length scale of exponential decay of
the kernel $\kappa(\vy,\vz,\vz')$ with $\cbz-\by$.  Definitions 
\eqref{eq:R4}, \eqref{eq:prop2.2} and assumption \eqref{eq:R2} give that 
\[
X_e = O(X_d) \ll a,
\]
so the resolution is worse than in homogeneous media,
\begin{equation}
R \gg \frac{L}{k_o a}.
\label{eq:R6}
\end{equation}
This is due to the smoothing needed to stabilize statistically the image
\cite{borcea2006adaptive}. The goal of the convex optimization
\eqref{eq:f12} is to overcome this blurring and localize better the
sources in cross-range.

The parameter $R_3$ is the CINT range resolution limit. Because we are
in the narrowband regime, we obtain from definition \eqref{eq:R4} and
\eqref{eq:R2} that $ \Omega_e \approx B,$
and therefore $R_3$ is similar to the range resolution  in homogeneous media,
\begin{equation}
R_3 = \frac{c_o}{\Omega_e}  \approx \frac{c_o}{B}.
\end{equation}  
The results obtained in \cite{borcea2015resolution} for imaging with
$l_1$ optimization in homogeneous media show that it is not possible
to improve the $c_o/B$ range resolution, unless the sources are very far
apart in range.  We cannot expect to do better in random media, so we
do not seek any super-resolution in range, in the narrowband regime.

Note that by the first inequality in \eqref{eq:as4p} we have $\widetilde{R} \ll X_d$, so 
the kernel decays with the offsets $\tbz$ and $\tz$ on the length scales $\widetilde{R}$ and $\widetilde{R}_3$.   
These scales are, up to a constant of order one,  the resolution limits of imaging in homogeneous media.

\subsection{The CINT kernel in the  broadband regime}
\label{sect:CINT2}
The expression of the CINT kernel is stated in the next proposition, proved in appendix \ref{sect:apCINT}.
\begin{proposition}
\label{prop.3bb}
Consider the same points and notation as in Proposition \ref{prop.3}. Under the assumptions \eqref{eq:as1}-\eqref{eq:as5} 
and \eqref{eq:as4bb}-\eqref{eq:as3bb}, the CINT kernel is given by 
\begin{align}
\kappa (\vy,\vz,\vz') &\approx \frac{C}{\sqrt{1 + \frac{|\cbz-\by|^2}{2 \theta^2 R^2}}}\exp\left[-\frac{|\cbz-\by|^2}{2 R^2} - \frac{
    \big(\cz_3-y_3 + \frac{|\cbz|^2-|\by|^2}{2 L}\big)^2}{2 R_3^2\big[1 + \frac{|\cbz-\by|^2}{2 \theta^2 R^2}\big]}
  \right]\, \mathfrak{M}(\vy,\vz,\vz'),
\label{eq:CINTKernBB}
\end{align}
where $C$ is a constant, 
\begin{equation}
\theta = \frac{6\om_o X_e}{\Omega_e a},
\label{eq:defTheta}
\end{equation}
and $\mathfrak{M}(\vy,\vz,\vz')$ is a complex multiplicative factor with absolute value 
\begin{align}
\left|\mathfrak{M}(\vy,\vz,\vz')\right| = \exp \left\{ -\frac{\tz_3^2}{2 \widetilde{R}_3^2} - \frac{|\tbz|^2}{2} \left[ \frac{1}{\gamma X_d^2 }
+ \frac{1}{\widetilde{R}^2} \right]+ \frac{\left|\frac{(\cbz-\by)}{R} \cdot\frac{\tbz}{\widetilde{R}}\right|^2 }{4 \theta^2\big[1 + \frac{|\cbz-\by|^2}{2 \theta^2 R^2}\big]} \right\}
\label{eq:KernBB}
\end{align}
\end{proposition}

Because $X_e = O(X_d)$ and $\Omega_e = \Omega_d$, we obtain from definitions \eqref{eq:defTheta} and \eqref{eq:prop2.2} that 
\begin{equation}
\theta = O\left(\frac{\ell}{a} \right) = O(1),
\label{eq:KernBB1}
\end{equation}
where we used the assumption \eqref{eq:as3bb} on the aperture.\footnote{In the narrowband case the aperture may be much larger than $\ell$,
as in \eqref{eq:as1}. It is only  in the broadband case that we take $\ell = O(a)$ to simplify the expression of the CINT kernel. } 
The first term in the exponential in \eqref{eq:CINTKernBB} gives the focusing in cross-range, which is the same as in the narrowband case:
$|\cbz-\by| = O(R)$. This means that the denominators in \eqref{eq:CINTKernBB} are order one,
\[
1 + \frac{|\cbz-\by|^2}{2 \theta^2 R^2} = O(1).
\]
The second term in the exponential in \eqref{eq:CINTKernBB} gives the focusing in range. In our setting we have by the  
paraxial approximation that 
\[
\cz_3 - y_3 + \frac{|\cbz|^2-|\by|^2}{2L} \approx |(\cbz,\cz_3)|-|(\by,y_3)|,
\]
so CINT estimates the distance from the center of the array to $(\cbz,\cz_3)$ with resolution 
of order $R_3$. Since $\Omega_e \ll B$, this resolution is worse than in homogeneous media 
\[
R_3 = \frac{c_o}{\Omega_e} = O\left(\frac{c_o}{B}\right) = O(\widetilde{R}_3),
\]
so in the broadband regime it makes sense to seek an improvement  of both the range and cross-range  resolution with optimization.

Equations \eqref{eq:CINTKernBB} and \eqref{eq:KernBB} show that the kernel decays with the offset $\tz_3$ on the same scale $\widetilde{R}_3$ as before. To see the 
decay with the offset $\tbz$, we note  that the last two terms in \eqref{eq:KernBB} satisfy
\begin{align*}
\frac{|\tbz|^2}{2} \left[ \frac{1}{\gamma X_d^2 }
+ \frac{1}{\widetilde{R}^2} \right]- \frac{\left|\frac{(\cbz-\by)}{R} \cdot\frac{\tbz}{\widetilde{R}}\right|^2 }{4 \theta^2\big[1 + \frac{|\cbz-\by|^2}{2 \theta^2 R^2}\big]} &= \frac{|\tbz|^2}{2 R^2 \big[1 + \frac{|\cbz-\by|^2}{2 \theta^2 R^2}\big]} \left[ 1 + \frac{\widetilde{R}^2 \big[1 + \frac{|\cbz-\by|^2}{2 \theta^2 R^2}\big]}{\gamma X_d^2} \right] 
+ \frac{\frac{|\cbz-\by|^2}{R^2} \frac{|\tbz|^2}{\widetilde{R}^2} - \left|\frac{(\cbz - \by)}{R}\cdot\frac{\tbz}{\widetilde {R}}\right|^2}{4 \theta^2 \big[1 + \frac{|\cbz-\by|^2}{2 \theta^2 R^2}\big]} \\
&\gtrsim \frac{|\tbz|^2}{2 R^2 \big[1 + \frac{|\cbz-\by|^2}{2 \theta^2 R^2}\big]},
\end{align*}
where we used that $\widetilde{R} \ll X_d$, as explained in the previous section. 
This shows that the kernel decays with the cross-range offsets on the same scale $\widetilde{R}$ as before.

\subsection{The approximate convolution model}
\label{sect:CINTConv}
Let us discretize the imaging region $\mathfrak{D}$ defined in \eqref{eq:imreg} on a mesh 
with size $\vec{\bm{h}} = (h,h,h_3)$.  In principle, the steps $h$ and $h_3$ may be chosen arbitrarily small, to avoid discretization error due to 
sources being off the mesh. However, we know from \cite{borcea2015resolution} and the analysis below and the numerical simulations that 
we cannot expect reconstructions at scales that are finer than the resolution limits in homogeneous media. This motivates 
us to formulate the inversion using the assumption  that the sources are further apart than $3 \widetilde{R}$ in cross-range and $3 \widetilde{R}_3$ in range. This leads to a simpler optimization problem because by  Propositions 
\ref{prop.3} and \ref{prop.3bb} we have 
\[
|\mathfrak{M}(\vy,\vz,\vz')| \le 
\exp\left(-\frac{9}{2}\right) \ll 1, \quad \mbox{if} ~ |\bz-\bz'| \ge 3 \widetilde{R} ~ \mbox{or}~ |z_3-z_3'| \ge 3 \widetilde{R}_3,
\]
and we may work only with the diagonal part of the CINT kernel.

We obtain the linear system of equations \eqref{eq:f11}, with vector $\bu$ of components $|\rho(\vz)|^2$ at the
$N_z$ mesh points in $\mathfrak{D}$.  The ``data'' vector $\bd$
consists of the samples of the CINT image at $N_y < N_z$ equidistant
points in $\mathfrak{D}$, and in the narrowband regime the $N_y \times N_z$ matrix
$\bm{\mathcal{M}}$ has entries
\begin{equation}
m_{\vy,\vz} = C\exp\left[-\frac{|\bz-\by|^2}{2 R^2} - \frac{
    (\cz_3-y_3)^2}{2 R_3^2} \right], \qquad B \ll \Omega_d,
    \label{eq:R8}
\end{equation}
with constant $C$. This
depends only on $\vy-\vz$, so we have a convolution as stated in
section \ref{sect:OPT}. In the broadband regime, the entries of $\bm{\mathcal{M}}$ are 
\begin{equation}
m_{\vy,\vz} = \frac{C}{\sqrt{1+\frac{|\bz-\by|^2}{2 \theta^2 R^2}}} \exp\left[-\frac{|\bz-\by|^2}{2 R^2} - \frac{
    | z_3-y_3 + \frac{|\bz-\by|^2}{2 L } + \frac{\by \cdot (\bz-\by)}{L}|^2}{2 R_3^2\big[1+\frac{|\bz-\by|^2}{2 \theta^2 R^2}\big]} \right], \qquad B \gg \Omega_d,
    \label{eq:R8bb}
\end{equation}
with constant $C$. Were it not for the last term in \eqref{eq:R8bb}, we would have 
a convolution. This term is large only at points $\vy = (\by,y_3)$ with  $\by$ near the boundary 
of the imaging region ($|\by| \sim D < a$), because by definition \eqref{eq:defTheta} and the assumption
$\theta = O(1)$ we get 
\[
\left| \frac{\by \cdot (\bz-\by)/L}{R_3}\right| = O\left(\frac{|\by|}{L} \frac{R}{R_3}\right) = 
O\left(\frac{|\by|}{a \theta}\right) = O\left(\frac{|\by|}{a}\right).
\]
For points with $|\by| \ll D < a$ the right hand side in \eqref{eq:R8bb} is approximately a function of 
$\vy-\vz$, corresponding to a convolution model.
\section{Resolution analysis}
\label{sect:res}
In this section we analyze the reconstruction of the vector $\bu$ of
source intensities using the convex optimization formulation described
in section \ref{sect:OPT}. To simplify the analysis, we treat the
approximation in \eqref{eq:f11} as an equality, and study the $l_1$
optimization
\begin{equation}
\min_{\bu \in \mathbb{R}^{N_z}} \|\bu\|_{_1} ~ ~ \mbox{such that} ~~
\bm{\mathcal{M}}\, \bu = \bd.
\label{eq:Res1}
\end{equation}
This neglects additive noise and random fluctuations of the CINT
function, which are small in our scaling. It also implies that the
sources are on the reconstruction mesh, so that the equality
constraint in \eqref{eq:Res1} holds for the true discretized source
intensity. Naturally, in practice the sources may lie anywhere in
$\mathfrak{D}$, and noise and distortions due to the random medium
play a role. This is why we use the more robust formulation
\eqref{eq:f12} in the numerical simulations in section \ref{sect:num}.

We expect from the study \cite{candes2014towards} of deconvolution
using $l_1$ optimization that the solution of \eqref{eq:Res1} should
be a good approximation of the unknown vector of intensities if the
sources are well separated. We show in this section that this is
indeed the case. We also consider the case of clusters of nearby
sources, and show that the $l_1$ solution is useful when the clusters
are well separated. The analysis is built on our recent results in
\cite{borcea2015resolution}.

\subsection{Definitions}
\label{sect:T1Def}
Let $\bm{Y} = \left\{\vy_s, ~ ~ s = 1, \ldots, N_s\right\}$ be the set
that supports the unknown, point-like sources in $\mathfrak{D}$. We
quantify the spatial separation between them using the following
definition:

\vspace{0.05in}\begin{definition}
  \label{def.1} The points in $\bm{Y}$ are separated
  by at least $\vH = (H,H,H_3)$, if the intersection of $\bm{Y}$ with
  any rectangular prism of sides less then $H$ in
  cross-range and $H_3$ in range consists of at most one point.
\end{definition}

\vspace{0.05in} \noindent
For example, if the sources are all in the same cross-range plane, and
the minimum distance between any two of them is $H_{min}$, we may
take $H = H_{min}$ and $H_3 = D_3$. 

We search the sources on a mesh with $N_z$ points denoted generically
by $\vz$. The mesh discretizes $\mathfrak{D}$, and we call it
$\mathfrak{D}_z$. For simplicity we let $\bm{Y} \subset
\mathfrak{D}_z$. To any $\vz \in \mathfrak{D}_z$, we associate the
column vector $\bm{m}_{\vz} \in \mathbb{R}^{N_y}$ of the matrix
$\bm{\mathcal{M}}$.  Its entries are given in \eqref{eq:R8} in the narrowband regime and 
by \eqref{eq:R8bb} in the broadband regime, for $N_y
< N_z$ points $\vy$ at which we sample the CINT image.

\vspace{0.05in}
\begin{definition}
  \label{def.2}
We quantify the interaction between two presumed sources at $\vz,\vz'
\in \mathfrak{D}_{\vz}$ using the cross-correlation of the associated
column vectors in $\bm{\mathcal{M}}$,
\begin{equation}
  \mathcal{I}_{\vz,\vz'} = \frac{\left| \left< \bm{m}_{\vz},
    \bm{m}_{\vz'} \right>\right|}{\|\bm{m}_{\vz}\|_2
    \|\bm{m}_{\vz'}\|_2}.
  \label{eq:Res2}
\end{equation}
Here $\left< \cdot, \cdot \right>$ is the Euclidian inner product and
$\| \cdot \|_2$ is the Euclidian norm.
\end{definition}

\vspace{0.05in} \noindent Note that \eqref{eq:Res2} is symmetric and
non-negative, and attains its maximum at $\vz' = \vz$, where
$\mathcal{I}_{\vz,\vz} = 1$. We will show below that \eqref{eq:Res2}
decreases  as the points $\vz$ and $\vz'$ grow apart.
This motivates the next definition which uses $\mathcal{I}_{\vz,\vz'}$
to measure the distance between $\vz$ and $\vz'$.

\vspace{0.05in}
\begin{definition}
  \label{def.3}
We define the semimetric $\Delta:\mathfrak{D}_z \times
\mathfrak{D}_z \to [0,1]$ by
\begin{equation}
 \Delta(\vz,\vz') = 1 - \mathcal{I}_{\vz,\vz'}, \qquad \forall
 \vz,\vz' \in \mathfrak{D}_z,
 \label{eq:Res3}
\end{equation}
and let 
\begin{equation}
  \mathscr{B}_r(\vz) = \left\{ \vz' \in \mathfrak{D}_z ~ ~ \mbox{s.t.}
  ~ ~ \Delta(\vz,\vz') < r \right\}
  \label{eq:Res4}
\end{equation}
be the open balls defined by $\Delta$.
\end{definition}

\vspace{0.05in} \noindent We will show that $\mathcal{I}_{\vz,\vz'}$ decays as $\|\vz-\vz'\|_2$ grows. Thus, 
we say that points $\vz'$ outside $\mathscr{B}_r(\vz)$ have a weaker interaction with $\vz$ than points
inside $\mathscr{B}_r(\vz)$.  Moreover, we may relate intuitively $\Delta(\vz,\vz')$ to the Euclidian distance
$\|\vz-\vz'\|_2$.

\vspace{0.05in}
\begin{definition}
  \label{def.4}
We define the interaction coefficient of the set $\bm{Y}$ of source
locations by
\begin{equation}
  \mathcal{I}(\bm{Y}) = \max_{\vz \in \mathfrak{D}_z} \sum_{\vy \in
    \bm{Y}\setminus \mathscr{N}(\vz)} \mathcal{I}_{\vz,\vy},
  \label{eq:Res5}
\end{equation}
where $\mathscr{N}(\vz)$ is the closest point to $\vz$ in $\bm{Y}$, as
measured with the semimetric $\Delta$.
\end{definition}

\vspace{0.05in} \noindent In general more than one point may be
closest to $\vz$. If this is so, we let $\mathscr{N}(\vz)$ be any such
point.

\subsection{Results}
The results stated here describe the relation between the reconstruction $\bu_\star$, the
solution of the convex optimization problem \eqref{eq:Res1}, and the
true unknown vector $\bu$ of source intensities. The next theorem
shows that $\bu_\star$ is essentially supported in the set $\bm{Y}$,
when the points there are well separated.

\vspace{0.05in}
\begin{theorem}
  \label{thm.1}
Suppose that the source locations in $\bm{Y}$ are separated by at
least $\vH = (H,H,H_3)$ in the sense of Definition \ref{def.1}, where
\begin{equation}
\frac{H}{R} = \alpha, ~ ~ \frac{H_3}{R_3} = \alpha_3, \label{eq:asH}
\end{equation}
for 
$\alpha, \alpha_3 > 1$. Take $r \in (0,1)$ small enough so that the balls
$\mathscr{B}_r(\vy_s)$ centered at $\vy_s \in \bm{Y}$, for $s = 1,
\ldots, N_s$, are disjoint. Let $\bu_\star$ be the $l_1$ minimizer in
\eqref{eq:Res1}, and decompose it as $ \bu_\star = \bu_\star^{(i)} +
\bu_\star^{(o)}, $ where $\bu_\star^{(i)}$ is supported in the union
$\bigcup_{s=1}^{N_s}\mathscr{B}_r(\vy_s)$ of balls centered at the
points in $\bm{Y}$, and $\bu_\star^{(o)}$ is supported in the
complement of this union. Then, there exists a constant $C$ that is
independent of $\alpha$ and $\alpha_3$, such that
\begin{equation}
  \|\bu_\star^{(o)}\|_1 \le \frac{C}{r} \mathfrak{F}(\alpha,\alpha_3) 
  \|\bu_\star\|_1,
\label{eq:Res6}
\end{equation}
where $\mathfrak{F}(\alpha,\alpha_3)$ is function that decays with $\alpha$ and $\alpha_3$. 
In the narrowband case it is given by 
\begin{equation}
\mathfrak{F}(\alpha,\alpha_3)= \frac{ \exp \left[- \left(\frac{\min\{\alpha,\alpha_3\}}{4}
         \right)^2\right]}{\alpha^2 \alpha_3}, 
\label{eq:defFnb}
\end{equation}
for arbitrary $\alpha, \alpha_3 > 1$. In the broadband case $\alpha_3 > 8 \alpha/\theta$, and 
 \begin{equation}
\mathfrak{F}(\alpha,\alpha_3)= \frac{ \exp \left[-\frac{1}{2} \left(\frac{\alpha}{4}\right)^2\right] + \exp[-\alpha]}{\alpha^3}. 
\label{eq:defFbb}
\end{equation}

\end{theorem}

\vspace{0.0in} \noindent Note that the scales of separation between
the sources are the resolution parameters $R$ and $R_3$ of CINT. The
parameters $\alpha$ and $\alpha_3$ in the separation assumption may be any
non-negative real numbers, but the statement of the theorem is useful
only when the coefficient in front of $\|\bu_\star\|_1$ in
\eqref{eq:Res6} is smaller than one.  This happens for large enough
$\alpha$ and $\alpha_3$. The larger the separation between the sources, the smaller
the right hand side in \eqref{eq:Res6} is, and the better the
concentration of the support of $\bu_\star$ near the points in
$\bm{Y}$. The next corollary gives an estimate of the error of the
reconstruction.

\vspace{0.05in}
\begin{corollary}
\label{cor.1}
Let $\bu \in \mathbb{R}^{N_z}$ be the vector of true source
intensities, and use the same assumptions and notation as in Theorem
\ref{thm.1}. Denote the entries of the $l_1$ minimizer $\bu_\star$ by
$u_\star(\vz)$, where $\vz$ are the $N_z$ points on the mesh
$\mathfrak{D}_z$. Define the effective reconstructed source intensity
vector $\overline{\bu}_\star \in \mathbb{R}^{N_z}$, with entries
\begin{equation}
  \overline{u}_\star(\vz) = \left\{ \begin{array}{ll} \sum_{\vz' \in
      \mathscr{B}_r(\vz)} u_\star(\vz') \,
    \mathcal{I}_{\vz,\vz'}, \quad &\mbox{if} ~ \vz \in \bm{Y}, \\ 0,
    &\mbox{otherwise}. \end{array} \right.
    \label{eq:Res7}
\end{equation}
Then, we have the following  estimate of the relative error
\begin{equation}
  \frac{\| \bu - \overline{\bu}_\star\|_1}{\|\bu\|_1} \le \frac{C}{ r} \mathfrak{F}(\alpha,\alpha_3),
  \label{eq:Res8}
\end{equation}
with constant $C$ independent of $\alpha$ and $\alpha_3$.
\end{corollary}

\vspace{0.05in} \noindent This result says that when the sources are
far apart, the effective intensity vector $\overline{\bu}_\star$ is
close to the true solution $\bu$. By definition, the support of
$\overline{\bu}_\star$ is at the source points in $\bm{Y}$.  Its
entries $\overline{u}_\star(\vz)$ at $\vz \in \bm{Y}$ are weighted
averages of the entries of $\bu_\star$ at points $\vz' \in
\mathscr{B}_r(\vz)$, with weights $\mathcal{I}_{\vz,\vz'}$. When $r$
is small, these weights are close to one, so $\overline{u}_\star(\vz)$
is approximately the sum of the entries of $\bu_\star$ supported in
the ball $\mathscr{B}_r(\vz)$.

Theorem \ref{thm.1} and its corollary are not useful when the sources
are clustered together. The next result deals with this case, when the
clusters are well separated.

\vspace{0.05in}
\begin{theorem}
\label{thm.2}
Let $\epsilon \in (0,1)$ be such that there exists a subset
$\bm{Y}_\epsilon$ of $\bm{Y}$, satisfying
\begin{equation}
  \bm{Y} \subset \bigcup_{\vz \in \bm{Y}_\epsilon}\,
  \mathscr{B}_\epsilon(\vz), \qquad \mathscr{B}_\epsilon(\vz) \bigcap
  \mathscr{B}_\epsilon(\vz') = \emptyset, ~ ~ \forall \,\vz,\vz' \in
  \bm{Y}_\epsilon, ~ ~ \vz \ne \vz'.
  \label{eq:Res9}
\end{equation}
Suppose that the points in $\bm{Y}_\epsilon$ are separated by at least
least $\vH = (H,H,H_3)$ in the sense of Definition \ref{def.1}, where
$H$ and $H_3$ satisfy \eqref{eq:asH}, for some $\alpha, \alpha_3 >
1$. Let $r$ satisfy $\epsilon < r < 1$, and decompose the $l_1$
minimizer $\bu_\star$ in \eqref{eq:Res1} as $ \bu_\star =
\bu_\star^{(i)} + \bu_\star^{(o)}, $ where $\bu_\star^{(i)}$ is
supported in the union $\bigcup_{\vz \in
  \bm{Y}_\epsilon}\mathscr{B}_r(\vz)$ of balls centered at the points
in $\bm{Y}_\epsilon$, and $\bu_\star^{(o)}$ is supported in the
complement of this union. There exists a constant $C$ that is
independent of $\alpha$ and $\alpha_3$, such that
\begin{equation}
  \|\bu_\star^{(o)}\|_1 \le \frac{C}{ r} \mathfrak{F}(\alpha,\alpha_3)\|\bu_\star\|_1 +
  \frac{\epsilon}{r} \|\bu\|_1,
\label{eq:Res10}
\end{equation}
where $\bu$ is the vector of true source intensities.
\end{theorem}

\vspace{0.05in} \noindent Equation \eqref{eq:Res9} says that we can
cover the sources with disjoint balls of radius $\epsilon$, centered
at the points in $\bm{Y}_\epsilon$. Thus, we call $\bm{Y}_\epsilon$
the effective support of the sources, and $\epsilon$ the radius of the
clusters. The statement of the theorem is that when the clusters are
well separated, and they have small radius, the $l_1$ minimizer will
be supported in their vicinity. As expected, \eqref{eq:Res10}
converges to \eqref{eq:Res6} in the limit $\epsilon \to 0$.
\subsection{Proofs}
\label{sect:proofs}
We use \cite[Theorem 4.1 and Corollary 4.2]{borcea2015resolution}
which state that for the decomposition of the $l_1$ minimizer
$\bu_\star$ as in Theorem \ref{thm.1}, and for the effective
reconstructed source intensity vector $\overline{\bu}_\star$ defined
in \eqref{eq:Res7}, we have
\begin{equation}
  \|\bu_\star^{(o)}\|_1 \le \frac{2 \mathcal{I}(\bm{Y})}{r}
  \|\bu_\star\|_1 ~ ~ \mbox{and} ~ ~ \frac{\|\bu -
    \overline{\bu}_\star\|_1}{\|\bu\|_1} \le \frac{2
    \mathcal{I}(\bm{Y})}{r}.
  \label{eq:Pf1}
\end{equation}
To determine the interaction coefficient $\mathcal{I}(\bm{Y})$ of the
sources, we estimate first the cross-correlations $\mathcal{I}_{\vz,
  \vz'}$:

\vspace{0.05in}
\begin{lemma} \label{lem.1}
The cross-correlations $\mathcal{I}_{\vz,\vz'}$ defined in
\eqref{eq:Res2} satisfy 
\begin{equation}
  \mathcal{I}_{\vz,\vz'} \approx \exp\left[-\frac{|\bz-\bz'|^2}{4 R^2}
    -\frac{(z_3-z_3')^2 }{4R_3^2}\right], \label{eq:Pf2}
\end{equation}
in the narrowband regime and 
\begin{equation}
  \mathcal{I}_{\vz,\vz'} \le C \exp\left[-\frac{|\bz-\bz'|^2}{8 R^2}
    -\frac{\theta\left|z_3-z_3' +
      \frac{|\bz|^2-|\bz'|^2}{2L}\right|}{R_3}\right],
  \label{eq:Pf2bb}
\end{equation}
in the broadband regime, for all $ \vz,\vz' \in
  \mathfrak{D}_z$,  with $\vz = (\bz,z_3)$ and $ \vz' =
  (\bz',z_3')$. The constant $C$ in \eqref{eq:Pf2bb} is given by 
 \begin{equation}
 C =\frac{3 e^{-\theta^2}}{2 \mbox{erfc}(\sqrt{2} \, \theta)},
 \end{equation}
 with $\theta$ defined in \eqref{eq:defTheta} of order one.
\end{lemma}

\vspace{0.05in} \noindent The proof is in Appendix
\ref{ap:PfLem1}. The next lemma, proved in sections  \ref{sect:pfLemnb} and \ref{sect:pfLembb}, gives the estimate of
$\mathcal{I}(\bm{Y})$, that combined with \eqref{eq:Pf1} proves
Theorem \ref{thm.1} and Corollary \ref{cor.1}.

\vspace{0.05in}
\begin{lemma} \label{lem.2}
Suppose that the points in $\bm{Y}$ are separated by at least $\vH =
(H,H,H_3)$ in the sense of Definition \ref{def.1}, where $H$ and $H_3$
are as in \eqref{eq:asH}, for some $\alpha,\alpha_3 > 1$. The interaction
coefficient satisfies
\begin{equation}
  \mathcal{I}(\bm{Y}) \le C \mathfrak{F}(\alpha,\alpha_3)  \label{eq:Pf3}
\end{equation}
for a constant $C$ that is independent of $\alpha$ and $\alpha_3$ and $\mathfrak{F}(\alpha,\alpha_3)$ as defined in Theorem \ref{thm.1}.
\end{lemma}

\vspace{0.05in}
To prove Theorem \ref{thm.2}, we use \cite[Theorem
  4.4]{borcea2015resolution} which states that
\begin{equation}
\|\bu_\star^{(o)}\|_1 \le \frac{2 \mathcal{I}(\bm{Y}_\epsilon)}{r}
\|\bu_\star\|_1 + \frac{\|\bu\|-\|\overline{\bu}\|_1}{r},
\label{eq:PfT2.1}
\end{equation}
for $\overline{\bu}$ defined by 
\begin{equation}
 \overline{\bu}(\vz) = \left\{ \begin{array}{ll} \sum_{\vz' \in
      \mathscr{B}_\epsilon(\vz)\cap \bm{Y}} u(\vz') \,
    \mathcal{I}_{\vz,\vz'}, \quad &\mbox{if} ~ \vz \in \bm{Y}_\epsilon, \\ 0,
    &\mbox{otherwise}. \end{array} \right.
    \label{eq:PfT2.2}
\end{equation}
The interaction coefficient of the effective support $\bm{Y}_\epsilon$
is as in Lemma \ref{lem.2}, so it remains to estimate the last term in
\eqref{eq:PfT2.1}. Let us define the set
\[
S_{\vz,\epsilon} = \left\{ \vz' \in \bm{Y} ~ ~ \mbox{s.t.} ~ ~ \vz'
\in \mathscr{B}_\epsilon(\vz)\right\},
\]
so that with definition \eqref{eq:PfT2.2} we can write
\[
\|\overline{\bu}\|_1 = \sum_{\vz \in \bm{Y}_\epsilon} \left|
\sum_{\vz'
  \in S_{\vz,\epsilon}} u(\vz') \, \mathcal{I}_{\vz,\vz'} \right| = 
\sum_{\vz \in \bm{Y}_\epsilon} \sum_{\vz'
  \in S_{\vz,\epsilon}} |u(\vz')| \, \mathcal{I}_{\vz,\vz'},
\]
where we used that by definition $ \mathcal{I}_{\vz,\vz'} \ge 0$ and $u(\vz')  = |\rho(\vz')|^2 \ge 0$.  
Since the norm of the
vector of the true source intensities is given by
\[
\|\bu\|_1 = \sum_{\vz \in \bm{Y}} |u(\vz)| = \sum_{\vz \in
  \bm{Y}_\epsilon} \sum_{\vz' \in S_{\vz\,\epsilon}} |u(\vz')|,
\]
we obtain that 
\begin{equation}
\|\bu\|_1 - \|\overline{\bu}\|_1 = \sum_{\vz \in
  \bm{Y}_\epsilon} \sum_{\vz' \in S_{\vz,\epsilon}} |u(\vz')| \left(1 -
\mathcal{I}_{\vz,\vz'}\right) < \epsilon \sum_{\vz \in
  \bm{Y}_\epsilon} \sum_{\vz' \in S_{\vz,\epsilon}} |u(\vz')| = \epsilon
\|\bu\|_1.
\label{eq:PfT2.4}
\end{equation}
The inequality is because 
\[
\Delta(\vz,\vz') = 1 - \mathcal{I}_{\vz,\vz'} < \epsilon, \quad
\forall \, \vz' \in \mathscr{B}_\epsilon(\vz).
\]
The statement of Theorem \ref{thm.2} follows by substitution of
\eqref{eq:PfT2.4} in \eqref{eq:PfT2.1}, and using the estimate in
Lemma \ref{lem.2}, with $\bm{Y}$ replaced by
$\bm{Y}_\epsilon$. $\Box$.

\subsubsection{Proof of Lemma \ref{lem.2} in the narrowband regime}
\label{sect:pfLemnb}
\vspace{0.05in} Recall Definition \ref{def.4} of
$\mathcal{I}(\bm{Y})$, and let $\vz \in \mathfrak{D}_z$ be the
maximizer of the sum in \eqref{eq:Res5}, so that
\begin{equation}
  \mathcal{I}(\bm{Y}) =  \sum_{\vy \in
    \bm{Y}\setminus \mathscr{N}(\vz)} \mathcal{I}_{\vz,\vy}.
  \label{eq:Pf4}
\end{equation}
We denote the components of $\vz$ by $z_j$, for $j=1,2,3$, and
conclude from the source separation assumption in the lemma that the
set
\begin{equation}
  \mathscr{S}_{\vz} = \left\{ \vz'= (z_1',z_2',z_3') \in \mathfrak{D}
  ~ ~ \mbox{s.t.} ~ ~ |z_j-z_j'| < H, ~j = 1, 2, ~ ~ |z_3-z_3'| <
  H_3\right\}
  \label{eq:defSz}
\end{equation}
contains at most one point in $\bm{Y}$.  This may be
$\mathscr{N}(\vz)$, the closest point in $\bm{Y}$ to $\vz$ with
respect to the semimetric $\Delta$, satisfying
\begin{equation}
\mathcal{I}_{\vz,\mathscr{N}(\vz)} \ge \mathcal{I}_{\vz,\vy}.
    \qquad \forall \,
      \vy \in \bm{Y}.
      \label{eq:Pf5}
\end{equation}
Alternatively, $\mathscr{S}_{\vz}$ may be empty or contain another
point in $\bm{Y}$. In either case, we obtain from equations
\eqref{eq:Pf4} and \eqref{eq:Pf5} that
\[
  \mathcal{I}(\bm{Y}) \le \sum_{\vy \in
    \bm{Y}\setminus \mathscr{S}_{\vz}} \mathcal{I}_{\vz,\vy}, 
\]
and from the bound in Lemma \ref{lem.1},  
\begin{equation}
  \mathcal{I}(\bm{Y}) \le C \sum_{\vy \in \bm{Y}\setminus
    \mathscr{S}_{\vz}} \mathscr{E}_{\vz,\vy}, ~ ~\mbox{for} ~ ~
  \mathscr{E}_{\vz,\vy} = \exp\left[-\frac{|\bz-\by|^2}{8 R^2}
    -\frac{(z_3-y_3)^2}{R_3}\right].
  \label{eq:Pf6}
\end{equation}

Using again the source separation assumption in the lemma, we conclude
that for any $\vy \in \bm{Y}$, we can define a set
$\mathscr{H}_{\vy}$, in the form of a rectangular prism of sides $H/2$
in cross-range and $H_3/2$ in range, satisfying
\begin{equation}
  \vy \in \mathscr{H}_{\vy} ~ ~ \mbox{and} ~ ~ \mathscr{H}_{\vy}
  \bigcap \mathscr{H}_{\vy'} = \emptyset, \qquad \forall \, \vy \ne
  \vy' \in \bm{Y}.
  \label{eq:Pf.7}
\end{equation}
There are many such sets, but we make our choice so that $\vy$ is the
furthermost point to $\vz$ in  $\mathscr{H}_{\vy}$, satisfying
\begin{equation}
  \mathscr{E}_{\vz,\vy} \le \mathscr{E}_{\vz,\vz'}, \qquad \forall \,
  \vz' \in \mathscr{H}_{\vy}.
 \label{eq:Pf.8}
\end{equation}
This allows us to write
\begin{equation}
  \mathscr{E}_{\vz,\vy} \le \frac{8}{H^2 H_3} \int_{\mathscr{H}_{\vy}} d \vz'
  \mathscr{E}_{\vz,\vz'}, \qquad \forall \, \vy \in \bm{Y},
 \label{eq:Pf.9}
\end{equation}  
and obtain from \eqref{eq:Pf6} that
\begin{equation}
  \mathcal{I}(\bm{Y}) \le \frac{8 C^2}{H^2 H_3} \sum_{\by \in \bm{Y}
    \setminus \mathscr{S}_{\vz}} \int_{\mathscr{H}_{\vy}} d \vz'
  \mathscr{E}_{\vz,\vz'} \le \frac{8 C^2}{H^2 H_3} \int_{\mathbb{R}^3
    \setminus \mathscr{S}_{\vz,\frac{1}{2}}} d \vz' \mathscr{E}_{\vz,\vz'},
  \label{eq:Pf10}
\end{equation}
with $\mathscr{S}_{\vz, \frac{1}{2}}$ defined as in \eqref{eq:defSz}, with
half the values of $H$ and $H_3$,
\begin{equation}
  \mathscr{S}_{\vz,\frac{1}{2}} = \left\{ \vz'= (z_1',z_2',z_3') \in
  \mathfrak{D} ~ ~ \mbox{s.t.} ~ ~ |z_j-z_j'| < H/2, ~j = 1, 2, ~ ~
  |z_3-z_3'| < H_3/2\right\}.
  \label{eq:defSzhalf}
\end{equation}
The last inequality in \eqref{eq:Pf10} is because the integrand is
positive, the sets $\mathscr{H}_{\vy}$ are disjoint, and
\[
\bigcup_{\vy \in \bm{Y}\setminus \mathscr{S}_{\vz}} \mathscr{H}_{\vy} \subset
\mathbb{R}^3
\setminus \mathscr{S}_{\vz,\frac{1}{2}}.
\]

We estimate the integral in \eqref{eq:Pf10} by decomposing the set
$\mathscr{S}_{\vz,\frac{1}{2}}^c = \mathbb{R}^3\setminus
\mathscr{S}_{\vz,\frac{1}{2}}$ in three components denoted by
$\mathscr{C}_{\vz,j}$, where
\[
\mathscr{C}_{\vz,j} = \left\{\vz' \in \mathbb{R}^3 ~ ~ \mbox{s.t.} ~ ~ |z_j-z_j'|
\ge H/2 \right\}, \quad j = 1,2,
\]
and
\[
\mathscr{C}_{\vz,3} = \left\{\vz' \in \mathbb{R}^3 ~ ~ \mbox{s.t.} ~
~ |z_j-z_j'| < H/2, ~ |z_3-z_3'| \ge H_3/2 \right\}.
\]
We have
\begin{align}
  \frac{8}{H^2 H_3} \int_{\mathscr{C}_{\vz,1}} d \vz' \, \mathscr{E}_{\vz,\vz'}
  &= \frac{8}{H^2 H_3} \int_{|z_1'-z_1| \ge H/2} d z_1'\,
  e^{-\frac{(z_1'-z_1)^2}{4 R^2}} \int_{-\infty}^\infty d z_2'\,
  e^{-\frac{(z_2'-z_2)^2}{4 R^2}} \int_{-\infty}^\infty d z_3' \,
  e^{-\frac{(z_3'-z_3)^2}{4 R^2_3}} \nonumber \\ &=\frac{32\pi R
    R_3}{H^2 H_3} 2 \int_{H/2}^\infty dt \, e^{-\frac{t^2}{4 R^2}}
  \nonumber \\ &= \frac{64 \sqrt{\pi}}{\alpha^2 \alpha_3} \mbox{erfc}
  \left(\frac{\alpha}{4}\right),
  \label{eq:Pf11}
\end{align}
where we evaluated the integrals over $z_3'$ and $z_2'$ in the second
line, and used \eqref{eq:asH} in the last line.  The integral over
$\mathscr{C}_{\vz,2}$ is the same, so it remains to estimate
\begin{align}
  \frac{8}{H^2 H_3} \int_{\mathscr{C}_{\vz,3}} d \vz' \, \mathscr{E}_{\vz,\vz'}
   = \frac{8}{H^2 H_3} \int_{|z_1'-z_1| < H/2} d z_1'\,
  e^{-\frac{(z_1'-z_1)^2}{4 R^2}} \int_{|z_2'-z_2| < H/2} d z_2'\,
  e^{-\frac{(z_2'-z_2)^2}{4 R^2}}\nonumber \\
  \times \int_{|z_3'-z_3| \ge H_3/2} d z_3' \,
  e^{-\frac{(z_3'-z_3)^2}{4 R^2_3}}. \label{eq:Pf12}
\end{align}
We bound the integrals over $z_1'$ and $z_2'$ by those of the real line, and 
rewrite the integral over $z_3'$ in terms of the complementary error function,
to obtain 
\begin{align}
\frac{8}{H^2 H_3 } \int_{\mathscr{C}_{\vz,3}} d \vz' \,
\mathscr{E}_{\vz,\vz'} &\le \frac{64 \sqrt{\pi}}{\alpha^2 \alpha_3} \mbox{erfc}
  \left(\frac{\alpha_3}{4}\right).
\label{eq:Pf14}
\end{align}
The statement of Lemma \ref{lem.2} follows from \eqref{eq:Pf10}, with
right hand side given by the sum of the integrals over $\mathscr{C}_{\vz,1}$ and
$\mathscr{C}_{\vz,2}$ estimated in \eqref{eq:Pf11}, and over $\mathscr{C}_{\vz,3}$,
estimated in \eqref{eq:Pf14}. $\Box$

\subsubsection{Proof of Lemma \ref{lem.2} in the broadband regime}
\label{sect:pfLembb}
We obtain from Lemma \ref{lem.1}, the same way as above, and with the same notation, that 
\begin{equation}
  \mathcal{I}(\bm{Y}) \le C \sum_{\vy \in \bm{Y}\setminus
    \mathscr{S}_{\vz}} \mathscr{E}_{\vz,\vy}, ~ ~\mbox{for} ~ ~
  \mathscr{E}_{\vz,\vy} = \exp\left[-\frac{|\bz-\by|^2}{8 R^2}
    -\frac{\theta\left|z_3-y_3 +
      \frac{|\bz|^2-|\by|^2}{2L}\right|}{R_3}\right].
  \label{eq:Pf6bb}
\end{equation}
We also define as before, using  the source separation assumption in the lemma, the set
$\mathscr{H}_{\vy}$, satisfying \eqref{eq:Pf.7} and \eqref{eq:Pf.8}.
This leads us to the bound \eqref{eq:Pf10}, with the set $\mathscr{S}_{\vz,\frac{1}{2}}$ defined in 
\eqref{eq:defSzhalf}.

We estimate the integral in \eqref{eq:Pf10} by decomposing the set
$\mathbb{R}^3\setminus \mathscr{S}_{\vz,\frac{1}{2}}$ in three
parts $S_{\vz,j}$, where
\[
S_{\vz,j} = \left\{\vz' \in \mathbb{R}^3 ~ ~ \mbox{s.t.} ~ ~ |z_j-z_j'|
\ge H/2 \right\}, \quad j = 1,2,
\]
and
\[
S_{\vz,3} = \left\{\vz' \in \mathbb{R}^3 ~ ~ \mbox{s.t.} ~ ~ |z_j-z_j'| < H/2, ~
|z_3-z_3'| \ge H_3/2 \right\}.
\]
We have
\begin{align}
  \frac{8}{H^2 H_3} \int_{S_{\vz,1}} d \vz' \, \mathscr{E}_{\vz,\vz'}
  &= \frac{8}{H^2 H_3} \int_{|z_1'-z_1| \ge H/2} d z_1'\,
  e^{-\frac{(z_1'-z_1)^2}{8 R^2}} \int_{-\infty}^\infty d z_2'\,
  e^{-\frac{(z_2'-z_2)^2}{8 R^2}} \int_{-\infty}^\infty d z_3' \,
  e^{-\frac{\theta}{ R_3} \left|z_3'-z_3 +
    \frac{|\bz|^2-|\bz'|^2}{2 L} \right|} \nonumber \\ &=\frac{128
    \pi R^2 R_3}{H^2 H_3 \theta} \mbox{erfc}\left(\frac{H}{4
    \sqrt{2} R}\right)\nonumber \\ &\le \frac{128\pi }{\alpha^2 \alpha_3 \theta}e^{-\left(\frac{h}{4 \sqrt{2} R}\right)^2}
  \nonumber \\ &\le \frac{16 \pi}{\alpha^3} e^{-\left(\frac{\alpha}{4
      \sqrt{2}}\right)^2},
  \label{eq:Pf11bb}
\end{align}
where we evaluated the integrals over $z_3'$ and $z_2'$ in the second
line. The first inequality is because the complementary error function satisfies $\mbox{erfc}(x) \le \exp(-x^2)$,
and the second inequality is by the assumption on $\alpha_3$. The
integral over $S_{\vz,2}$ is the same, so it remains to estimate
\begin{align}
  \frac{8}{H^2 H_3} \int_{S_{\vz,3}} d \vz' \, \mathscr{E}_{\vz,\vz'}
   \approx \frac{8}{H^2 H_3} \int_{|z_1'-z_1| < H/2} d z_1'\,
  e^{-\frac{(z_1'-z_1)^2}{8 R^2}} \int_{|z_2'-z_2| < H/2} d z_2'\,
  e^{-\frac{(z_2'-z_2)^2}{8 R^2}}\nonumber \\
  \times \int_{|z_3'-z_3| \ge H_3/2} d z_3' \,
  e^{-\frac{\theta}{R_3} \left|z_3'-z_3 +
    \frac{|\bz'|^2-|\bz|^2}{2 L} \right|}. \label{eq:Pf12bb}
\end{align}
Because $|z_j'-z_j| < H/2$ for $j = 1,2$, and therefore $|\bz'-\bz| <
H/\sqrt{2}$, we have
\begin{align}
  \theta \left|\frac{|\bz'|^2-|\bz|^2}{2 L R_3}\right| \le
  \frac{\theta|\bz'-\bz| |\bz'+\bz|}{2 L R_3} \le \frac{\theta H D}{2
   L R_3} = \frac{\theta \alpha D R}{2  L R_3}
  = \frac{3 D \alpha }{a} \lesssim {3 \alpha}.
\end{align}
Here we used that $|\bz'+\bz| \le D \sqrt{2}$ for all $\bz,\bz' \in
[-D/2,D/2] \times [-D/2,D/2]$, and substituted $H = R \alpha$ and
definitions \eqref{eq:prop3.1} and \eqref{eq:defTheta}. The last inequality is by assumption
\eqref{eq:as5}. With this bound we can estimate the integral
over $z_3'$ as follows
\begin{align}
  \int_{|z_3'-z_3| \ge H_3/2} d z_3' \, e^{-\frac{\theta}{
      R_3} \left|z_3'-z_3 + \frac{|\bz'|^2-|\bz|^2}{2 L} \right|} &=
  \frac{R_3}{\theta} \left[\int_{\frac{\theta H_3}{2R_3}}^\infty dt \,
    e^{-\left|-t +
      \frac{\theta(|\bz'|^2-|\bz|^2)}{2 L R_3}\right|} +
    \int_{\frac{\theta H_3}{2 R_3}}^\infty dt \,
    e^{-\left|t + \frac{\theta(|\bz'|^2-|\bz|^2)}{2
        LR_3}\right|}\right] \nonumber \\ & \le \frac{2 R_3}{\theta}
  \int_{\frac{\theta H_3}{2 R_3}}^\infty dt \, e^{-(t-3
      \alpha)} \nonumber \\ &= \frac{2 R_3}{\theta}
  e^{- \left(\frac{\theta\alpha_3}{2 } - 3 \alpha
    \right)} \int_0^\infty dt e^{-t} \nonumber \\ &\le
    \frac{2 R_3}{\theta}
    e^{-\alpha},
    \label{eq:Pf13bb}
\end{align}
where the last inequality is by the assumption $H_3/R_3 = \alpha_3 \ge 8 \alpha/\theta$. Substituting in
\eqref{eq:Pf12bb} we get
\begin{align}
\frac{8}{H^2 H_3 } \int_{S_{\vz,3}} d \vz' \,
\mathscr{E}_{\vz,\vz'} &\le \frac{16 R_3}{H^2 H_3
  \theta}e^{-\alpha} \int_{|z_1'-z_1| < H/2} d z_1'\,
e^{-\frac{(z_1'-z_1)^2}{8 R^2}} \int_{|z_2'-z_2| < H/2} d z_2'\,
e^{-\frac{(z_2'-z_2)^2}{8 R^2}} \nonumber \\ &\le \frac{128
  \pi R^2 R_3}{H^2 H_3 \theta} e^{-\alpha} = \frac{128 \pi}{\alpha^2 \alpha_3 \theta} \le
\frac{16 \pi}{\alpha^3} e^{-\alpha}.
\label{eq:Pf14bb}
\end{align}
Here we bounded each Gaussian integral by $2 \sqrt{2 \pi} R$, which is
the integral over the real line, and used again the assumption on
$\alpha_3$.

The statement of Lemma \ref{lem.2} follows from \eqref{eq:Pf10}, with
right hand side given by the sum of the integrals over $S_{\vz,1}$ and
$S_{\vz,2}$ estimated in \eqref{eq:Pf11bb}, and over $S_{\vz,3}$,
estimated in \eqref{eq:Pf14bb}. $\Box$

\section{Numerical simulations}
\label{sect:num}
We present here numerical simulations obtained with the wave propagation model described in equation 
\eqref{eq:a6}, with random travel time fluctuations  computed by the line integrals in \eqref{eq:a7p}, in  one 
realization of the random process $\mu$. We  generate it numerically using random Fourier 
series \cite{devroye2006nonuniform}, for the Gaussian autocorrelation \eqref{eq:a2}.  All the length scales are 
normalized by $\ell$ in the simulations, and are chosen to satisfy marginally the  assumptions in section \ref{sect:sc4}.
Specifically, we take $\la_o = 1.75 \cdot 10^{-4} \,\ell $ and  $L = 800\,  \ell,$
so that 
\[
\sqrt{\la_o L} = 0.12 <  \ell <  9.72 \,  \ell = (\la_o L^3)^{1/4},  
\]
and the aperture is $a = 16 \ell$. This is  slightly larger than the bound in \eqref{eq:as1}, but  we also have the
apodization \eqref{eq:R3}.  We verify that 
\[
\frac{\la_o}{\sqrt{\ell L}} = 6.17 \cdot 10^{-7} \ll  \frac{\sqrt{\la_o \ell}}{L} = 5.22 \cdot 10^{-6} \ll \left(\frac{\ell}{L}\right)^{3/2} = 4.42 \cdot 10^{-5},
\]
and we take the strength of the fluctuations $\sigma = 1.5 \cdot 10^{-6}$. With this choice we obtain
\[
\frac{\Omega_d}{\om_o} = 0.083 \quad  \mbox{and} \quad X_d = 0.14 \ell.
\]
We show results in two dimensions, for a linear array and a narrowband regime with bandwidth $B = 0.0032 \om_o$. Since in this regime  we can only expect improvements in the cross-range localization of the sources, we focus attention at a given range, 
and display cross-range sections of the images.

The migration and CINT images are calculated as in equations \eqref{eq:f5} and \eqref{eq:f6} from the data contaminated with $5\%$ additive, uncorrelated, Gaussian noise. The thresholding parameters in the CINT image formation are $\Omega = B/2$ and $X = X_d/2$, 
and the sources are off the reconstruction mesh. The mesh size is  $H = \la_o L/(6X)$, unless stated otherwise 
The optimization formulation is 
\begin{equation}
\min_{\bu} \|\bu\|_1 \quad \mbox{such that } \quad \|\bm{\mathcal{M}}\, \bu - \bd\|_{2} \le \delta,
\label{eq:OPTForm}
\end{equation}
with tolerance $\delta = 0.05 \|\bd\|_2$. The entries of matrix $\bm{\mathcal{M}}$ are as defined in \eqref{eq:R8}, 
with constant $C$ given in appendix \ref{sect:apCINT}. For comparison, we also present the results of a 
direct application of $L_1$ optimization to the array data, without using the CINT image formation, as in \cite{borcea2015resolution}. We 
call this method "direct $l_1$ optimization" and  refer 
to \cite[Appendix A]{borcea2015resolution} for details. We also refer to \eqref{eq:OPTForm} as "$l_1$ optimization" and 
solve it with the software package \cite{cvx}.

\begin{figure}[t]
\begin{centering}
{{\includegraphics[width=0.35\textwidth]{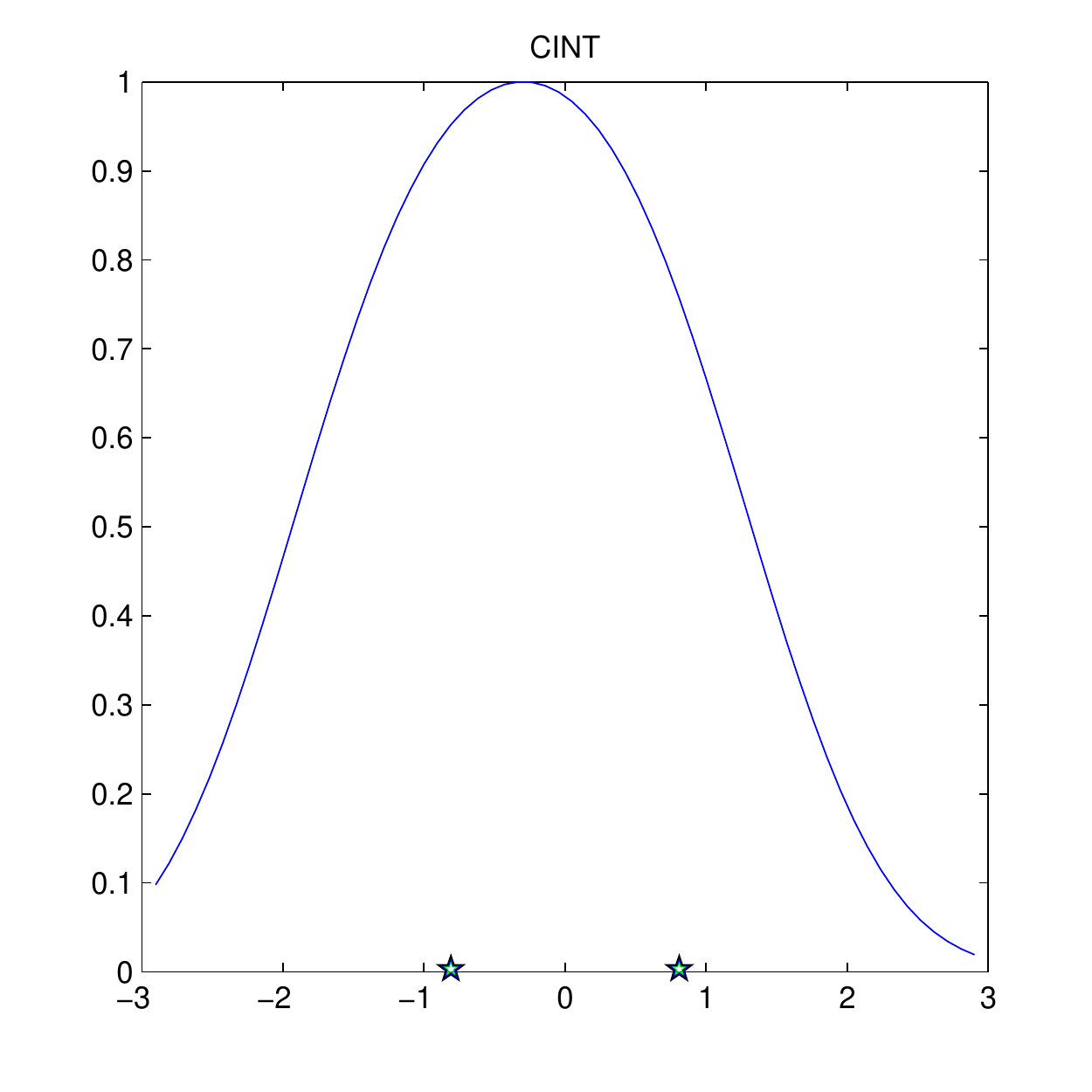}}}
{{\includegraphics[width=0.35\textwidth]{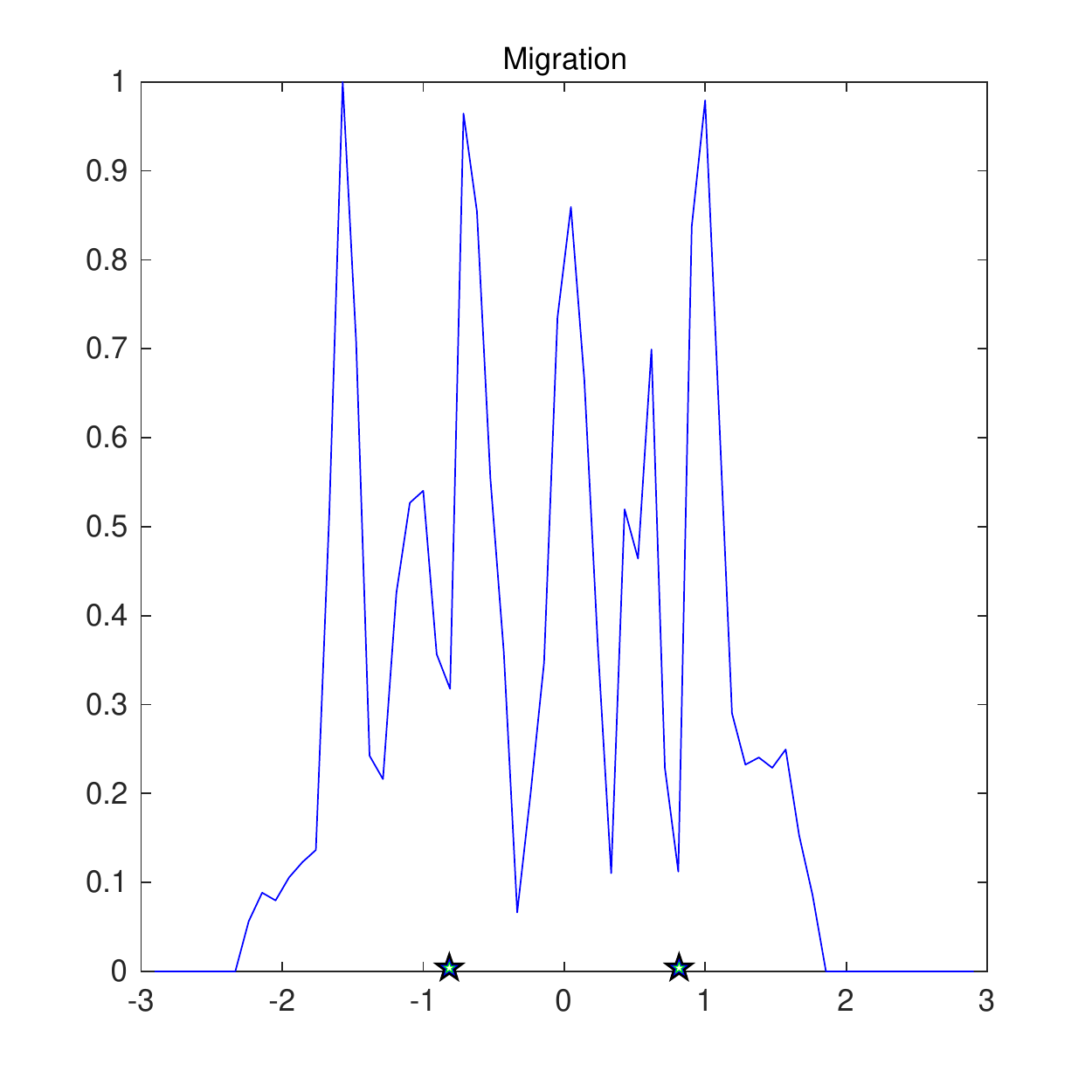}}}

\vspace{-0.15in}
 {{\includegraphics[width=0.35\textwidth]{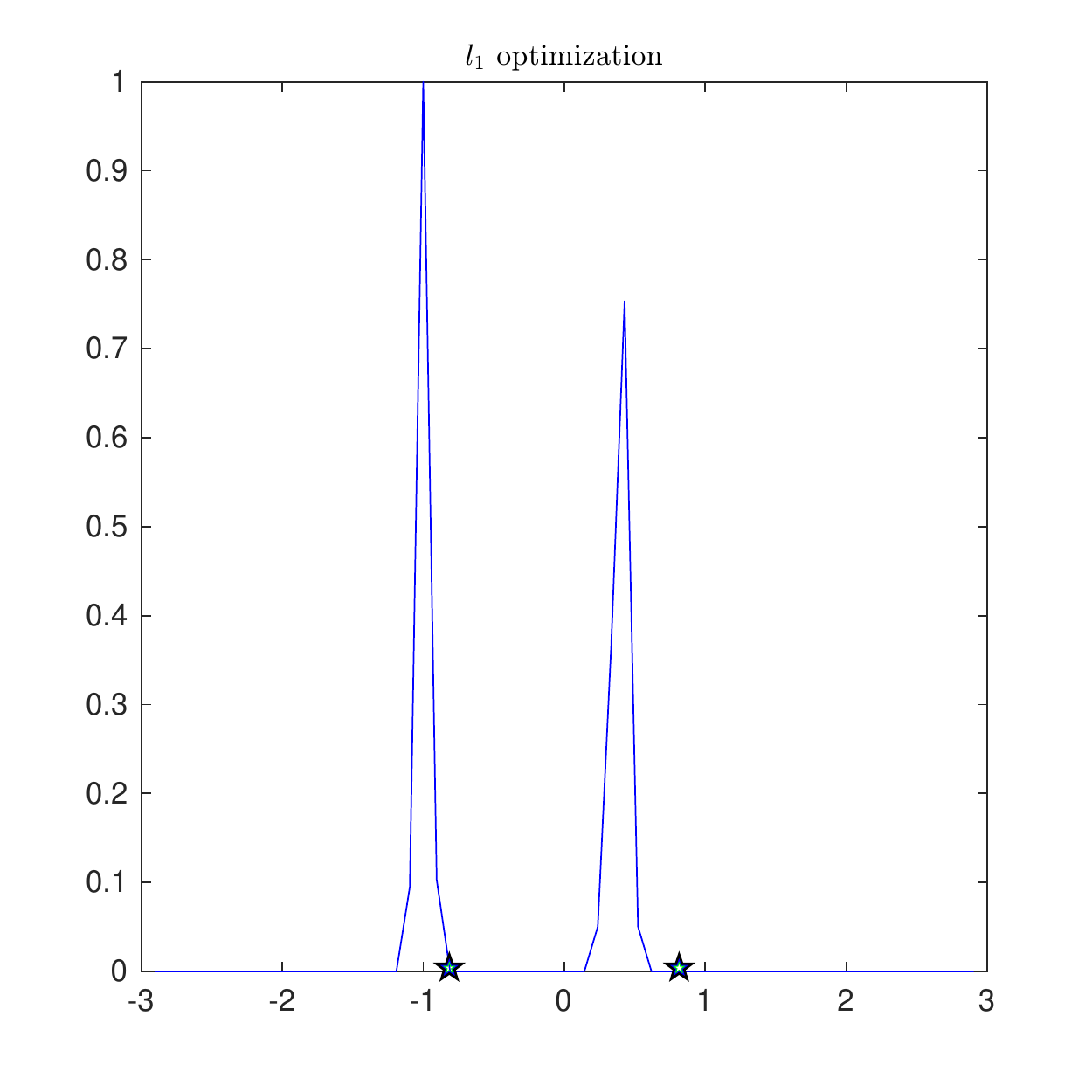}}}
{{\includegraphics[width=0.35\textwidth]{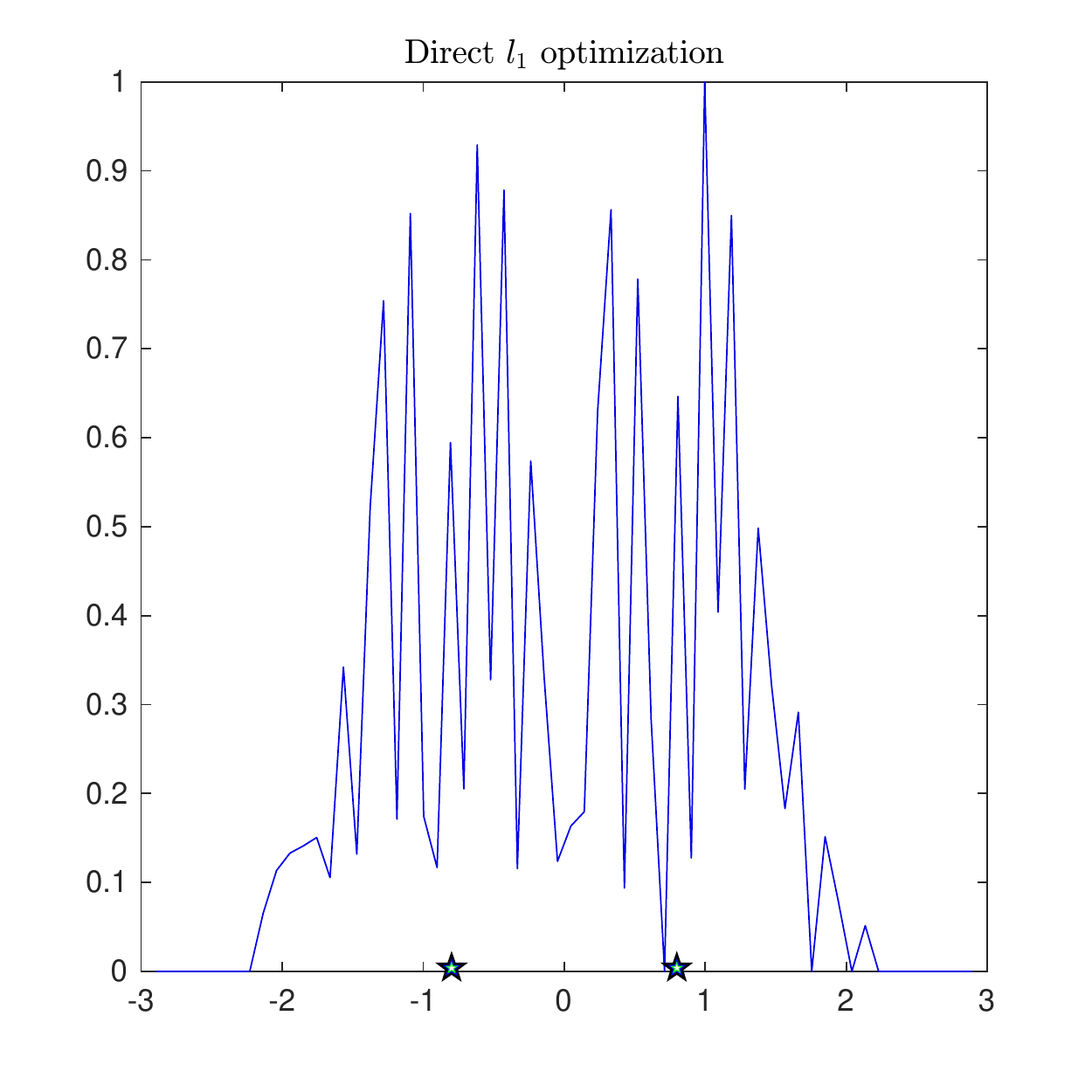}}} 
\par\end{centering}

\vspace{-0.2in}
\caption{Results in one realization of the random medium. The abscissa is the cross-range in 
units $\la_o L/X$, and the source locations are shown with the stars. 
We display the CINT and migration images in the top row, and the 
$l_1$ and direct $l_1$ optimization results in the bottom row.}
\label{fig:1}
\end{figure}

\begin{figure}[h]
\begin{centering}
\includegraphics[width=0.35\textwidth]{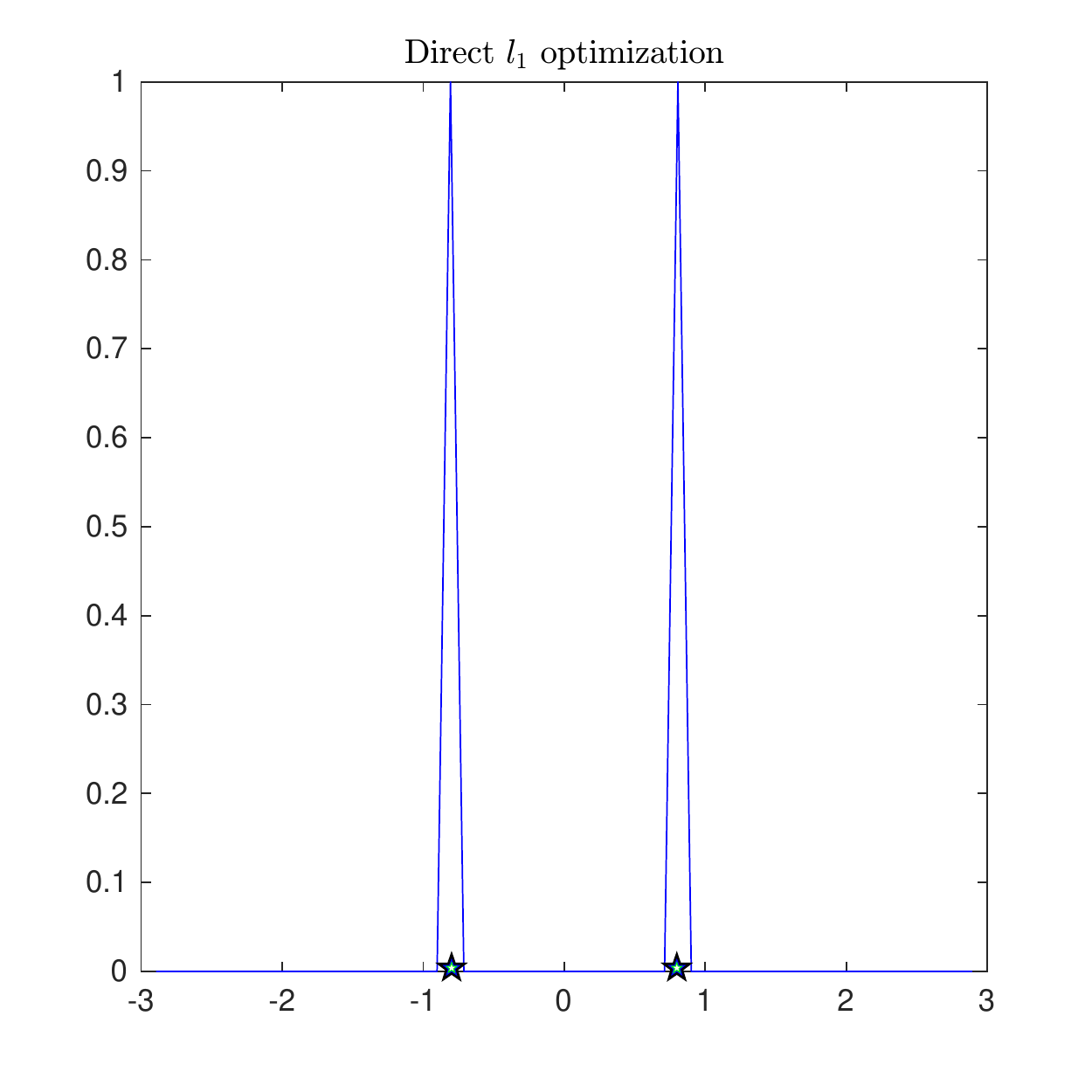}
\par\end{centering}
\vspace{-0.2in}
\caption{Result of direct $l_1$ optimization in the homogeneous medium. The abscissa is the cross-range in 
units $\la_o L/X$, and the source locations are shown with the stars. }
\label{fig:2}
\end{figure}

We begin with the results in Figure \ref{fig:1}, for two sources that are at about $2 \la_o L/X$ apart. The exact source locations
are indicated on the abscissa in the plots, where the units are in $\la_o L/X$. We display in the top 
row the CINT and migration images, and in the bottom row the solutions of the $l_1$ optimization and the 
direct $l_1$ optimization. Both migration and direct $l_1$ optimization give spurious peaks, due 
to the random medium. To confirm this, we show in Figure \ref{fig:2} the results of the direct $l_1$ optimization for the 
same sources in the homogeneous medium, where the reconstruction is very good. The CINT image shown in the top 
left plot is blurry, and it cannot distinguish the two sources. The $l_1$ optimization improves the result, although there 
is a small shift in the estimate of the source locations. This shift changes from one realization to another, and it is due 
to the small random fluctuations of the CINT image. 

\begin{figure}[h]
\begin{centering}
{{\includegraphics[width=0.35\textwidth]{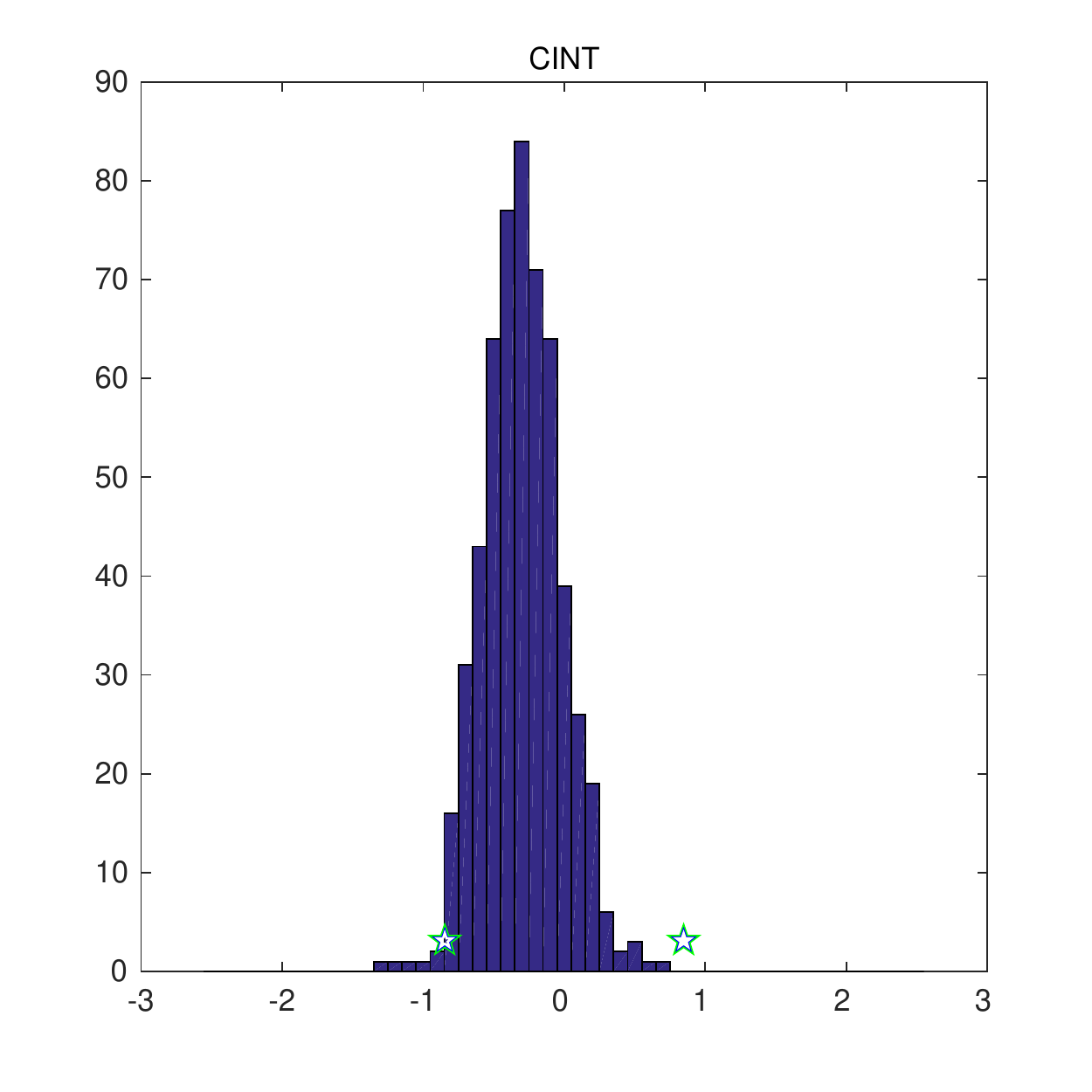}}}
{{\includegraphics[width=0.35\textwidth]{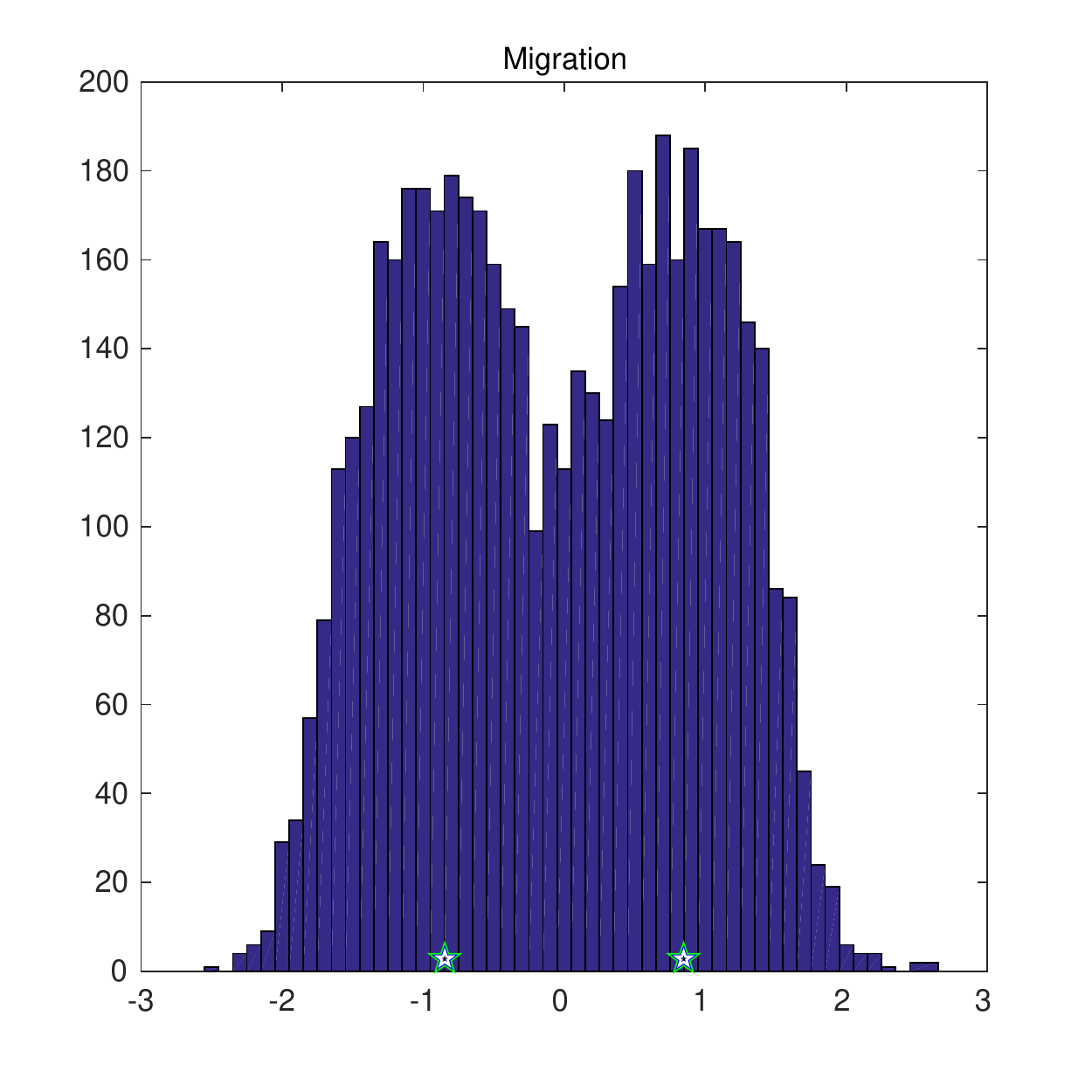}}}

\vspace{-0.16in}
 {{\includegraphics[width=0.35\textwidth]{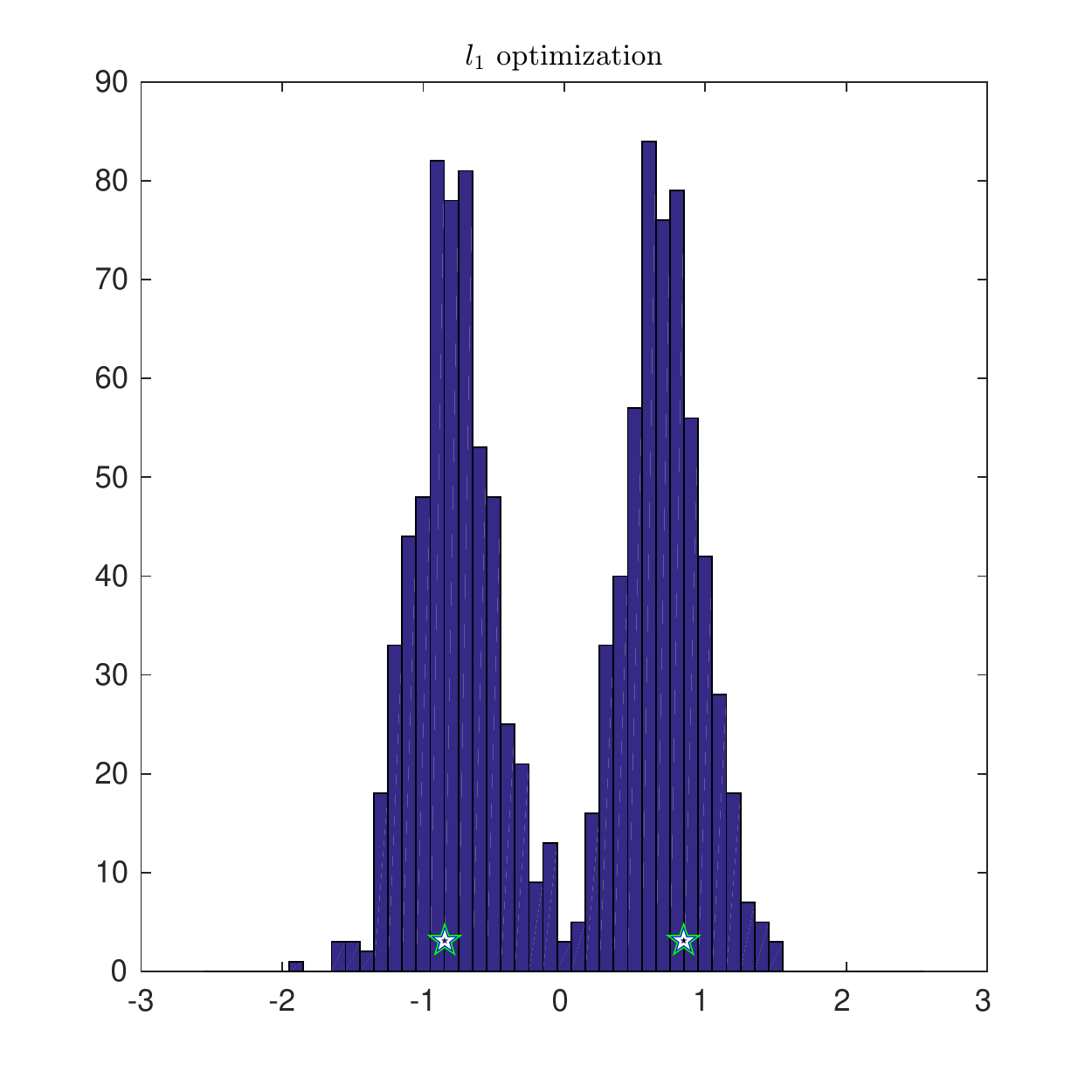}}}
{{\includegraphics[width=0.35\textwidth]{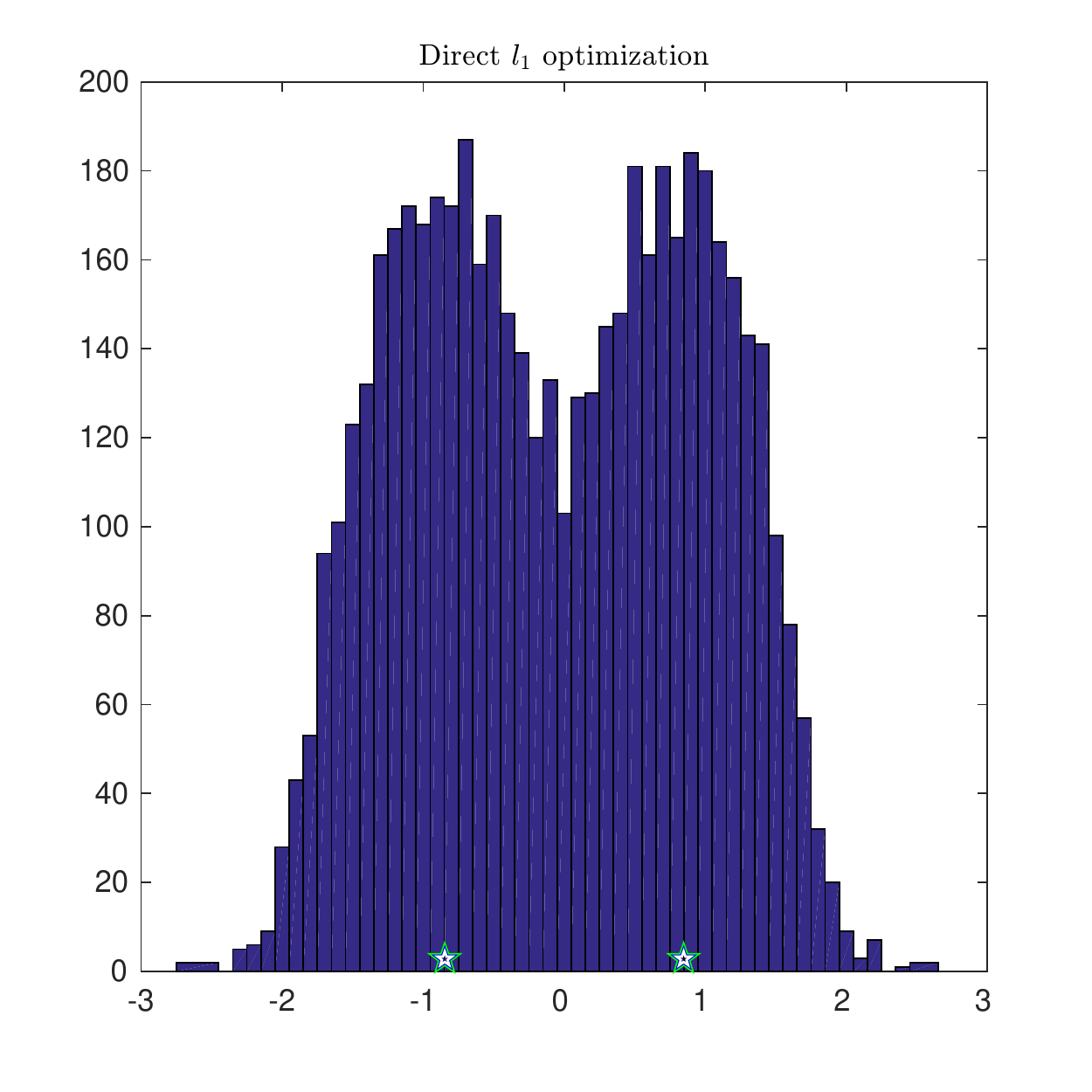}}} 
\par\end{centering}
\vspace{-0.2in}
\caption{Histogram of the number of peaks obtained in $500$ realizations of the random medium. The abscissa is the cross-range in 
units $\la_o L/X$, and the source locations are shown with the stars. 
We display the results for the CINT and migration images in the top row, and the 
$l_1$ and direct $l_1$ optimization results in the bottom row.
The heights of the histograms
are different because each method finds a different number of peaks.
On the average, the number of peaks found in simulations are\;\; Migration:\ 9.5,\;\;\;
CINT: 1.01,\;\;\; $l_1$ optimization:\ 2.04,\;\;\; Direct $l_1$ optimization:\ 9.65. 
}
\label{fig:3}
\end{figure}

To illustrate the robustness of the methods to different realizations of the random medium, we display in Figure \ref{fig:3} the histograms 
of the number of peaks found by each method at a given cross-range location. We define filtered peaks as local maxima whose values are above $33\%$
of the maximum of the image. The height of the histograms varies among the plots in Figure \ref{fig:3}  because each method finds a different number of peaks. On the average, the migration images have $9.5$ peaks, the direct $l_1$ method finds $9.65$ peaks, the CINT image has $1.01$ peaks and the 
$l_1$ optimization finds $2.04$ peaks. While both migration and direct $l_1$ find many spurious peaks, that are far from the source
locations, the $l_1$ optimization separates well the two sources and almost always peaks at their true locations. 

\begin{figure}[h]
\begin{centering}
\includegraphics[width=0.35\textwidth]{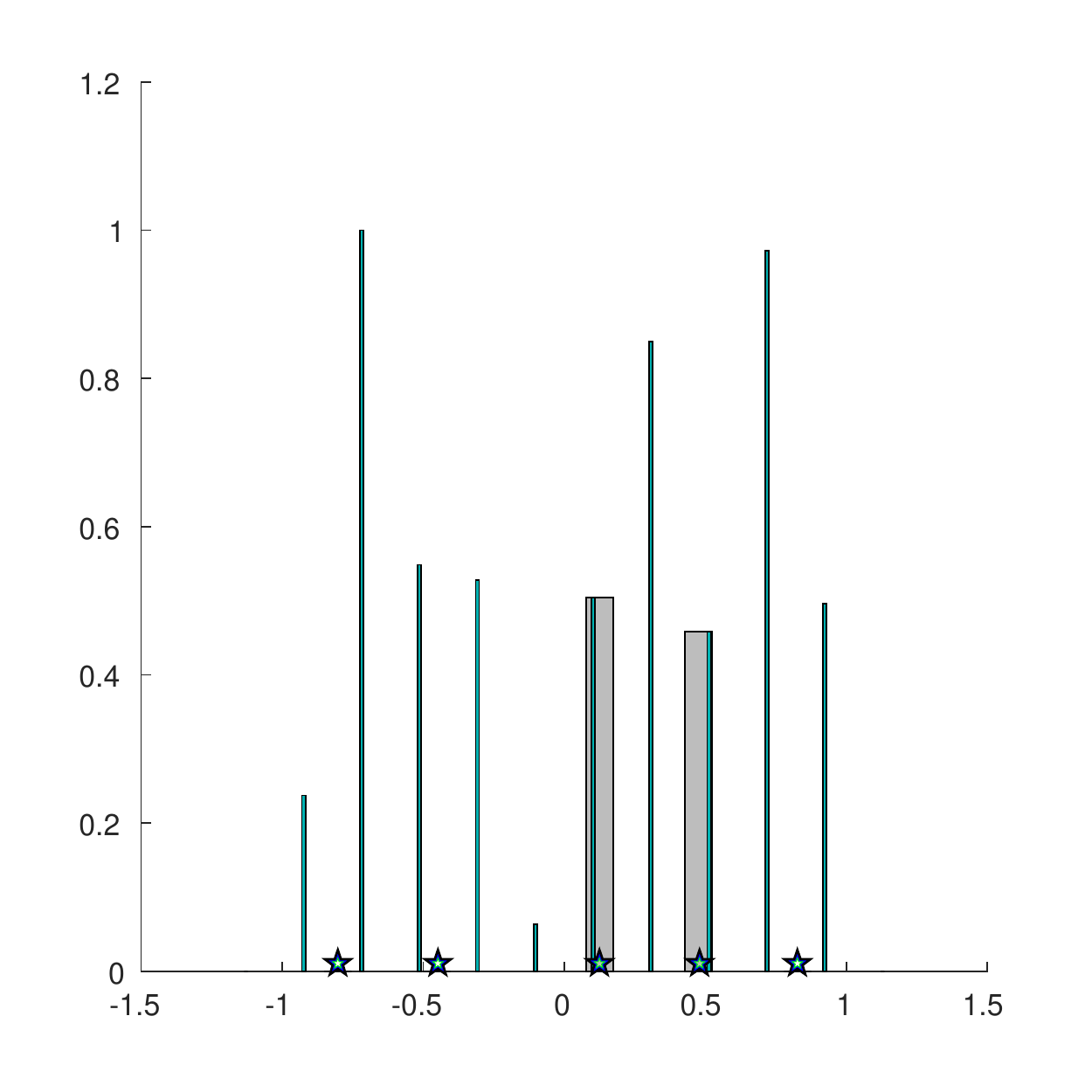}\includegraphics[width=0.35\textwidth]{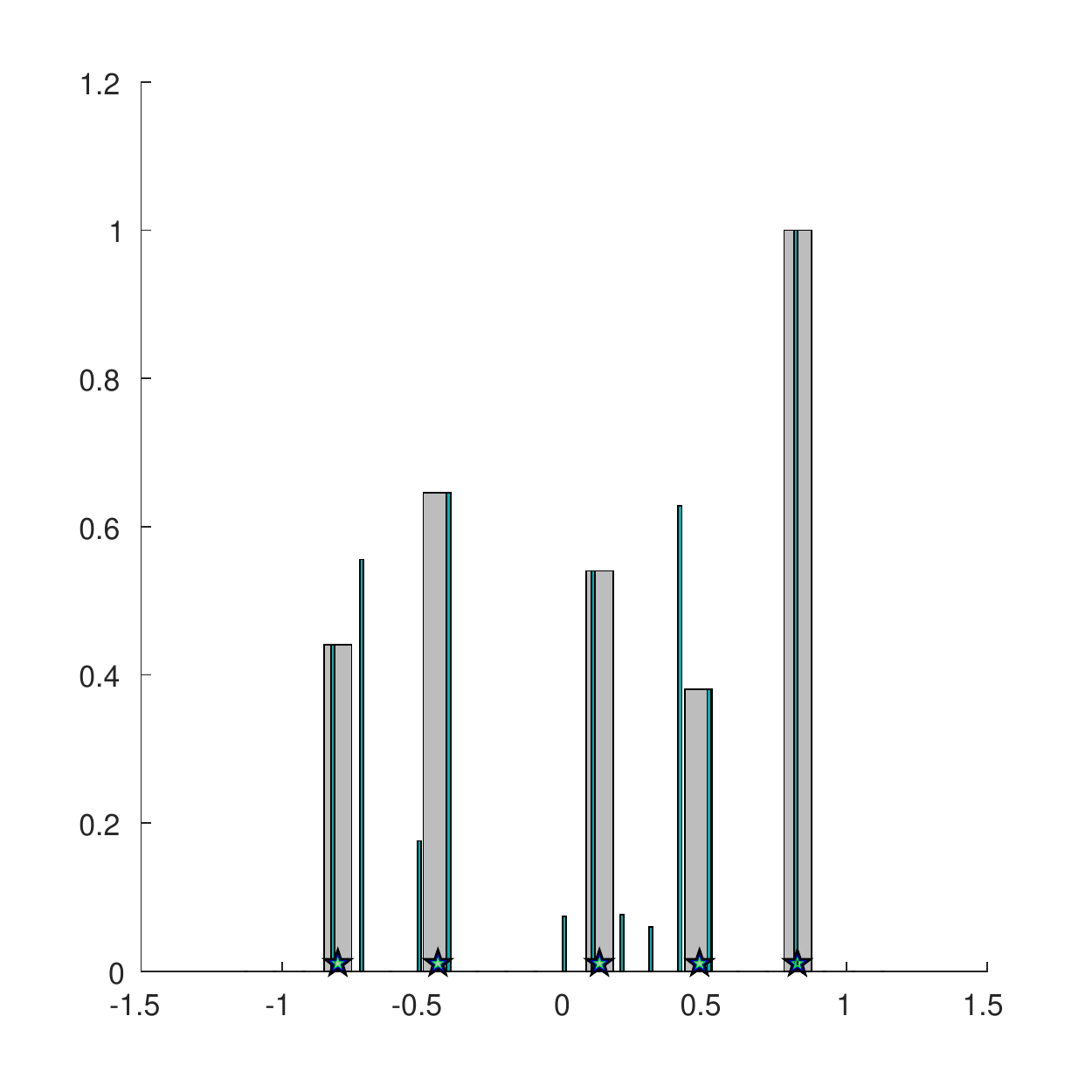}

\vspace{-0.15in}
\includegraphics[width=0.35\textwidth]{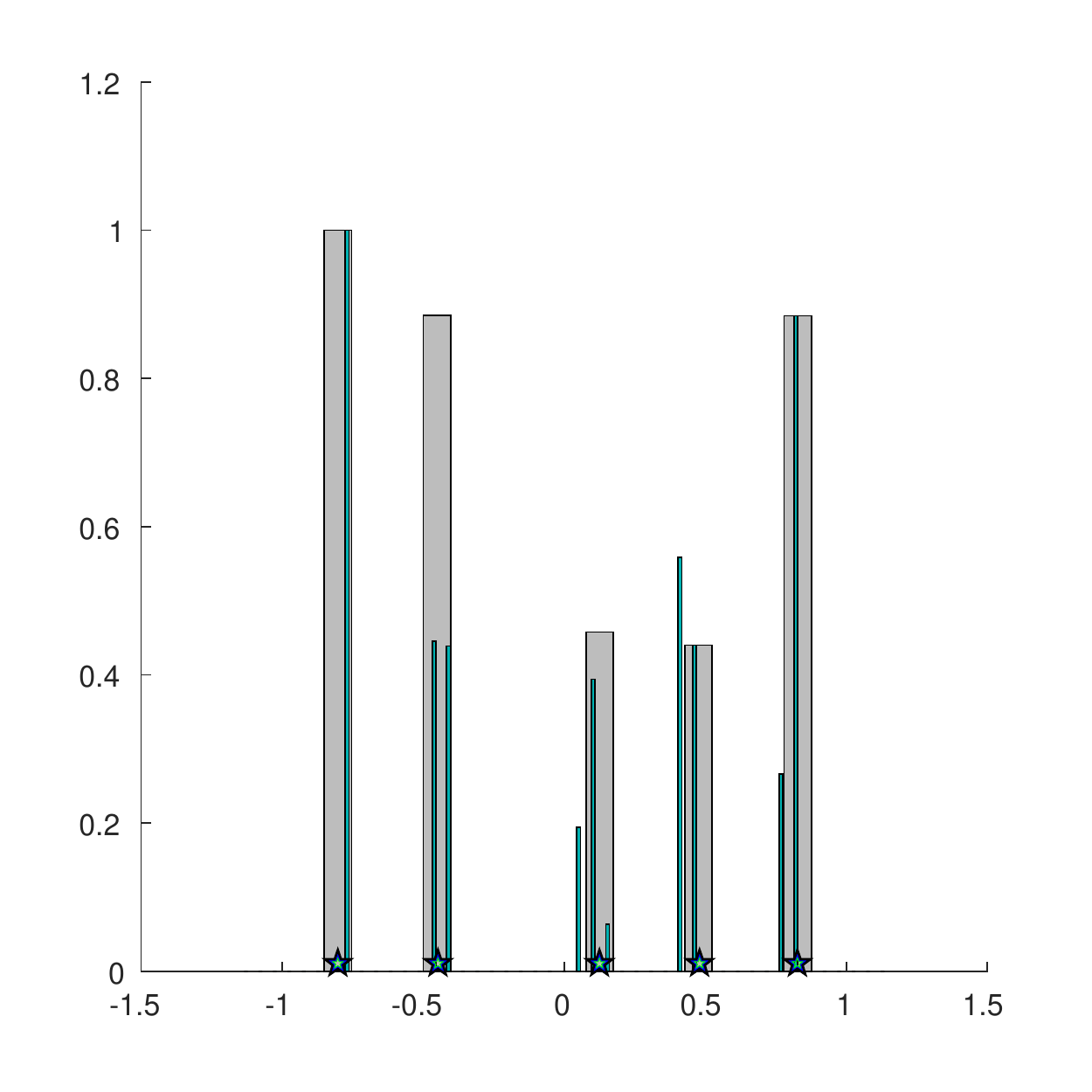}\includegraphics[width=0.35\textwidth]{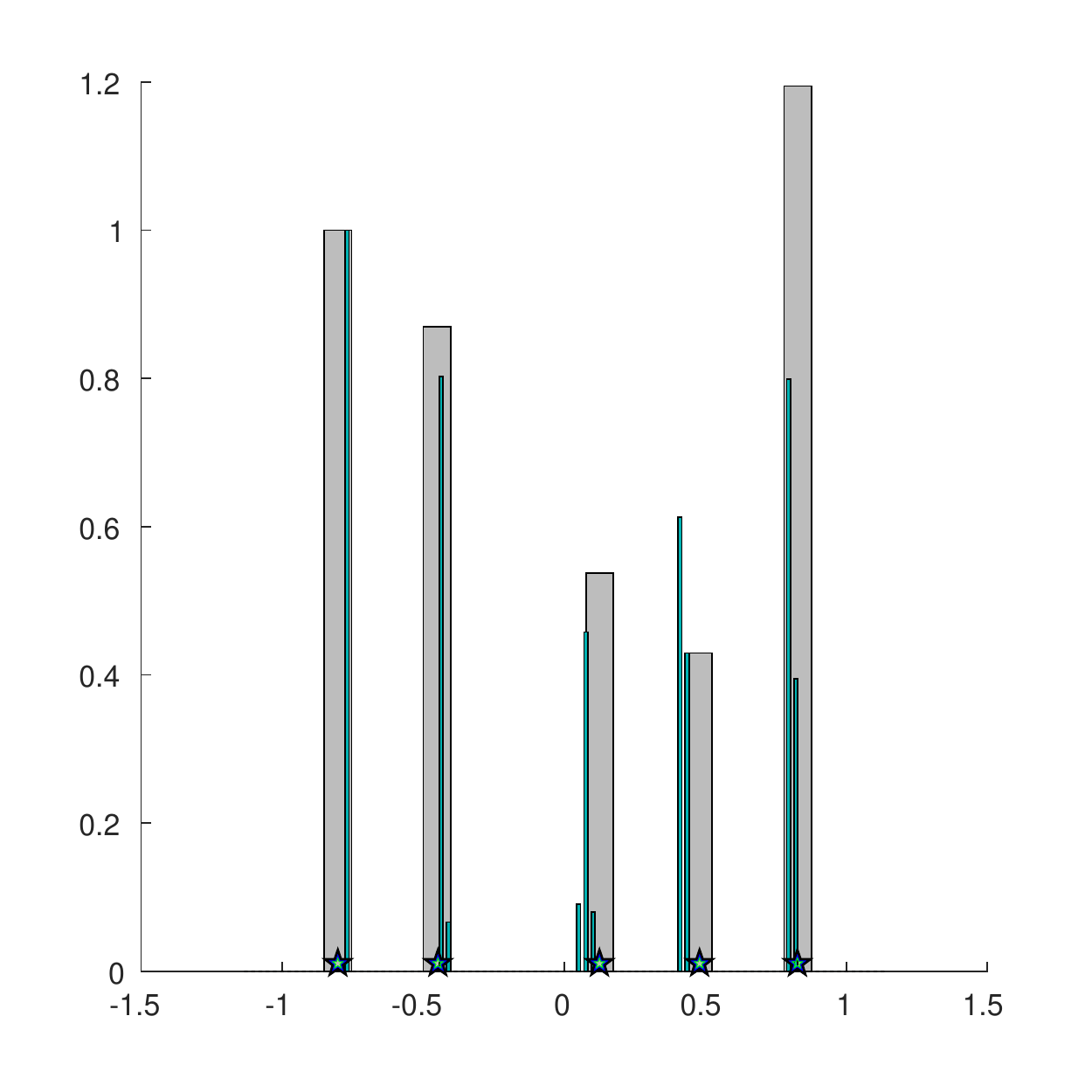}
\par\end{centering}
\vspace{-0.2in}
\caption{We display $l_1$ optimization results in one realization of the random medium, 
for $5$ sources with locations indicated by the stars in the abscissa. The units of the 
abscissa are in $\la_o L/X$. The dark  thin
bars show the reconstruction and the light gray bars give the aggregated values of the reconstruction 
in the intervals of length $r = L\lambda_{o}/(4X)$, centered at the source locations. The mesh size is, 
clockwise, starting from the top left, $H = 1, 1/2, 1/8$ and $1/4$ of $\la_o L/X$. The results improve as we 
refine the mesh.}
\label{fig:4}
\end{figure}
We display in Figure \ref{fig:4} the effect of the mesh size $H$ on the $l_1$ optimization results. 
Because we have $5$ sources in this simulation, that are closer apart than $\la_o L/X$, we do not expect 
a nearly exact reconstruction with the $l_1$ optimization. Thus, we display in addition to the actual reconstructions the
aggregated values recovered in the intervals of length $r = \la_o L/(4X),$ centered at the true source locations. 
We observe that the results improve as we decrease the mesh size from $H = \la_o L/X$ to $\la_o L/(8X)$. This is due to the 
fact that the sources are off the grid, and the discretization error decreases as we reduce $H$. 

\begin{figure}[h]
\centering
\includegraphics[scale=0.41]{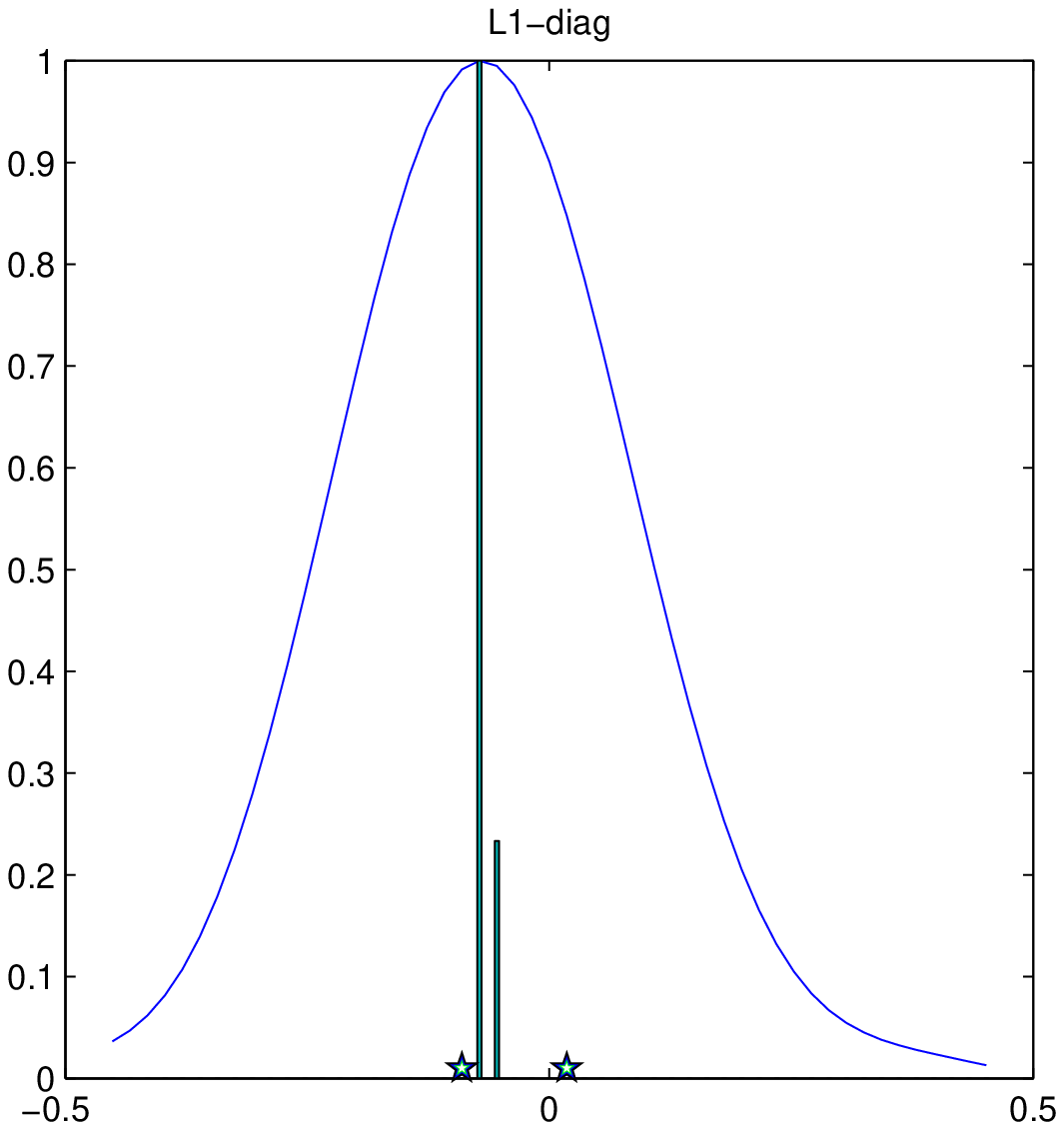}\includegraphics[scale=0.41]{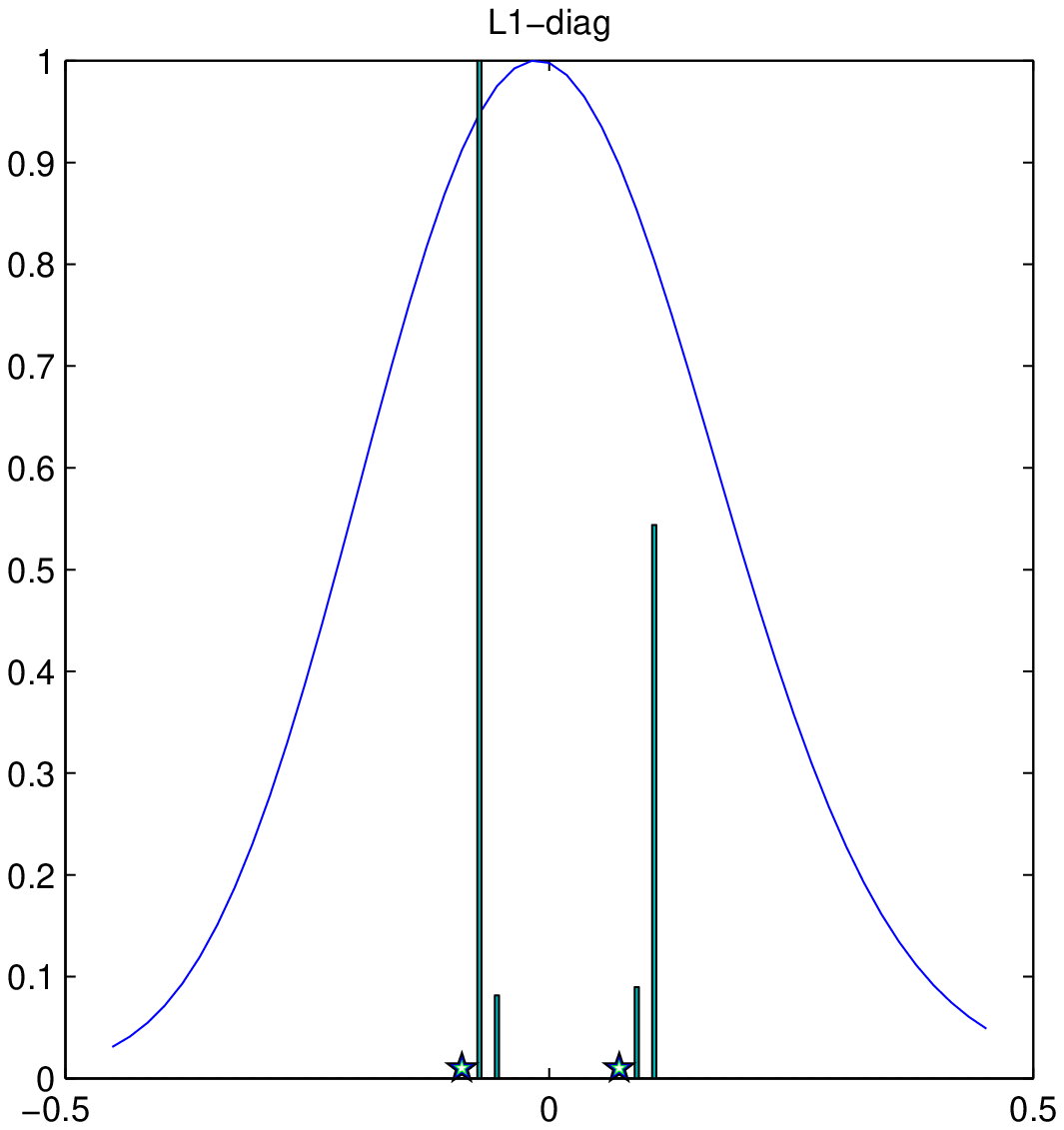}
\includegraphics[scale=0.41]{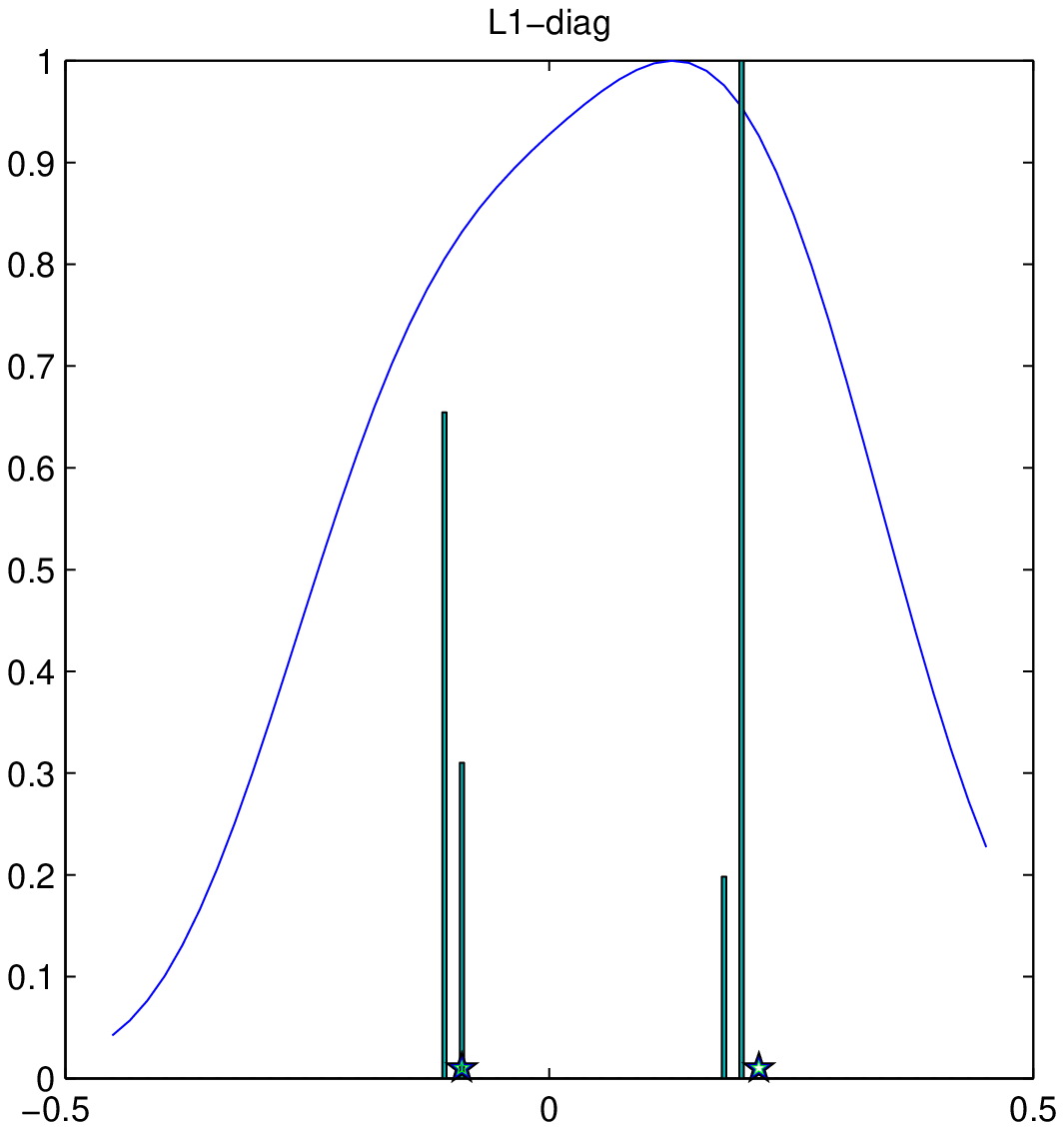}

\vspace{-0.2in}
\caption{We display $l_1$ optimization results in one realization of the random medium, 
for $2$ sources separated by, from left to right, $3/4$, $1$ and $2$ $\la_o L/X$. The units of the 
abscissa are in $\la_o L/X$.}
\label{fig:5}
\end{figure}
The last illustration, in Figure \ref{fig:5}, shows the effect of the source separation on the quality of the reconstructions. As expected from the 
results in section \ref{sect:res}, the reconstruction is better when the sources are further apart. 

\section{Summary}
\label{sect:sum}
We  studied receiver array imaging of remote localized sources in random media, using convex optimization. 
The scattering regime is defined by precise scaling assumptions, and  leads to large random wavefront distortions of the waves 
measured at the array.  Conventional imaging methods like reverse time migration, also known as backprojection 
\cite{Biondi,munson1983tomographic},
or standard $l_1$ optimization \cite{chai2013robust,fannjiang2010compressed}, cannot deal with such distortions and produce poor and unreliable results.  
We base our imaging on the coherent interferometric (CINT) method \cite{borcea2006adaptive} which mitigates random media effects like wavefront distortions at the expense of image resolution. The goal of the 
convex optimization is to remove the blur in the CINT images and thus improve the source localization.  
We show with a detailed analysis that under generic conditions the CINT imaging function is approximately a 
convolution of a blurring kernel with the discretized unknown source intensity on the imaging  mesh. The kernel has 
a generic expression, that depends on the known CINT resolution limits obtained in \cite{borcea2006adaptive,borcea2007asymptotics,borcea2011enhanced}
for various wave propagation models.   The optimization seeks to undo this convolution.  The analysis and 
numerical simulations show that it gives very good estimates of the source locations when they 
are sufficiently far apart. This is in agreement with  the results in \cite{candes2014towards}. We also show that 
when the sources are clustered together, the estimates are not close to the true locations pointwise, 
but they are supported in their vicinity.

\section*{Acknowledgments}
We gratefully acknowledge support from ONR Grant N00014-14-1-0077 and the AFOSR Grant FA9550-15-1-0118.

\appendix
\section{The paraxial approximation}
\label{ap:DerPar}
Equation \eqref{eq:a13} follows easily from the Taylor expansion 
\[
|\vx-\vy| = \sqrt{y_3^2 + |\bx-\by|^2} = L\left[ 1 +
  O\left(\frac{D_3}{L}\right) + O\left(\frac{a^2}{L^2}\right)\right],
\]
where we used that $y_3 = L + O(D_3)$, $|\bx| = O(a)$ and $|\by| =
O(D) \le O(a)$.  For the phase we obtain similarly 
\begin{equation*}
k_o |\vx-\vy| = k_o \sqrt{y_3^2 + |\by-\bx|^2} = k_o \left( y_3 +
\frac{|\bx-\by|^2}{2 y_3} \right) +
O\left(\frac{k_o|\bx-\by|^4}{y_3^3}\right),
\end{equation*}
where $k_o = \om_o/c_o = 2 \pi /\la_o$.  Expanding the square in the
right hand side and using that $y_3 = L + O(D_3)$,
\begin{equation*}
  k_o |\vx-\vy| = k_o \left( y_3 + \frac{|\bx|^2 - 2 \bx \cdot \by +
    |\by|^2}{2 L} \right) + O\left(\frac{a^4}{\la_o L^3}\right) +
  O\left(\frac{a^2 D_3}{\la_o L^2}\right) + O\left(\frac{a D
    D_3}{\la_o L^2}\right) + O\left(\frac{D^2 D_3}{\la_o L^2}\right).
\end{equation*}
The last two terms in the remainder are dominated by the previous
ones, because $D < a$, and approximation \eqref{eq:a12} follows.

\section{Derivation of the statistical moments}
\label{ap:moments}
We begin with the second moments of the Gaussian process
\eqref{eq:a7}, with Gaussian autocorrelation \eqref{eq:a2}, and then
prove  Propositions \ref{prop.1} and \ref{prop.2}.

Consider points $\vy$ and $\vy'$ in the imaging region, so that
$|\vx-\vy| \ge |\vx'-\vy'|$, and write
\begin{align*}
\EE\left[\nu(\vx,\vy) \nu(\vx',\vy')\right] &= \frac{\sqrt{
    |\vx'-\vy'||\vx-\vy|}}{\sqrt{2\pi} \ell} \iint_0^1 d \vartheta d
\vartheta' \, e^{-\frac{1}{2 \ell^2} \left|
  (\vartheta'-\vartheta)(\vx-\vy) + \vartheta' (\vx'-\vx)+
  (1-\vartheta')(\vy'-\vy) \right|^2} \nonumber \\ &=
\sqrt{\frac{|\vx'-\vy'|}{|\vx-\vy|}} \int_0^1 d \vartheta'
\int_{-(1-\vartheta')|\vx-\vy|/\ell}^{\vartheta'|\vx-\vy|/\ell} \frac{d \widetilde
  \vartheta}{\sqrt{2 \pi}} \, e^{-\frac{1}{2} \left| \widetilde
  \vartheta \frac{(\vx-\vy)}{|\vx-\vy|} + \vartheta'
  \frac{(\vx'-\vx)}{\ell} + (1-\vartheta') \frac{(\vy'-\vy)}{\ell}
  \right|^2},
\end{align*}
where we changed variables
\[
(\vartheta, \vartheta') \leadsto (\widetilde \vartheta, \vartheta'), ~
~ \widetilde \vartheta = (\vartheta'-\vartheta) |\vx-\vy|/\ell.
\]
Let $\vec{\bm{m}} = (\vx-\vy)/|\vx-\vy|$ and note  that  since
$|\vx-\vy| \sim L \gg \ell$ and the Gaussian is negligible for
$|\widetilde \vartheta| > 3$, we can extend the $\widetilde \vartheta$
integral to the real line and obtain
\begin{align}
\EE\left[\nu(\vx,\vy) \nu(\vx',\vy')\right] \approx \sqrt{
  \frac{|\vx'-\vy'|}{|\vx-\vy|}} \int_0^1 d\vartheta \, e^{-\frac{1}{2
    \ell^2} \left| (I-\vec{\bm{m}}\vec{\bm{m}}^T)\left[\vartheta (\vx'-\vx) + (1-\vartheta) (\vy'-\vy)\right]
  \right|^2}.
  \label{eq:B1}
\end{align}
In the system of coordinates centered at the array, we calculate
 \[
 \vec{\bm{m}} = \frac{(\bx,0)-(\by,L+y_3)}{\sqrt{(L+y_3)^2 + |\bx-\by|^2}} = \left[-\ve_3 + \frac{(\bx-\by,0)}{L+y_3}\right] \left[1 + O\left(\frac{a^2}{L^2}\right)\right],
\] 
with $\ve_3$ the unit vector in the range direction, and  obtain 
\[
|\vec{\bm{m}} + \ve_3| = O\left(\frac{a}{L}\right) \ll 1.
\]
The projections on the plane orthogonal to $\vec{\bm m}$ are 
\begin{align}
(I-\vec{\bm{m}}\vec{\bm{m}}^T)\frac{(\vx'-\vx)}{\ell} &= \frac{(\bx'-\bx,0)}{\ell} + O\left(\frac{a^2}{\ell L}\right), \label{eq:B1p} \\
(I-\vec{\bm{m}}\vec{\bm{m}}^T)\frac{(\vy'-\vy)}{\ell} &= \frac{(\by'-\by,0)}{\ell} + O\left(\frac{a D}{\ell L}\right) + O\left(\frac{a D_3}{\ell L}\right), \label{eq:B1pp}
\end{align}
with negligible residuals by assumptions \eqref{eq:as1}-\eqref{eq:as2} and \eqref{eq:as5}, which give 
\[
\frac{a^2}{\ell L} \ll \frac{ \sqrt{\la_o L}}{\ell} \ll 1, \qquad 
\frac{a D_3}{\ell L} \ll \frac{\la_o L}{\ell a} < \frac{\la_o L}{\ell^2} \ll 1.
\]
Thus, we can approximate \eqref{eq:B1} as 
\begin{align}
\EE\left[\nu(\vx,\vy) \nu(\vx',\vy')\right] \approx \sqrt{
  \frac{|\vx'-\vy'|}{|\vx-\vy|}} \int_0^1 d\vartheta \, e^{-\frac{1}{2
    \ell^2} \left| \vartheta (\bx'-\bx) + (1-\vartheta) (\by'-\by)
  \right|^2}.
  \label{eq:B1n}
\end{align}
This expression can be simplified further  for small offsets satisfying $ |\by'-\by|
\ll \ell$ and $|\bx'-\bx| \ll \ell$, by expanding the exponential and
then integrating in $\vartheta$,
\begin{equation}
\EE\left[\nu(\vx,\vy) \nu(\vx',\vy')\right] \approx \sqrt{
  \frac{|\vx'-\vy'|}{|\vx-\vy|}} \left[ 1 - \frac{|\by'-\by|^2 +
    (\by'-\by)\cdot (\bx'-\bx) + |\bx'-\bx|^2}{6\ell^2} \right].
\label{eq:B2}
\end{equation}
Note that the small receiver offset condition is consistent with 
$|\bx'-\bx| \ll X \sim X_d \ll \ell$ used in the CINT image formation.

The proof of Proposition \ref{prop.1} follows easily from
\eqref{eq:B2} and definitions \eqref{eq:a6}, \eqref{eq:a7}. Because
the process $\nu(\vx,\vy)$ is Gaussian, we have
\begin{align}
\EE\left[ \hat G(\vx,\vy,\om)\right] &\approx \hat G_o (\vx,\vy,\om)
\EE\left[ \exp\left(i \frac{(2 \pi)^{1/4}}{2} k \sigma \sqrt{\ell |\vx-\vy|}
  \nu(\vx,\vy)\right)\right] \nonumber \\ &\approx \hat G_o
(\vx,\vy,\om) \exp \left\{ -\frac{\sqrt{2 \pi}}{8} k^2 \sigma^2 \ell
  |\vx-\vy| \EE \left[\nu^2(\vx,\vy)\right]\right\} \nonumber \\
& = \hat G_o
(\vx,\vy,\om) \exp \left[ -\frac{\sqrt{2 \pi}}{8} k^2 \sigma^2 \ell
  |\vx-\vy|\right].
\end{align}
This is equation \eqref{eq:prop1.1} with scattering mean free path
$\cS(\om)$ defined as in \eqref{eq:prop1.2}. $\Box$

To prove Proposition \ref{prop.2} we use again definitions
\eqref{eq:a6} and \eqref{eq:a7}, the Gaussianity of the process $\nu$,
and result \eqref{eq:B1} to write
\begin{equation}
\EE\left[ \hat G(\vx,\vy,\om+\tom/2) \overline{\hat
    G(\vx',\vy',\om-\tom/2)}\right] \approx \hat
G_o(\vx,\vy,\om+\tom/2) \overline{\hat G_o(\vx',\vy',\om-\tom/2)}
\, \mathcal{E},
\label{eq:B3}
\end{equation}
with 
\begin{align}
\mathcal{E} = \exp &\left\{ -\frac{\sqrt{2 \pi}}{8}
\left(\frac{\om+\tom/2}{c_o}\right)^2 \sigma^2 \ell |\vx-\vy| -
\frac{\sqrt{2 \pi}}{8} \left(\frac{\om-\tom/2}{c_o}\right)^2 \sigma^2
\ell |\vx'-\vy'| \right. \nonumber \\ &\left. + \frac{\sqrt{2
    \pi}(\om^2 - \tom^2/4)}{4c_o^2} \sigma^2 \ell |\vx'-\vy'| \int_0^1
d \vartheta \, e^{-\frac{1}{2 \ell^2} \left| \vartheta (\bx'-\bx) +
  (1-\vartheta) (\by'-\by) \right|^2 }\right\}.
\label{eq:B4p}
\end{align}
Note that 
\begin{align*}
\frac{\sqrt{2 \pi}}{8} \left(\frac{\om+\tom/2}{c_o}\right)^2 \sigma^2
\ell |\vx-\vy| &= \frac{|\vx-\vy|}{\cS (\om+\tom/2)} \gg 1,
\\ \frac{\sqrt{2 \pi}}{8} \left(\frac{\om-\tom/2}{c_o}\right)^2
\sigma^2 \ell |\vx'-\vy'| &= \frac{|\vx'-\vy'|}{\cS(\om-\tom/2)} \gg
1,
\end{align*}
so $\mathcal{E}$ is exponentially small unless the last term in
\eqref{eq:B4p} compensates the first two.  This happens only when
$|\by-\by'| \ll \ell$. For other offsets the integral over $\vartheta$
is small and, consequently, $\mathcal{E} \approx 0.$ 

When $|\by-\by'|
\ll \ell$, we can approximate the $\vartheta$ integral as in
\eqref{eq:B2}, and obtain
\begin{align}
\mathcal{E} \approx \exp &\left\{ -\frac{\sqrt{2 \pi}}{8}
\left(\frac{\om+\tom/2}{c_o}\right)^2 \sigma^2 \ell |\vx-\vy| -
\frac{\sqrt{2 \pi}}{8} \left(\frac{\om-\tom/2}{c_o}\right)^2 \sigma^2
\ell |\vx'-\vy'| \right. \nonumber \\ &\left. + \frac{\sqrt{2
    \pi}(\om^2 - \tom^2/4)}{4c_o^2} \sigma^2 \ell |\vx'-\vy'|\left[1 -
  \frac{|\by'-\by|^2 + (\by'-\by)\cdot (\bx'-\bx) +
    |\bx'-\bx|^2}{6\ell^2} \right]\right\}.
\label{eq:B4}
\end{align}
Rearranging the terms and using definition \eqref{eq:prop1.2} of the
scattering mean free path we get
\begin{align}
\mathcal{E} \approx \exp &\left\{ -\frac{|\vx-\vy| - |\vx'-\vy'|}{\cS(\om)}
- \frac{\sqrt{2 \pi} \sigma^2 \ell \tom^2}{8c_o^2} 
\left(\frac{|\vx-\vy| + 3 |\vx'-\vy'|}{4} \right) - \frac{\sqrt{2
    \pi} \sigma^2 \ell \om \tom}{8c_o^2}
(|\vx-\vy|-|\vx'-\vy'| ) \right. \nonumber \\ &\left. -\frac{\sqrt{2
    \pi}(\om^2 - \tom^2/4)\sigma^2 |\vx'-\vy'|}{24c_o^2 \ell}
\left(|\by'-\by|^2 + (\by'-\by)\cdot (\bx'-\bx) + |\bx'-\bx|^2\right)
\right\}.
\label{eq:B5}
\end{align}
Recall that $|\vx-\vy| \ge |\vx'-\vy'|$, and conclude from the decay
of the first term that $\mathcal{E}$ is large if
\[
|\vx-\vy| - |\vx'-\vy| = O(\cS(\om)).
\]
Substituting in \eqref{eq:B5}, and using the scales
\begin{equation}
\Omega_d = \frac{2 c_o}{(2 \pi)^{1/4} \sigma \sqrt{\ell |\vx-\vy|}},
\quad X(\om) = \frac{\sqrt{3} \ell \Omega_d}{\om},
\label{eq:B6}
\end{equation} 
we get
\begin{align}
\mathcal{E} \approx \exp &\left\{ -\frac{|\vx-\vy| - |\vx'-\vy'|}{\cS(\om)}
- \frac{\tom^2}{2 \Omega_d^2} \left[1 + O
  \left(\frac{\cS(\om)}{L}\right)\right] + O\left(\frac{\sigma^2 \ell
  \om \tom \cS(\om)}{c_o^2}\right) \right. \nonumber \\ &\left. -
\frac{(|\by'-\by|^2 + (\by'-\by)\cdot (\bx'-\bx) + |\bx'-\bx|^2)}{2
  X(\om)^2} \left[ 1 + O \left(\frac{\cS(\om)}{L}\right) + O
  \left(\frac{\Omega_d^2}{\om^2}\right)\right] \right\}.
\label{eq:B7}
\end{align}
This equation simplifies because $\cS(\om) \ll L$ and
$\Omega_d \ll \om = O(\om_o)$. Moreover, \eqref{eq:B7} is large only
when $|\tom| \lesssim \Omega_d$, so we estimate using definitions
\eqref{eq:prop1.2} and \eqref{eq:B6} and the assumption \eqref{eq:as3}
that
\[
\frac{\sigma^2 \ell \om \tom \cS}{c_o^2} \lesssim
O\left(\frac{\sigma^2 \ell \om_o \Omega_d \cS}{c_o^2}\right) =
O\left(\frac{\la_o}{\sigma \sqrt{\ell L}} \right) \ll 1.
\]
Similarly, $\mathcal{E}$ is large for cross-range offsets of at most 
$O(X)$, so we can write  
\[
|\vx-\vy| - |\vx'-\vy'| \approx y_3 - y_3' + \frac{|\bx|^2 -
  |\bx'|^2}{2 L} + \frac{|\by|^2 - |\by'|^2}{2 L} - \frac{\bx \cdot
  \by - \bx' \cdot \by'}{L} = y_3 - y_3' + O \left(\frac{a X}{L}
  \right).
\]
We also have $X \sim X_d$, and
\[
\frac{a X_d}{\cS L} = O\left( \frac{a X_d}{L} \frac{\sigma^2
  \ell}{\la_o^2} \right) = O \left(\frac{a \sigma \ell^{3/2}}{\la_o
  L^{3/2}} \right) \ll O\left[\frac{\sigma \ell^{3/2}}{(\la_o
  L)^{3/4}}\right] \ll O \left(\frac{\ell^2}{\la_o^{1/4} L^{1+3/4}}
\right) \ll O\left[ \left(\frac{\la_o}{L}\right)^{1/4}\right] \ll 1,
\]
where the two equalities are by definition \eqref{eq:prop1.2} and
\eqref{eq:prop2.2} of $\cS$ and $X_d$, the first inequality is by
\eqref{eq:as1}, the second by \eqref{eq:as3}, the third by
\eqref{eq:as2}, and the last by \eqref{eq:a3}.  Gathering the results
we arrive at the approximation
\begin{align}
\mathcal{E} \approx \exp &\left[-\frac{|y_3-y_3'|}{\cS(\om)} -
  \frac{\tom^2}{2 \Omega_d^2} - \frac{|\by'-\by|^2 + (\by'-\by)\cdot
    (\bx'-\bx) + |\bx'-\bx|^2}{2 X^2(\om)}\right].
\label{eq:B9}
\end{align}
The statement of Proposition \ref{prop.2} follows from this equation
and $|\om-\om_o| \lesssim \pi B \ll \om_o$. $\Box$
\section{Consistency of scaling}
\label{ap:verify}
The scaling \eqref{eq:as3} is consistent because 
\[
\frac{(\ell/L)^{3/2}}{\sqrt{\ell \la_o}/L} = \frac{\ell}{\sqrt{\la_o
    L}} \gg 1 \quad \mbox{and} \quad \frac{\sqrt{\ell
    \lambda_o}/{L}}{\la_o/\sqrt{\ell L}} = \frac{\ell}{\sqrt{\lambda_o
    L}} \gg 1,
\]
where we used  \eqref{eq:as2}.

The scaling \eqref{eq:as4} is consistent because by assumption
\eqref{eq:as1} and \eqref{eq:prop2.2},
\[
\frac{(a/L)^2}{\la_o L/(a X_d)} = \frac{a^4}{\la_o L^3} \frac{X_d}{a}
\ll \frac{\ell}{a} < 1.
\]
We also have 
\[
\frac{\la_o L/(a X_d)}{\Omega_d/\om_o} = O\left(\frac{\la_0 L}{a \ell
  (\Omega_d/\om_o)^2}\right) = O \left(\frac{\sigma^2 L^2}{a \la_o}
\right) \ll \frac{\ell}{a} < 1,
\]
so \eqref{eq:as4} implies $ B \ll \Omega_d. $ 

Definition \eqref{eq:prop2.2} and assumptions \eqref{eq:as2},
\eqref{eq:as3} give
\[
\frac{\la_o L}{X_d} = O\left(\frac{\la_o L}{\ell}
\frac{\om_o}{\Omega_d} \right) = O\left(\sigma
\frac{L^{3/2}}{\ell^{1/2}}\right) \ll \sqrt{\la_o L} \ll \ell.
\]
This verifies that the assumptions \eqref{eq:as5} on $D$ are consistent. 

The assumptions \eqref{eq:as5} on $D_3$ are consistent in the narrowband regime, when
\eqref{eq:as4} holds, because by \eqref{eq:as4pp} we have $\Omega/B = O(1)$ and 
\[
\frac{c_o/B}{\la_o L^2/a^2} =O \left( \frac{ \om_o}{B}
\frac{a^2}{L^2}\right) \ll 1.
\]
In the broadband regime $\Omega = \Omega_d$ and assumptions \eqref{eq:as5} on $D_3$ are consistent because
\[
\frac{c_o/\Omega_d}{\la_o L^2/a^2} = O\left(\frac{\sigma \sqrt{\ell L}}{\la_o L^2/a^2}\right) \ll \frac{\ell}{L} \ll 1,
\]
by definition \eqref{eq:prop2.2} and assumptions \eqref{eq:as1} and \eqref{eq:as3}. Moreover, assumption \eqref{eq:as3bb} is consistent because 
\[
\frac{\la_o^{2/3} \ell^{1/6}/L^{5/6}}{\la_o/\sqrt{\ell L}} = \left(\frac{\ell}{\sqrt{\la_o L}}\right)^{2/3} \gg 1, \qquad 
\frac{\la_o^{2/3} \ell^{1/6}/L^{5/6}}{\sqrt{\la_o \ell}/L} = \left(\frac{\sqrt{\la_o L}}{\ell} \right)^{1/3} \ll 1,
\]
by assumption \eqref{eq:as2}.

\section{Derivation of the CINT kernel}
\label{sect:apCINT}
We begin with the expression \eqref{eq:f9} of the CINT kernel, and use
the Gaussian pulse, thresholding windows and apodization to obtain
\begin{align}
\kappa(\vy,\vz,\vz') \approx &\frac{N_r^2\sqrt{2 \pi}}{a^4 B } \int d
\bx \, e^{-\frac{|\bx|^2}{(a/6)^2}} \int d \tbx \,
e^{-\frac{|\tbx|^2}{2 X^2}-\frac{|\tbx|^2}{4(a/6)^2}} \int d \om \,
e^{-\frac{(\om-\om_o)^2}{2 B^2}} \int d \tom \, e^{-\frac{\tom^2}{2
    \Omega^2}-\frac{\tom^2}{8 B^2}} \nonumber \\ & \times \hat
G\left(\left(\bx + \tbx/2,0\right),\vz,\om +
\tom/2\right)\overline{\hat G\left(\left(\bx -
  \tbx/2,0\right),\vz',\om - \tom/2\right)} \nonumber \\ & \times
\exp\left[ - i \frac{(\om + \tom/2)}{c_o} \left| (\bx +
  \tbx/2,0)-\vy\right| + i \frac{(\om - \tom/2)}{c_o} \left| (\bx -
  \tbx/2,0)-\vy\right| \right], \label{eq:D0}
\end{align}
for points $\vy = (\by,y_3)$, $\vz = (\bz,z_3)$ and $\vz' =
(\bz',z_3')$ in the imaging region. The integration over $\bx$ and
$\tbx$ extends to the whole plane $\mathbb{R}^2$, and those over $\om$
and $\tom$ to the real line, with the aperture and bandwidth
restrictions ensured by the Gaussians.
Because CINT is statistically stable in our scaling, we may
approximate the right hand side of \eqref{eq:D1} by its
expectation. Using \eqref{eq:a17} and the paraxial approximation
\eqref{eq:a14}, we obtain
\begin{align}
\kappa(\vy,\vz,\vz') \approx &\frac{N_r^2\sqrt{2 \pi}}{a^4 B (4 \pi L)^2
} e^{-\frac{|\bz-\bz'|^2}{2 X_d^2}} \int d \bx \,
e^{-\frac{|\bx|^2}{(a/6)^2}} \int d \tbx \, e^{-\frac{|\tbx|^2}{2
    X_e^2} - \frac{\tbx\cdot(\bz-\bz')}{2 X_d^2}} \int d \om \,
e^{-\frac{(\om-\om_o)^2}{2 B^2}} \int d \tom \, e^{-\frac{\tom^2}{2
    \Omega_e^2}} \nonumber \\ & \times \exp \left\{ i \frac{\tom}{c_o}
\left[ \frac{z_3+z_3'}{2} - y_3 + \frac{1}{2 L} \left(\frac{|\bz|^2
    + |\bz'|^2}{2}- |\by|^2\right) - \frac{\bx}{L}
  \cdot\left(\frac{\bz+\bz'}{2}-\by \right) - \frac{\tbx}{4 L} \cdot
  (\bz-\bz') \right] \right\} \nonumber \\ & \times \exp \left\{ i
\frac{\om}{c_o} \left[ z_3-z_3' + \frac{|\bz|^2-|\bz'|^2}{2 L} -
  \frac{\bx \cdot(\bz-\bz')}{L} - \frac{\tbx}{L} \cdot
  \left(\frac{\bz+\bz'}{2}-\by\right)\right]\right\}, \label{eq:D1}
 \end{align}
with $X_e = O(X_d)$ and $\Omega_e = O(B)$ defined in
\eqref{eq:R4}. Note that the last term in the second line of
\eqref{eq:D1} is negligible, because by assumptions \eqref{eq:as1},
\eqref{eq:as4} and definition \eqref{eq:prop2.2},
\[
\frac{\tom \tbx \cdot (\bz-\bz')}{4 L c_o} = O\left(\frac{B X_d^2}{c_o
  L} \right) = O\left(\frac{B}{\om_o} \frac{a X_d}{\la_o L}
\frac{X_d}{a}\right) \ll 1,
\]
in the narrowband case. Moreover, in the broadband case  
\[
\frac{\tom \tbx \cdot (\bz-\bz')}{4 L c_o} = O\left(\frac{\Omega_d X_d^2}{c_o  L} \right) = 
O\left(\frac{\ell^2}{\la_o L} \frac{\Omega_d^3}{\om_o^3} \right) = O\left(\frac{\la_o^2 \ell^{1/2}}{\sigma^3 L^{5/2}}\right) \ll 1,
\]
where the first equalities are by definitions \eqref{eq:prop2.2}, and the bound is by \eqref{eq:as3bb}. 
Let us introduce the center and difference coordinates 
\begin{equation}
\label{eq:D3}
\cbz = \frac{\bz+\bz'}{2}, \qquad \tbz = \bz-\bz', \qquad \cz_3 =
\frac{z_3+z_3'}{2}, \qquad \tz_3 = z_3 - z_3',
\end{equation}
and note that 
\[
\frac{\tom}{c_o L} \left(\frac{|\bz|^2 + |\bz'|^2}{2}- |\by|^2\right) = \frac{\tom}{c_o} \frac{(|\cbz|^2-|\by|^2)}{L} + 
\frac{\tom |\tbz|^2}{4 c_o L} \approx \frac{\tom}{c_o}\frac{(|\cbz|^2-|\by|^2)}{L},
\]
because 
\[
\frac{\tom}{c_o} \frac{|\tbx|^2}{L} = O\left(\frac{\Omega_d X_d^2}{c_o L}\right) \ll 1.
\]
The kernel \eqref{eq:D1} simplifies as 
\begin{align}
\kappa(\vy,\vz,\vz') \approx &\frac{N_r^2\sqrt{2 \pi}}{a^4 B (4 \pi L)^2} e^{-\frac{|\tbz|^2}{2 X_d^2}} \int d\tom  \, 
\exp\left[-\frac{\tom^2}{2 \Omega_e^2} + i \tom \left(\frac{\cz_3-y_3}{c_o} + \frac{|\cbz|^2-|\by|^2}{2 L c_o} \right)\right] \nonumber \\
&\times \int d \om \, 
\exp \left[-\frac{(\om-\om_o)^2}{2 B^2} + i \om \left(\frac{\tz_3}{c_o} + \frac{\cbz \cdot \tbz}{Lc_o} \right)\right] \nonumber \\
&\times \int d \tbx \, \exp \left[ - \frac{|\tbx|^2}{2 X_e^2} - \tbx \cdot \left(\frac{\tbz}{2 X_d^2} + \frac{i \om (\cbz-\by)}{L c_o}
\right)\right] \nonumber \\
& \times \int d \bx \, \exp \left[ -\frac{|\bx|^2}{(a/6)^2} - i \bx \cdot \left(\frac{\tom(\cbz-\by)}{L c_o} + \frac{\om \tbz}{L c_o}\right)
\right], \label{eq:DD1}
\end{align}
with $\Omega_e$ and $X_e$ defined in \eqref{eq:R4}, and after  evaluating the  last two integrals, we get 
\begin{align}
\kappa(\vy,\vz,\vz') \approx &\frac{N_r^2 \sqrt{2 \pi} X_e^2}{288 B a^2 L^2} e^{-\frac{|\tbz|^2}{2 X_d^2}} \int d\tom  \, 
\exp\left[-\frac{\tom^2}{2 \Omega_e^2} + i \tom \left(\frac{\cz_3-y_3}{c_o} + \frac{|\cbz|^2-|\by|^2}{2 L c_o} \right)\right] \nonumber \\
&\times \int d \om \, 
\exp \left[-\frac{(\om-\om_o)^2}{2 B^2} + i \om \left(\frac{\tz_3}{c_o} + \frac{\cbz \cdot \tbz}{L c_o} \right)\right] \nonumber \\
&\times \exp \left[
-\frac{X_e^2}{2} \left|\frac{\om(\cbz-\by)}{Lc_o} - \frac{i \tbz}{2 X_d^2} \right|^2 - \frac{1}{2} \left(\frac{a}{6\sqrt{2}}\right)^2 
\left|\frac{\tom (\cbz-\by)}{L c_o} 
+\frac{\om \tbz}{L c_o} \right|^2 \right]. \label{eq:DD2}
\end{align}
Let us change variables $\w = \om - \om_o$ and use the notation 
\begin{equation}
\cbZ = \frac{\cbz-\by}{L/(k_o X_e)}, \qquad \ctZ = \frac{\tbz}{6 \sqrt{2} L/(k_o a)}, \qquad \be = \frac{\cz_3-y_3}{c_o/\Omega_e} + 
\frac{|\cbz|^2-|\by|^2}{2L c_o/\Omega_e}, \qquad \theta = \frac{\om_o X_e}{\Omega_e (a/6)}.
\label{eq:DD3}
\end{equation} 
Define also 
\begin{equation}
\frac{1}{\gamma} = 1 - \frac{X_e^2}{4 X_d^2}  > \frac{3}{4},
\label{eq:DD4}
\end{equation}
with the inequality implied by the definition of $X_e$.  Substituting in \eqref{eq:DD2} we get 
\begin{align}
\kappa(\vy,\vz,\vz') \approx &\frac{N_r^2 \sqrt{2 \pi} X_e^2}{288 B a^2 L^2} \exp \left[-\frac{|\tbz|^2}{2 \gamma X_d^2} - 
\frac{|\cbZ|^2}{2} - \frac{|\ctZ|^2}{2} + i k_o \Big(\tz_3 + \frac{\tbz \cdot \cbz}{L} + \frac{X_e^2 \tbz \cdot (\cbz-\by)}{2 X_d^2 L}\Big) \right] 
\nonumber \\ & \times \int d \w \exp \left[-\frac{\w^2}{2} \Big(\frac{1}{B^2} + \frac{|\cbZ|^2 + |\ctZ|^2}{\om_o^2} \Big) + 
\frac{i \w}{c_o} \Big( \tz_3+\frac{\cbz \cdot \tbz}{L} + \frac{X_e^2 \tbz \cdot (\cbz-\by)}{2 X_d^2 L}\Big)-\frac{\w (|\cbZ|^2 + 
|\ctZ|^2)}{\om_o}\right] \nonumber \\
& \times 
\int d\tom  \, 
\exp\left[-\frac{\tom^2}{2 \Omega_e^2} \Big(1 + \frac{|\cbZ|^2}{2 \theta^2}\Big) + \frac{i \tom}{\Omega_e} \Big(\beta + \frac{i (\om_o + \w) 
\cbZ \cdot \ctZ}{\sqrt{2} \om_o \theta} \Big)
 \right], \label{eq:DD5}
\end{align}
and integrate next over $\tom$. We obtain after rearranging the terms that 
\begin{align}
\kappa(\vy,\vz,\vz') \approx &\frac{\pi N_r^2 X_e^2 \Omega_e}{144 B a^2 L^2 \sqrt{1 + \frac{|\cbZ|^2}{2 \theta^2}}} 
\exp \left[-\frac{|\tbz|^2}{2 \gamma X_d^2} - \frac{\Delta^2}{2}  -\frac{\beta^2}{2(1 + \frac{|\cbZ|^2}{2 \theta^2})}  
\right] \nonumber \\
&\times \exp \left[i k_o \Big(\tz_3 + \frac{\tbz \cdot \cbz}{L} + \frac{X_e^2 \tbz \cdot (\cbz-\by)}{2 X_d^2 L}\Big)
- \frac{i \beta \cbZ \cdot \ctZ}{\sqrt{2} \theta (1 + \frac{|\cbZ|^2}{2 \theta^2})}
 \right] \nonumber \\
 & \times \int d \w \, \exp \left[ - \frac{\w^2}{2 B^2} \Big(1+ \frac{\Delta^2 B^2}{\om_o^2} \Big) + \frac{i \w \eta}{B} - \frac{\w \Delta^2}{\om_o}\right],
\label{eq:DD6}\end{align}
with notation 
\begin{align}
\eta &= \frac{B}{c_o} \left[ \tz_3 + \frac{\cbz \cdot \tbz}{L} + \frac{X_e^2 \tbz \cdot(\cbz-\by)}{2 X_d^2 L} - 
\frac{\beta c_o \cbZ \cdot \ctZ}{\sqrt{2} \om_o \theta (1 + \frac{|\cbZ|^2}{2 \theta^2})}\right], \label{eq:DD7}\\
\Delta^2 & = |\cbZ|^2 + |\ctZ|^2 - \frac{|\cbZ \cdot \ctZ|^2}{2 \theta^2 (1 + \frac{|\cbZ|^2}{2 \theta^2})}. \label{eq:DD8}
\end{align}
Note that \eqref{eq:DD8} is non-negative because 
\[
|\ctZ|^2 - \frac{|\cbZ \cdot \ctZ|^2}{2 \theta^2 (1 + \frac{|\cbZ|^2}{2 \theta^2})} = \frac{|\ctZ|^2}{(1 + \frac{|\cbZ|^2}{2 \theta^2})} + 
\frac{(|\cbZ|^2 |\ctZ|^2 - |\cbZ \cdot \ctZ|^2)}{(1 + \frac{|\cbZ|^2}{2 \theta^2})} \ge \frac{|\ctZ|^2}{(1 + \frac{|\cbZ|^2}{2 \theta^2})}.
\]
Now we integrate in $\w$ in equation \eqref{eq:DD6}, and obtain 
\begin{align}
\kappa(\vy,\vz,\vz') \approx &\frac{\sqrt{2}\pi^{3/2} N_r^2 X_e^2 \Omega_e}{144 a^2 L^2 \sqrt{1 + \frac{|\cbZ|^2}{2 \theta^2}} 
\sqrt{1 +\frac{\Delta^2 B^2}{\om_o^2}} } 
\exp \left[-\frac{|\tbz|^2}{2 \gamma X_d^2} - \frac{\beta^2}{2(1 + \frac{|\cbZ|^2}{2 \theta^2})}  - \frac{\eta^2}{2 (1 + \frac{B^2 \Delta^2}{\om_o^2})}
-\frac{\Delta^2}{2(1 + \frac{B^2 \Delta^2}{\om_o^2})}\right] \nonumber \\
&\times \exp \left[i k_o \Big(\tz_3 + \frac{\tbz \cdot \cbz}{L} + \frac{X_e^2 \tbz \cdot (\cbz-\by)}{2 X_d^2 L}\Big)
- \frac{i \beta \cbZ \cdot \ctZ}{\sqrt{2} \theta (1 + \frac{|\cbZ|^2}{2 \theta^2})} - \frac{i \eta B \Delta^2}{\om_o(1 + \frac{B^2 \Delta^2}{\om_o^2})}
 \right] .
\label{eq:DD9}
\end{align}

This expression simplifies, because the exponential decay in $\Delta^2$ ensures that the kernel is large only when 
$\Delta = O(1)$. Since $B/\om_o \ll 1$ by assumption, we see that 
\begin{equation}
\frac{B^2 \Delta^2}{\om_o^2} \ll 1.
\label{eq:DD10}
\end{equation}
By the same reasoning, the kernel is large when $|\eta| = O(1)$, but then 
\begin{equation}
\left|\frac{\eta B \Delta^2}{\om_o(1 + \frac{B^2 \Delta^2}{\om_o^2})}\right| \approx |\eta| \Delta^2 \frac{B}{\om_o}  = O\left(\frac{B}{\om_o}\right) \ll 1.
\label{eq:DD11}
\end{equation}
We also have from $\Delta = O(1)$ and definition \eqref{eq:DD8} that both $|\cbZ|$ and $|\ctZ|$ are $O(1)$. Then, using the 
definition of $\ctZ$ in \eqref{eq:DD3}, we can estimate 
\begin{equation}
\frac{B}{c_o} \left| \frac{\cbz \cdot \tbz}{L} \right| \le O\left(\frac{B}{c_o} \frac{ \la_o L/{a} a}{L} \right) = O\left(\frac{B}{\om_o}\right) \ll 1, 
\label{eq:DD12}
\end{equation}
and 
\begin{equation}
\left|\frac{ \beta B  \cbZ \cdot \ctZ}{\om_o \theta}\right| = O \left(\frac{B}{\om_o \theta}\right) = O \left(\frac{B}{\om_o}\right) \ll 1.
\label{eq:DD13}
\end{equation}
Here we used that $\theta^{-1}$ is at most of order 1, which follows from its definition in \eqref{eq:DD3}. Indeed, in the broadband case we 
obtain from  assumption \eqref{eq:as3bb} that 
\begin{equation}
\theta^{-1} = \frac{a \Omega_e}{6 \om_o X_e} = O \left(\frac{a \Omega_d}{\om_o X_d}\right) = O \left(\frac{a}{\ell}\right) = O(1).
\end{equation}
In the narrowband case we obtain 
\begin{equation}
\theta^{-1} = \frac{a \Omega_e}{6 \om_o X_e} = O \left(\frac{a B}{\om_o X_d}\right)  \ll \frac{\la_o L}{X_d^2} = 
O \left(\frac{\la_o L}{\ell^2 (\Omega_d/\om_o)^2}\right) = O \left(\frac{\sigma^2 L^2}{\ell \la_o}\right) \ll 1,
\label{eq:DD14p}
\end{equation}
where the second equality is because $\Omega_e = O(B)$, the first inequality is by assumption \eqref{eq:as4},
the following equalities are by the definitions \eqref{eq:prop2.2} of $X_d$ and $\Omega_d$ and the last inequality is 
by assumption \eqref{eq:as3}.
We also have the estimate 
\begin{align}
\frac{B}{c_o} \left|\frac{X_e^2 \tbz \cdot (\cbz-\by)}{X_d^2 L} \right| &= O\left( \frac{B}{c_o L} \frac{L}{k_o a} \frac{L}{k_o X_d} \right) 
= O\left(\frac{B}{\om_o} \frac{\la_o L}{a X_d} \right) = O \left(\frac{B}{\om_o} \frac{\la_o L}{a \ell \Omega_d/\om_o} \right)\nonumber \\
& = 
O\left( \frac{B}{\om_o} \frac{\sigma L^{3/2}}{a \ell^{1/2}}\right) \ll \frac{B}{\om_o} \frac{\sqrt{\la_o L}}{a} \ll 1,
\label{eq:DD14}
\end{align}
where we used again definitions \eqref{eq:prop2.2} and the assumption \eqref{eq:as3}. Note that estimates \eqref{eq:DD12}-\eqref{eq:DD14}
and definition \eqref{eq:DD7} imply that 
\begin{equation}
\eta \approx \frac{B \tz_3}{c_o}.
\end{equation}
Substituting all the results in \eqref{eq:DD9}, we get the kernel 
\begin{align}
\kappa(\vy,\vz,\vz') \approx &\frac{\sqrt{2}\pi^{3/2} N_r^2 X_e^2 \Omega_e}{144 a^2 L^2 \sqrt{1 + \frac{|\cbZ|^2}{2 \theta^2}} } 
\exp \left[-\frac{|\tbz|^2}{2 \gamma X_d^2} - \frac{\beta^2}{2(1 + \frac{|\cbZ|^2}{2 \theta^2})}  - \frac{\tz_3^2}{2 (c_o/B)^2}
-\frac{\Delta^2}{2}\right] \nonumber \\
&\times \exp \left[i k_o \Big(\tz_3 + \frac{\tbz \cdot \cbz}{L} + \frac{X_e^2 \tbz \cdot (\cbz-\by)}{2 X_d^2 L}\Big)
- \frac{i \beta \cbZ \cdot \ctZ}{\sqrt{2}  \theta (1 + \frac{|\cbZ|^2}{2 \theta^2})} 
 \right] .
\label{eq:DD15}
\end{align}
The statement of Proposition \ref{prop.3bb} follows. 

In the narrowband case we know from \eqref{eq:DD14p} that  $\theta \gg 1$. We also have 
\begin{align}
\frac{\Omega_e}{c_o} \left|\frac{(|\cbz|^2-|\by|^2)}{L}\right| = O\left(\frac{B |(\cbz - \by)\cdot(\cbz+\by)|}{c_o L}\right) = O\left(\frac{B}{c_o} \frac{L/(k_o X_e) a}{L} \right) = 
O\left(\frac{\la_o L}{X_d^2} \right) = O \left(\frac{\sigma^2 L^2}{\la_o \ell}\right) \ll 1,
\end{align}
where we used that $\Omega_e = O(B)$, $X_e = O(X_d)$ and $|\cbZ| = O(1)$ i.e., $|\cbz-\by| = O(L/(k_o X_d)$. This estimate follows from 
definition \eqref{eq:prop2.2} and assumptions \eqref{eq:as3} and \eqref{eq:as4}.
The expression \eqref{eq:CINTK} in Proposition \ref{prop.3} follows from \eqref{eq:DD15} and definition \eqref{eq:DD3} of $\beta$. $\Box$

\section{Proof of Lemma \ref{lem.1}}
\label{ap:PfLem1}
We estimate  the inner product of the columns of the matrix using the continuum approximation
\begin{equation}
  \left<\bm{m}_{\vz},\bm{m}_{\vz'}\right> = \sum_{\vy \in \mathfrak{D}_z}
  m_{\vy,\vz} \, m_{\vy,\vz'} \approx \frac{1}{h^2 h_3}
  \int_{-\frac{D}{2}}^{\frac{D}{2}} d y_1
  \int_{-\frac{D}{2}}^{\frac{D}{2}} d y_2
  \int_{-\frac{D_3}{2}}^{\frac{D_3}{2}} m_{\vy,\vz}\, m_{\vy,\vz'},
\end{equation}
where $\vy = (\by,y_3)$, with $\by = (y_1,y_2)$, and $h$ and
$h_3$ are the mesh sizes in cross-range and range. 
\subsection{The narrowband regime}
With the expression \eqref{eq:R8} of $m_{\vy,\vz}$ we get
\begin{align}
  \left<\bm{m}_{\vz},\bm{m}_{\vz'}\right> \approx &\,\frac{C^2}{h^2
    h_3} \int_{\mathbb{R}^2} d \by \, \exp \left[ -
      \frac{(|\by-\bz|^2 + |\by-\bz'|^2)}{2 R^2}\right] 
  \int_{-\infty}^\infty d y_3 \, 
    \exp \left[  -
    \frac{ \left(y_3-z_3\right)^2}{2
      R_3^2} -
    \frac{ \left(y_3-z_3'\right)^2}{2
      R_3^2}
    \right],
\end{align}
where we used the paraxial approximation and extended the integrals to
the whole space with negligible error\footnote{This is assuming that
  $\vz$ and $\vz'$ (i.e., the sources) are not near the edge of the
  imaging region.}, due to the Gaussians. Evaluating the integrals,
\begin{align}
\left<\bm{m}_{\vz},\bm{m}_{\vz'}\right> \approx & \,\frac{C^2  \pi^{3/2} R^2 R_3}{h^2 h_3} 
\exp \left[ - \frac{|\bz-\bz'|^2}{4 R^2} - \frac{(z_3-z_3')^2}{4 R_3^2} \right].
\label{eq:ApE3}
\end{align}
Obviously, the maximum of \eqref{eq:ApE3} is attained at $\vz = \vz'$,
where
\[
\|\bm{m}_{\vz}\|_2^2 = \left<\bm{m}_{\vz},\bm{m}_{\vz}\right> \approx
\,\frac{C^2 \pi^{3/2} R^2 R_3}{h^2 h_3}.
\]
The result in Lemma \ref{lem.1} follows.  $\Box$

\subsection{The broadband regime}
 With the expression \eqref{eq:R8bb} of $m_{\vy,\vz}$ we get
\begin{align}
  \left<\bm{m}_{\vz},\bm{m}_{\vz'}\right> \approx &\,\frac{C^2}{h^2
    h_3} \int_{\mathbb{R}^2} d \by \, \frac{\exp \left[ -
      \frac{(|\by-\bz|^2 + |\by-\bz'|^2)}{2 R^2}\right]}{\left[1 +
      \frac{|\by-\bz|^2}{2 (\theta R)^2}\right]^{1/2} \left[1 +
      \frac{|\by-\bz'|^2}{2 (\theta R)^2}\right]^{1/2}}  \nonumber \\
  & \times
  \int_{-\infty}^\infty d y_3 \,
    \exp \left[  -
    \frac{ \left(y_3-z_3 + \frac{|\by|^2-|\bz|^2}{2 L} \right)^2}{2
      R_3^2 \left(1 + \frac{|\by-\bz|^2}{2 \theta^2 R^2}\right)} -
    \frac{ \left(y_3-z_3' + \frac{|\by|^2-|\bz'|^2}{2 L} \right)^2}{2
      R_3^2 \left(1 + \frac{|\by-\bz'|^2}{2 \theta^2 R^2}\right)}
    \right],
\end{align}
where we used the paraxial approximation and extended the integrals to
the whole space with negligible error\footnote{This is assuming that
  $\vz$ and $\vz'$ (i.e., the sources) are not near the edge of the
  imaging region.}, due to the Gaussians. Evaluating the integral over
$y_3$ and renaming the constant, we obtain
\begin{align}
\left<\bm{m}_{\vz},\bm{m}_{\vz'}\right> \approx & \,C R_3
\int_{\mathbb{R}^2} d \by \,
\frac{\exp \left[ -
\frac{(|\by-\bz|^2 + |\by-\bz'|^2)}{2 R^2} - \frac{\left(z_3-z_3' +
  \frac{|\bz|^2-|\bz'|^2}{2 L} \right)^2}{4 R_3^2 \left(1 +
  \frac{|\by-\bz|^2+|\by-\bz'|^2}{4 \theta^2 R^2}\right]}\right]}{\left(1 +
  \frac{|\by-\bz|^2+|\by-\bz'|^2}{4 \theta^2 R^2}\right)^{1/2}}.
\label{eq:ApE3bb}
\end{align}
Note that
\[
\frac{|\by-\bz|^2 + |\by-\bz'|^2}{2} = |\by -\cbz|^2 +
\frac{|\bz-\bz'|^2}{4},
\]
where $\cbz = (\bz+\bz')/2$. Substituting in \eqref{eq:ApE3} and
changing variables as $\bm{v} = {(\by-\cbz)}/{R},$
we get
\begin{align*}
\left<\bm{m}_{\vz},\bm{m}_{\vz'}\right> \approx & \,C R^2 R_3
e^{-\frac{|\bz-\bz'|^2}{4 R^2}} \int_{\mathbb{R}^2} d \bm{v}\,
\, \frac{\exp \left[ - |\bm{v}|^2 - \frac{\left(z_3-z_3' +
      \frac{|\bz|^2-|\bz'|^2}{2 L} \right)^2}{4 R_3^2 \left(1 +
      \frac{|\bz-\bz'|^2}{8 \theta^2 R^2} + \frac{|\bm{v}|^2}{2
        \theta^2}\right)}\right]}{\left(1 + \frac{|\bz-\bz'|^2}{8
    \theta^2 R^2} + \frac{|\bm{v}|^2}{2 \theta^2}\right)^{1/2}}.
\end{align*}
The integrand depends only on $v = |\bm{v}|$, so we can write the
integral in polar coordinates and obtain
\begin{align}
\left<\bm{m}_{\vz},\bm{m}_{\vz'}\right> \approx & \,2 \pi C R^2 R_3
e^{-\frac{|\bz-\bz'|^2}{4 R^2}} \int_{0}^\infty  d v\, v
\, \frac{\exp \left[ - v^2 - \frac{\left(z_3-z_3' +
      \frac{|\bz|^2-|\bz'|^2}{2 L} \right)^2}{4 R_3^2 \left(1 +
      \frac{|\bz-\bz'|^2}{8 \theta^2 R^2} + \frac{v^2}{2
        \theta^2}\right)}\right]}{\left(1 + \frac{|\bz-\bz'|^2}{8
    \theta^2 R^2} + \frac{v^2}{2 \theta^2}\right)^{1/2}}.
\label{eq:ApE6}
\end{align}
To eliminate the algebraic factors, let us change coordinates again
\[
t = \theta \left[2 + \frac{|\bz-\bz'|^2}{4
    \theta^2 R^2} + \frac{v^2}{ \theta^2}\right]^{1/2} ~ ~ \mbox{s.t.} ~ ~
dt = \frac{1}{\theta} \left[2 + \frac{|\bz-\bz'|^2}{4
    \theta^2 R^2} + \frac{v^2}{ \theta^2}\right]^{-1/2} v dv.
\]
Equation \eqref{eq:ApE6} becomes
\begin{align*}
\left<\bm{m}_{\vz},\bm{m}_{\vz'}\right> \approx & \,2 \pi \theta C R^2
R_3 e^{2 \theta^2} \int_{\sqrt{2
    \theta^2 + \frac{|\bz-\bz'|^2}{4 R^2}}}^\infty dt \,
\, \exp \left[ - \frac{\left(z_3-z_3' + \frac{|\bz|^2-|\bz'|^2}{2 L}
    \right)^2}{2 R_3^2 t^2} - t^2\right],
\end{align*}
with integral over $t$ evaluated below, in terms of the complementary
error function
\begin{align}
\left<\bm{m}_{\vz},\bm{m}_{\vz'}\right> \approx &
\,\frac{\pi^{3/2}}{4} \theta C R^2 R_3 e^{2 \theta^2} \left\{
e^{-\sqrt{2}\theta\left|\frac{z_3-z_3'}{R_3} +
  \frac{|\bz|^2-|\bz'|^2}{2 L R_3} \right|}\mbox{erfc} \left[ \sqrt{2
    \theta^2 + \frac{|\bz-\bz'|^2}{4 R^2}} - \frac{\theta
    \left|\frac{z_3-z_3'}{R_3} + \frac{|\bz|^2-|\bz'|^2}{2 L R_3}
    \right|}{\sqrt{2\left(2 \theta^2 + \frac{|\bz-\bz'|^2}{4
        R^2}\right)}}\right] \right. \nonumber \\
 &+ \left. e^{\sqrt{2}\theta\left|\frac{z_3-z_3'}{R_3} +
  \frac{|\bz|^2-|\bz'|^2}{2 L R_3} \right|}\mbox{erfc} \left[ \sqrt{2
    \theta^2 + \frac{|\bz-\bz'|^2}{4 R^2}} + \frac{\theta
    \left|\frac{z_3-z_3'}{R_3} + \frac{|\bz|^2-|\bz'|^2}{2 L R_3}
    \right|}{\sqrt{2\left(2 \theta^2 + \frac{|\bz-\bz'|^2}{4
        R^2}\right)}}\right]\right\}.
\label{eq:ApE8}
\end{align}

We are interested in the cross-correlation defined in \eqref{eq:Res2}.
The norm $\|\bm{m}_{\vz}\|_2$ is obtained by letting  $\vz = \vz'$ in
\eqref{eq:ApE8}, and the result is
\begin{align}
  \mathcal{I}_{\vz,\vz'} \approx & \, \frac{1}{2 \mbox{erfc}(\sqrt{2} \theta)}
    \left\{
e^{-\sqrt{2}\theta\left|\frac{z_3-z_3'}{R_3} +
  \frac{|\bz|^2-|\bz'|^2}{2 L R_3} \right|}\mbox{erfc} \left[ \sqrt{2
    \theta^2 + \frac{|\bz-\bz'|^2}{4 R^2}} - \frac{\theta
    \left|\frac{z_3-z_3'}{R_3} + \frac{|\bz|^2-|\bz'|^2}{2 L R_3}
    \right|}{\sqrt{2\left(2 \theta^2 + \frac{|\bz-\bz'|^2}{4
        R^2}\right)}}\right] \right. \nonumber \\
 &+ \left. e^{\sqrt{2}\theta\left|\frac{z_3-z_3'}{R_3} +
  \frac{|\bz|^2-|\bz'|^2}{2 L R_3} \right|}\mbox{erfc} \left[ \sqrt{2
    \theta^2 + \frac{|\bz-\bz'|^2}{4 R^2}} + \frac{\theta
    \left|\frac{z_3-z_3'}{R_3} + \frac{|\bz|^2-|\bz'|^2}{2 L R_3}
    \right|}{\sqrt{2\left(2 \theta^2 + \frac{|\bz-\bz'|^2}{4
        R^2}\right)}}\right]\right\}.
\label{eq:ApE9}
\end{align}
We can bound  the right hand side using the elementary inequality
$
\mbox{erfc}(t) \le e^{-t^2}$,  for all  $ t \ge 0.$
This gives
\begin{align}
 e^{\sqrt{2}\theta\left|\frac{z_3-z_3'}{R_3} +
   \frac{|\bz|^2-|\bz'|^2}{2 L R_3} \right|}&\mbox{erfc} \left[ \sqrt{2
     \theta^2 + \frac{|\bz-\bz'|^2}{4 R^2}} + \frac{\theta
     \left|\frac{z_3-z_3'}{R_3} + \frac{|\bz|^2-|\bz'|^2}{2 L R_3}
     \right|}{\sqrt{2\left(2 \theta^2 + \frac{|\bz-\bz'|^2}{4
         R^2}\right)}}\right] \nonumber \\
&\le  \exp \left[- 2 \theta^2 -
   \frac{|\bz-\bz'|^2}{4 R^2}-
   \frac{\theta^2\left(\frac{z_3-z_3'}{R_3} + \frac{|\bz|^2-|\bz'|^2}{2 L R_3}
     \right)^2}{2\left(2 \theta^2 + \frac{|\bz-\bz'|^2}{4
       R^2}\right)}\right] \nonumber \\
 &\le \exp \left[ -\theta^2 - \frac{|\bz-\bz'|^2}{8 R^2}-
  \theta \left|\frac{z_3-z_3'}{R_3} + \frac{|\bz|^2-|\bz'|^2}{2
      L R_3} \right|\right], \label{eq:ApE10}
\end{align}
where the last inequality is because
\begin{align*}
  \left[2 \theta^2 + \frac{|\bz-\bz'|^2}{4 R^2}+
  \frac{\theta^2\left(\frac{z_3-z_3'}{R_3} + \frac{|\bz|^2-|\bz'|^2}{2
      L R_3} \right)^2}{2\left(2 \theta^2 + \frac{|\bz-\bz'|^2}{4
      R^2}\right)}\right] -\left[ \theta^2 + \frac{|\bz-\bz'|^2}{8 R^2}+
  \theta \left| \frac{z_3-z_3'}{R_3} + \frac{|\bz|^2-|\bz'|^2}{2
      L R_3} \right|\right] = \nonumber \\
      \left[ \sqrt{\theta^2 + \frac{|\bz-\bz'|^2}{8 R^2}}- \frac{\theta \left| \frac{z_3-z_3'}{R_3} + \frac{|\bz|^2-|\bz'|^2}{2
      L R_3} \right|}{2 \sqrt{\theta^2 + \frac{|\bz-\bz'|^2}{8 R^2}} }\right]^2 \ge 0.
\end{align*}
For  the other term in \eqref{eq:ApE9} the bound is the same when the argument of the complementary 
error function is non-negative. If the argument is negative, then using that $\mbox{erfc}(t) \le 2$ for all
$t$, we get
\begin{align}
 e^{-\sqrt{2}\theta\left|\frac{z_3-z_3'}{R_3} +
   \frac{|\bz|^2-|\bz'|^2}{2 L R_3} \right|}&\mbox{erfc} \left[ \sqrt{2
     \theta^2 + \frac{|\bz-\bz'|^2}{4 R^2}} - \frac{\theta
     \left|\frac{z_3-z_3'}{R_3} + \frac{|\bz|^2-|\bz'|^2}{2 L R_3}
     \right|}{\sqrt{2\left(2 \theta^2 + \frac{|\bz-\bz'|^2}{4
         R^2}\right)}}\right] \le 2 e^{-\sqrt{2}\theta\left|\frac{z_3-z_3'}{R_3} +
   \frac{|\bz|^2-|\bz'|^2}{2 L R_3} \right|} \label{eq:ApE11}
\end{align}
But since in this case
\[
\theta
     \left|\frac{z_3-z_3'}{R_3} + \frac{|\bz|^2-|\bz'|^2}{2 L R_3} \right|> 2 \sqrt{2} \left(\theta^2 + \frac{|\bz-\bz'|^2}{8R^2}
     \right),
\]
we can write
\[
(\sqrt{2}-1) \theta  \left|\frac{z_3-z_3'}{R_3} + \frac{|\bz|^2-|\bz'|^2}{2 L R_3} \right| > 2\sqrt{2}(\sqrt{2}-1)
\left(\theta^2 + \frac{|\bz-\bz'|^2}{8R^2}
     \right) > \theta^2 + \frac{|\bz-\bz'|^2}{8R^2}.
\]
Substituting in \eqref{eq:ApE11} we get that
\begin{align}
 e^{-\sqrt{2}\theta\left|\frac{z_3-z_3'}{R_3} +
   \frac{|\bz|^2-|\bz'|^2}{2 L R_3} \right|}&\mbox{erfc} \left[ \sqrt{2
     \theta^2 + \frac{|\bz-\bz'|^2}{4 R^2}} - \frac{\theta
     \left|\frac{z_3-z_3'}{R_3} + \frac{|\bz|^2-|\bz'|^2}{2 L R_3}
     \right|}{\sqrt{2\left(2 \theta^2 + \frac{|\bz-\bz'|^2}{4
         R^2}\right)}}\right] \le \nonumber \\ &2 \exp \left[ -\theta^2 - \frac{|\bz-\bz'|^2}{8 R^2}-
  \theta \left|\frac{z_3-z_3'}{R_3} + \frac{|\bz|^2-|\bz'|^2}{2
      L R_3} \right|\right],
\end{align}
and using this result and \eqref{eq:ApE10} in \eqref{eq:ApE9} we get
\begin{align}
\mathcal{I}_{\vz,\vz'} \le \frac{3 e^{-\theta^2}}{2 \mbox{erfc}(\sqrt{2} \theta)} \exp \left[  - \frac{|\bz-\bz'|^2}{8 R^2}-
  \theta \left|\frac{z_3-z_3'}{R_3} + \frac{|\bz|^2-|\bz'|^2}{2
      L R_3} \right|\right].
\label{eq:ApE14}
\end{align}
This is the result in Lemma \ref{lem.1}. $\Box$


\bibliographystyle{siam} \bibliography{CINT.bib}

\begin{thebibliography}{10}

\bibitem{anitori2013design}
{\sc Laura Anitori, Ali Maleki, Matern Otten, Richard~G Baraniuk, and Peter
  Hoogeboom}, {\em Design and analysis of compressed sensing radar detectors},
  Signal Processing, IEEE Transactions on, 61 (2013), pp.~813--827.

\bibitem{arridge2009optical}
{\sc Simon~R Arridge and John~C Schotland}, {\em Optical tomography: forward
  and inverse problems}, Inverse Problems, 25 (2009), p.~123010.

\bibitem{beckers1993adaptive}
{\sc Jacques~M Beckers}, {\em Adaptive optics for astronomy-principles,
  performance, and applications}, Annual review of astronomy and astrophysics,
  31 (1993), pp.~13--62.

\bibitem{Biondi}
{\sc B.~Biondi}, {\em {3D seismic imaging}}, Society of Exploration
  Geophysicists, 2006.

\bibitem{bleistein2001mathematics}
{\sc N.~Bleistein, J.~K. Cohen, and J.J.W. Stockwell}, {\em Mathematics of
  multidimensional seismic imaging, migration, and inversion}, vol.~13,
  Springer, 2001.

\bibitem{borcea2011enhanced}
{\sc Liliana Borcea, Josselin Garnier, George Papanicolaou, and Chrysoula
  Tsogka}, {\em Enhanced statistical stability in coherent interferometric
  imaging}, Inverse problems, 27 (2011), p.~085004.

\bibitem{borcea2015resolution}
{\sc Liliana Borcea and Ilker Kocyigit}, {\em Resolution analysis of imaging
  with l1 optimization}, SIAM Journal on Imaging Sciences, 8 (2015),
  pp.~3015--3050.

\bibitem{borcea2016synthetic}
{\sc Liliana Borcea, Miguel Moscoso, George Papanicolaou, and Chrysoula
  Tsogka}, {\em Synthetic aperture imaging of direction-and frequency-dependent
  reflectivities}, SIAM Journal on Imaging Sciences, 9 (2016), pp.~52--81.

\bibitem{borcea2006adaptive}
{\sc Liliana Borcea, George Papanicolaou, and Chrysoula Tsogka}, {\em Adaptive
  interferometric imaging in clutter and optimal illumination}, Inverse
  Problems, 22 (2006), p.~1405.

\bibitem{borcea2007asymptotics}
\leavevmode\vrule height 2pt depth -1.6pt width 23pt, {\em Asymptotics for the
  space-time wigner transform with applications to imaging}, Stochastic
  Differential Equations: Theory and Applications (in Honor of Prof. Boris L.
  Rozovskii), Interdiscip. Math. Sci, 2 (2007), pp.~91--112.

\bibitem{candes2014towards}
{\sc Emmanuel~J Cand{\`e}s and Carlos Fernandez-Granda}, {\em Towards a
  mathematical theory of super-resolution}, Communications on Pure and Applied
  Mathematics, 67 (2014), pp.~906--956.

\bibitem{candes2013phaselift}
{\sc Emmanuel~J Candes, Thomas Strohmer, and Vladislav Voroninski}, {\em
  Phaselift: Exact and stable signal recovery from magnitude measurements via
  convex programming}, Communications on Pure and Applied Mathematics, 66
  (2013), pp.~1241--1274.

\bibitem{chai2010array}
{\sc Anwei Chai, Miguel Moscoso, and George Papanicolaou}, {\em Array imaging
  using intensity-only measurements}, Inverse Problems, 27 (2010), p.~015005.

\bibitem{chai2013robust}
{\sc A.~Chai, M.~Moscoso, and G.~Papanicolaou}, {\em Robust imaging of
  localized scatterers using the singular value decomposition and l1
  minimization}, Inverse Problems, 29 (2013), p.~025016.

\bibitem{Curlander}
{\sc John~C. Curlander and Robert~N. McDonough}, {\em Synthetic Aperture Radar:
  Systems and Signal Processing}, Wiley-Interscience, 1991.

\bibitem{cvx}
{\sc {CVX Research}}, {\em Cvx: matlab software for disciplined convex
  programming, version 2.0}.
\newblock {http://cvxr.com/cvx}{{\tt{http://cvxr.com/cvx}}}, August 2012.

\bibitem{devroye2006nonuniform}
{\sc Luc Devroye}, {\em Nonuniform random variate generation}, Handbooks in
  operations research and management science, 13 (2006), pp.~83--121.

\bibitem{fannjiang2010compressive}
{\sc A.~C. Fannjiang}, {\em {Compressive inverse scattering: I. High-frequency
  SIMO/MISO and MIMO measurements}}, Inverse Problems, 26 (2010), p.~035008.

\bibitem{fannjiang2010compressed}
{\sc A.~C. Fannjiang, T.~Strohmer, and P.~Yan}, {\em Compressed remote sensing
  of sparse objects}, SIAM Journal on Imaging Sciences, 3 (2010), pp.~595--618.

\bibitem{munson1983tomographic}
{\sc David~C Munson, James~Dennis O'Brien, and W~Kenneth Jenkins}, {\em A
  tomographic formulation of spotlight-mode synthetic aperture radar},
  Proceedings of the IEEE, 71 (1983), pp.~917--925.

\bibitem{papanicolaou2007self}
{\sc George Papanicolaou, Lenya Ryzhik, and Knut S{\o}lna}, {\em
  {Self-averaging from lateral diversity in the It{\^o}-Schr{\"o}dinger
  equation}}, Multiscale Modeling \& Simulation, 6 (2007), pp.~468--492.

\bibitem{rytov1989principle}
{\sc SM~Rytov, Yu~A Kravtsov, and VI~Tatarskii}, {\em Principle of statistical
  radiophysics iv: Wave propagation through random media. chapter 4}, 1989.

\bibitem{van1999multiple}
{\sc MCW~van van Rossum and Th~M Nieuwenhuizen}, {\em Multiple scattering of
  classical waves: microscopy, mesoscopy, and diffusion}, Reviews of Modern
  Physics, 71 (1999), p.~313.

\end{thebibliography}

\end{document}